\documentclass[11pt,prd,preprintnumbers,amsmath,amssymb,nofootinbib]{article}
\pdfoutput=1

\usepackage[utf8]{inputenc}
\usepackage{amsmath,amssymb,amscd}
\usepackage{listings}
\usepackage{caption}
\usepackage{dsfont}
\usepackage{slashed}
\usepackage{color}
\usepackage[caption=false]{subfig}
\usepackage[pdftex]{graphicx}
\usepackage{epstopdf}
\usepackage{listings}
\usepackage{epsfig}
\usepackage{listings}
\usepackage[font=small,labelfont=bf]{caption}
\usepackage{cite}
\usepackage{hyperref}
\usepackage{paralist}
\usepackage{placeins}
\usepackage{graphicx}
\usepackage{enumitem}
\usepackage{bm}
\usepackage{xcolor,colortbl}
\usepackage{multirow}

\usepackage{subfig}
\usepackage{color,graphicx,slashed,hyperref}
\usepackage[utf8]{inputenc}
\hypersetup{
   bookmarks=true,         
   unicode=true,          
   pdftoolbar=true,        
   pdfmenubar=true,        
   pdffitwindow=false,     
   pdfstartview={FitH},    
   pdftitle={Draft},    
   pdfauthor={Author},     
   pdfsubject={Subject},   
   pdfcreator={Creator},   
   pdfproducer={Producer}, 
   pdfkeywords={keyword1} {key2} {key3}, 
   pdfnewwindow=true,      
   colorlinks=true,       
   linkcolor=blue,          
   citecolor=blue,        
   filecolor=black,      
   urlcolor=blue           
}

\selectfont

\newcommand{\eq}[1]{Eq.~(\ref{#1})}

\newcommand{\beq} {\begin{equation}}
\newcommand{\eeq} {\end{equation}}
\newcommand{\bea} {\begin{eqnarray}}
\newcommand{\eea} {\end{eqnarray}}
\newcommand{\ba} {\begin{eqnarray*}}
\newcommand{\ea} {\end{eqnarray*}}

\definecolor{Celadon}{rgb}{0.67, 0.94, 0.82}
\definecolor{Pink}{rgb}{0.9, 0.0, 0.0}
\definecolor{darkred}{rgb}{0.5, 0.2, 0.13}
\definecolor{ForestGreen}{RGB}{34,139,34}

\renewenvironment{align}{
    \begin{equation}
    \begin{aligned}
}{
    \end{aligned}
    \end{equation}
    \ignorespacesafterend
}

\begin{document}

\thispagestyle{empty} 
\begin{flushright}
TTP21-007\\ 
P3H-21-017
\end{flushright}
\vspace*{0.5cm}\par
\begin{center}	
{\par\centering \textbf{  
\LARGE \bf Systematic approach to $B$-physics anomalies}} \\
\vskip .25cm\par
{\par\centering \textbf{
\LARGE  and $t$-channel dark matter}} \\
\vskip 1.cm\par
{\scalebox{.88}{\par\centering \large  
\sc Giorgio Arcadi$^{\,1,2}$, 
~Lorenzo Calibbi$^{\,3}$,  
~Marco Fedele$^{\,4}$, 
and Federico Mescia$^{\,5}$}} \\
{\par\centering \vskip 0.6 cm\par}
{\small
$^1$ \textit{Dipartimento di Matematica e Fisica, Universit\`a di Roma 3, Via della Vasca Navale 84, 00146, Roma, Italy}\\
$^2$ \textit{INFN Sezione Roma 3}\\
$^3$  \textit{School of Physics, Nankai University, Tianjin 300071, China} \\
$^4$  \textit{Institut f\"ur Theoretische Teilchenphysik, Karlsruhe Institute of Technology, D-76131 Karlsruhe, Germany} \\
$^5$  \textit{Dept.~de F\'{\i}sica Qu\`antica i Astrof\'{\i}sica, Institut de Ci\`encies del Cosmos (ICCUB), Universitat de Barcelona,\\Mart\'i i Franqu\`es 1, E-08028 Barcelona, Spain} } \\
{\vskip 1.cm\par}
\begin{abstract}
\noindent We study renormalisable models with minimal field content that can provide a viable
Dark Matter candidate through the standard freeze-out paradigm and, simultaneously, accommodate the observed anomalies in semileptonic $B$-meson decays at one loop.
Following the hypothesis of minimality, this outcome can be achieved by extending the particle spectrum of the Standard Model either with one vector-like fermion and two scalars or two vector-like fermions and one scalar. 
The Dark Matter annihilations are mediated by $t$-channel exchange of other new particles contributing to the $B$-anomalies, thus resulting in a correlation between flavour observables and Dark Matter abundance.
Again based on minimality, we assume the new states  to couple only with left-handed muons and second and third generation quarks. Besides an ad hoc symmetry needed to stabilise the Dark Matter, the interactions of the new states are dictated only by gauge invariance. We present here for the first time a systematic classification of the possible models of this kind, according to the quantum numbers of the new fields under the Standard Model gauge group. Within this general setup we identify a group of representative models that we systematically study, applying the most updated constraints from 
flavour observables, dedicated Dark Matter experiments,
and LHC searches of leptons and/or jets and missing energy, and of disappearing charged tracks.
\end{abstract}
{\vskip 3.cm\par}
\emph{E-mail:} 
\href{mailto:giorgio.arcadi@uniroma3.it}{giorgio.arcadi@uniroma3.it}, \href{mailto:calibbi@nankai.edu.cn}{calibbi@nankai.edu.cn}, \href{mailto:marco.fedele@kit.edu}{marco.fedele@kit.edu}, \href{mailto:mescia@ub.edu}{mescia@ub.edu} 
\end{center}

\newpage

\tableofcontents

\setcounter{footnote}{0}

\newenvironment{Appendix}
{
	\setcounter{section}{1}
	\setcounter{equation}{0}
	\renewcommand{\thesubsection}{\Alph{subsection}}
	\renewcommand{\theequation}{A.\arabic{equation}}
}


\section{Introduction}
The first decade of operation of the Large Hadron Collider (LHC) has resulted in the tremendous success represented by the discovery of the Higgs boson and provided us with a host of precise measurements and searches for new phenomena, finding no conclusive evidence of departures from the predictions of the Standard Model (SM).
Nevertheless, the SM leaves unanswered a number of  fundamental questions that provide strong motivation for its extension. The most compelling problem is that the SM lacks a candidate of dark matter~(DM), whose existence has been established by an impressive number of cosmological and astrophysical observations, spanning many orders of magnitude in redshift: from the Cosmic Microwave
Background~(CMB) to galactic rotation curves, see Ref.~\cite{Bertone:2016nfn} for a review. However, we do not have at the moment knowledge about the nature of dark matter nor about its mass and interactions with the SM sector, being all evidence based on its gravitational effects. All direct and indirect searches for particle dark matter have so far given negative results. Nevertheless, it is plausible that DM interacts to some extent to the SM fields, as a substantial DM abundance must be produced in the early universe. This is the case of the thermal freeze-out mechanism, which assumes that DM is a thermal relic, most commonly a weakly-interacting massive particle (WIMP). 
Sizeable interactions to SM particles are then required to keep DM in thermal equilibrium with the SM bath in the early universe and to ensure an efficient DM annihilation mechanism in order to avoid the WIMP relic density to be larger than the DM abundance that is observed today. In this work, we are going to assume that the observed DM density is accounted for by a thermal WIMP that interacts with SM quarks and leptons and other extra fields in a way that can address the so-called $B$-physics (or flavour) anomalies.

In fact, although direct searches performed by the LHC collaborations for the production of new particles have found no evidence of their existence, several experimental collaborations, with LHCb being the prominent one, have brought to light a persistent and coherent pattern of deviations from the SM predictions in semileptonic decays of $B$ mesons of the kind $b\to s\ell^+\ell^-$. This could very well be the first experimental hint for beyond the SM~(BSM) physics at energies not much larger than the electroweak scale.
In particular, LHCb and $B$-factory experiments observe a deviation from Lepton Flavour Universality~(LFU) predicted by the SM in the theoretically clean observables $R_{K^{(*)}}\equiv {\rm BR}\left(B \to K^{(*)} \mu^+\mu^-\right) /{ {\rm BR}\left(B \to K^{(*)}e^+e^-\right)}$~\cite{Aaij:2019wad,Aaij:2017vbb,Abdesselam:2019wac}. Moreover, a number of measurements
are in tension with the SM predictions for the branching ratios and angular distributions of several $b\to s\mu^+\mu^-$ modes~\cite{Aaij:2015oid,Aaij:2015dea,Aaij:2016flj,Wehle:2016yoi,Aaboud:2018krd,Khachatryan:2015isa,Sirunyan:2017dhj,Aaij:2020nrf,Aaij:2020ruw,Aaij:2015esa}.
All these anomalies could be explained by a deficit of 
$b\to s\mu^+\mu^-$ events compared to SM expectations due to the interference between SM and BSM amplitudes. The simplest way to achieve such an effect is to add non-standard contributions, $\delta C^{9,10}_\mu$, to the following operators 
\begin{align}
\label{eq:Heff}
\mathcal{H}_{\rm eff} \supset -\frac{4 G_F}{\sqrt{2}} \frac{e^2}{16\pi^2}V_{tb} V_{ts}^* \left[C^9_\mu \,(\overline{s}\gamma_\mu P_L b) (\overline{\mu} \gamma^\mu\mu)+ C^{10}_\mu\, (\overline{s}\gamma_\mu P_L b) (\overline{\mu} \gamma^\mu\gamma_5 \mu)	  +{\rm h.c.} \right].
\end{align}
While not providing the absolute the best fit to the anomalies, an interesting scenario, still in excellent agreement with the data, is represented by $\delta C^{9}_\mu=-\delta C^{10}_\mu \approx -0.5$, corresponding to the case of only left-handed (LH) currents entering Eq.~\eqref{eq:Heff}. According to global fits to $B$-physics data, such a scenario is preferred to the SM prediction at the $\sim 5\sigma$ level~\cite{Descotes-Genon:2013wba,Altmannshofer:2013foa,Ghosh:2014awa,DAmico:2017mtc,Ciuchini:2019usw,Alguero:2019ptt,Alok:2019ufo,Datta:2019zca,Aebischer:2019mlg,Kowalska:2019ley,Ciuchini:2020gvn,Hurth:2020ehu}.
This does not reflect of course an established breakdown of the SM: a combination of overlooked systematics, statistical fluctuations, and underestimated hadronic uncertainties could conspire to account for such a large deviation from the SM in the global fit. Nevertheless it is tempting to explore new physics~(NP) scenarios that could explain the anomalies and assess their capability of addressing other shortcomings of the SM, in particular the DM problem.  

In this paper we systematically build a set of simplified models that can explain the $B$-physics anomalies and
simultaneously provide a good DM candidate, 
and we study their phenomenology with a particular focus on the LHC limits on production of new heavy particles and the bounds from direct- and indirect-detection DM searches.
Our aim is to highlight the \emph{minimal} building blocks that a more complete theory may need to include. For the sake of minimality we are going to employ the following procedure.
\begin{figure}[!t]
\centering
\includegraphics[width=0.90\textwidth]{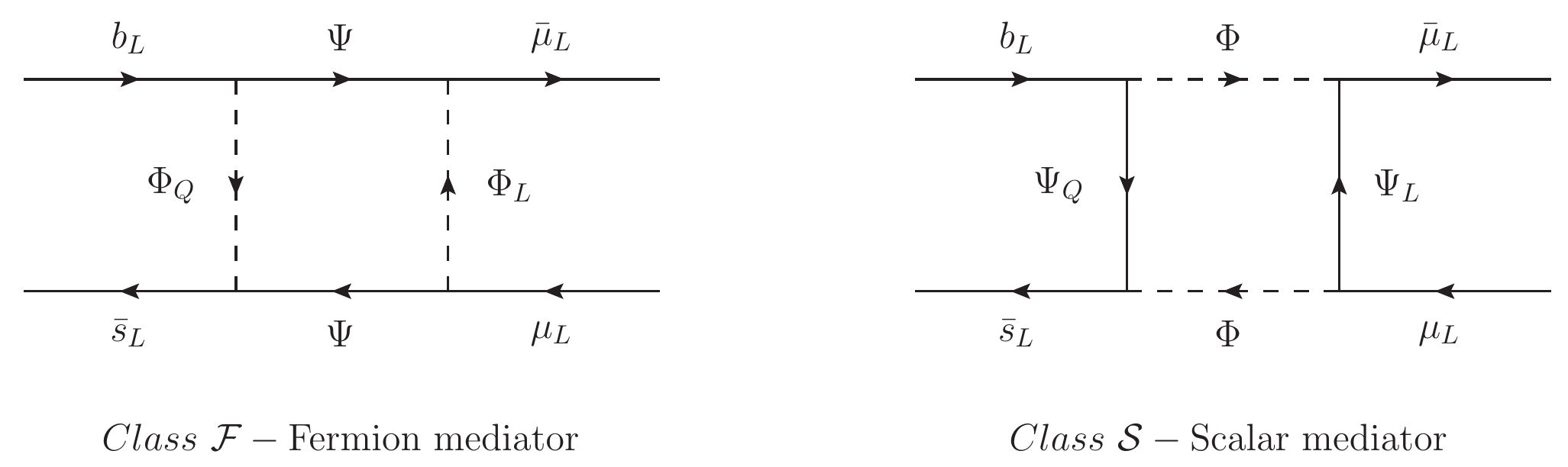}
\caption{Basic diagrams contributing to $b\to s\mu\mu$.
\label{fig:boxes}}
\end{figure}
\begin{itemize}
    \item We focus on minimal solutions of the $B$-physics anomalies of the $\delta C^{9}_\mu=-\delta C^{10}_\mu$ kind, hence we only introduce new fields\,---\,in the lowest possible number\,---\,that couple to left-handed quarks and leptons (the SM $SU(2)_L$ doublets).
    \item We require that at least one of the BSM fields contains a state which can be a good DM candidate, i.e.~neutral and colour singlet.
    \item We assume that DM stability is ensured by an unbroken symmetry (a ${\bf Z}_2$ parity or another global symmetry), which forbids interactions between a NP particle and two SM particles, as well as mixing between NP and SM fields. As a consequence the BSM contributions to $C^{9}_\mu$ and $C^{10}_\mu$ can only arise through one-loop diagrams like those shown in Figure~\ref{fig:boxes} with only BSM fields running in the loop, as in the framework studied in Refs.~\cite{Gripaios:2015gra,Arnan:2016cpy,Arnan:2019uhr}. Notice that for simplicity we only consider spin 0 and spin 1/2 fields and that only three new fields need to be added to the SM.
    \item $SU(3)_c\otimes SU(2)_L\otimes U(1)_Y$ gauge invariance and the requirement of a consistent DM candidate tightly constrain the possible quantum numbers of the BSM fields. Furthermore, imposing the predicted relic density to be at (or below) the observed value $\Omega_\text{DM}h^2\simeq 0.12$ results in non-trivial conditions on the spectrum and couplings of the new particles, such that DM efficiently annihilates into SM particles. An unavoidable 
    annihilation mode is given by the \emph{$t$-channel} exchange of 
    the other fields entering the loop of Figure~\ref{fig:boxes}, possibly alongside coannihilations and processes involving gauge interactions (if DM belongs to a non-trivial representation of the electroweak gauge group), see 
    Figure~\ref{fig:DMann}.
\end{itemize}

\begin{figure}[!t]
\centering
\includegraphics[width=0.98\textwidth]{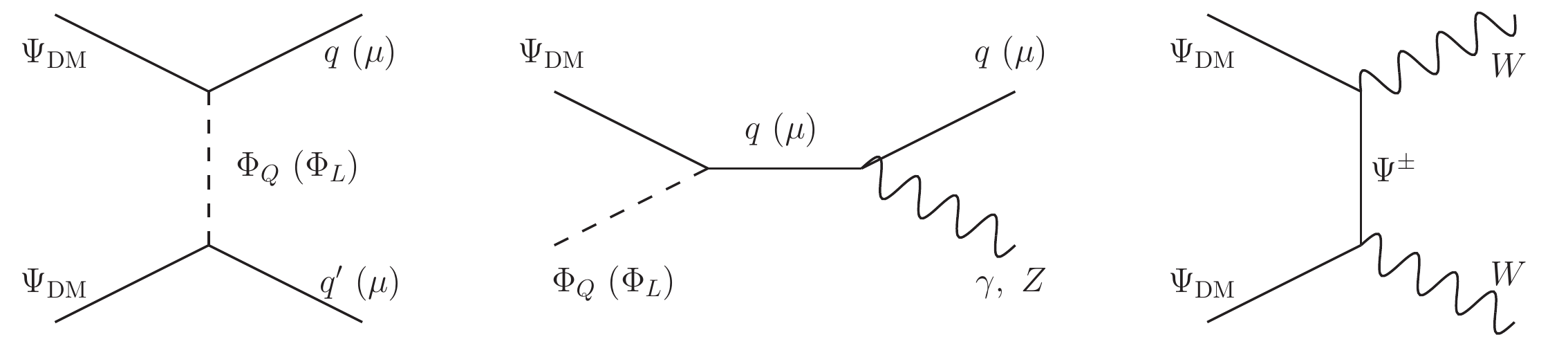}
\caption{Illustrative DM (co-)annihilation diagrams for the case of fermion DM belonging to the field denoted as $\Psi$ in Figure~\ref{fig:boxes}. Analogous diagrams arise in the other cases. Gauge diagrams such as the third one are only present if DM belongs to an $SU(2)_L$ multiplet ($\Psi^\pm$ are charged states in the same multiplet).
\label{fig:DMann}}
\end{figure}

To the best of our knowledge, this is the first systematic study of the connection between flavour anomalies and dark matter. However, several previous works in the literature proposed specific models that fulfill the conditions outlined above, which then will be included in our classification, see Refs.~\cite{Kawamura:2017ecz,Cline:2017qqu,Barman:2018jhz,Grinstein:2018fgb,Cerdeno:2019vpd,Huang:2020ris}. Other works that addressed simultaneously DM and the $B$-physics anomalies (among other observables) include Refs.~\cite{Sierra:2015fma,Belanger:2015nma,Celis:2016ayl,Altmannshofer:2016jzy,Ko:2017quv,Ko:2017yrd,Cline:2017lvv,Baek:2017sew,Cline:2017aed,Falkowski:2018dsl,Arcadi:2018tly,Baek:2018aru,Vicente:2018frk,Baek:2019qte,Calibbi:2019bay,Darme:2020hpo,Kachanovich:2020yhi,Guadagnoli:2020tlx,Borah:2020swo,Carvunis:2020exc,Chao:2021qxq}.

The outline of the paper is the following. 
In Section~\ref{sec:setup} we define our setup and show the set of minimal models that follows from the approach described above. In Section~\ref{sec:strategy} we describe our strategy and how we impose constraints from $B$-physics, LHC searches, and DM phenomenology. In Section~\ref{sec:results} we choose a number of representative models, for which the results of our analysis are presented in detail. Finally we summarise and conclude in Section~\ref{sec:conclusions}.

\begin{table}[t]
\centering
\renewcommand{\arraystretch}{1.1}
\begin{tabular}{ | c | c  c  c |  }
\hline
${SU\left( 3 \right)_c}$ & ${{\Phi_Q},{\Psi_Q}}$ & ${{\Phi_L},{\Psi_L}}$ & ${\Psi ,\Phi }$ \\
\hline
  A &\bf 3&\bf 1&\bf 1\\
  B &\bf 1& ${\bf \bar 3}$&\bf 3\\
\hline
\hline
${SU\left( 2 \right)_L}$ & ${{\Phi _Q},{\Psi _Q}}$ & ${{\Phi_L},{\Psi_L}}$ & $ {\Psi ,\Phi}$ \\
\hline
I    & \bf 2 &\bf 2 &\bf 1\\
II   & \bf 1 &\bf 1 &\bf 2\\
III  & \bf 3 &\bf 3 &\bf 2\\
IV   & \bf 2 &\bf 2 &\bf 3\\
V    & \bf 3 &\bf 1 &\bf 2\\
VI   & \bf 1 &\bf 3 &\bf 2\\
\hline
\hline
$U(1)_Y$ & ${{\Phi _Q},{\Psi _Q}}$ & ${{\Phi_L},{\Psi_L}}$ & $ {\Psi ,\Phi}$ \\
\hline
    & $1/6-X$ & $-1/2-X$ & $X$ \\
\hline  
\end{tabular}
\caption{Possible gauge quantum numbers of the new fields appearing in the loop diagrams in Figure~\ref{fig:boxes}. The displayed $SU(3)_c$ representations are all the possible allowing a DM candidate, while only $SU(2)_L$ representations with $d\le 3$ are shown.
\label{tab:reps}}
\end{table}

\section{Setup}
\label{sec:setup}
As discussed above, we consider models that can give rise to the contributions to $b\to s\mu\mu$ processes shown in Figure~\ref{fig:boxes}.\footnote{If the mediator field is a real scalar singlet or a Majorana fermion singlet, additional crossed box diagrams must be taken into account, see \cite{Arnan:2016cpy,Arnan:2019uhr}.} 
We classify our models in two classes according to the spin of the field that couple to both quarks and leptons\,---\,which we call ``flavour mediator'' independently of its quantum numbers\,---\,as follows.
\paragraph{Class $\mathcal{F}$.} These models feature a vector-like fermion $\Psi$ as flavour mediator and two extra scalars $\Phi_Q$ and $\Phi_L$ coupling to the SM left-handed fermions with interactions described by the following Lagrangian:
\begin{eqnarray}
{\cal L}_\mathcal{F}~ \supset ~ \Gamma^Q_i\,\bar Q_i\, P_R \Psi \, \Phi_Q  + \Gamma^L_i \,\bar L_i\, P_R \Psi \, \Phi_L  +
{\rm{h}}{\rm{.c.}}\,.
\label{eq:Linta}
\end{eqnarray}
\paragraph{Class $\mathcal{S}$.} In these models, we introduce a scalar flavour mediator $\Phi$ and two fermions $\Psi_Q$ and $\Psi_L$ in vector-like representations of the SM gauge group:
\begin{eqnarray}
{\cal L}_\mathcal{S} ~\supset ~ \Gamma^Q_i\, \bar Q_i \,P_R \Psi_Q \, \Phi + \Gamma^L_i\, \bar L_i\, P_R \Psi_L \, \Phi  +
{\rm{h}}{\rm{.c.}}\,.
\label{eq:Lintb}
\end{eqnarray}

\medskip
\noindent 
In the spirit of our simplified-model approach, we are considering non-zero couplings of the new fields only to second and third generation left-handed quarks ($\Gamma^Q_2$, $\Gamma^Q_3$) and muons ($\Gamma^L_2$). For more definiteness we will use, throughout the paper, the following notation: $\Gamma_2^Q=\Gamma_s^Q$, $\Gamma_3^Q=\Gamma_b^Q$ and $\Gamma_2^L=\Gamma_\mu^L$. These couplings are defined in the basis where the down-quark and charged-lepton mass matrices are flavour diagonal. 
Furthermore we assume a global symmetry, whose effect is 
to forbid mixing between extra fields and SM fields and ensure that the lightest new state is stable. This can be achieved by introducing an unbroken $\mathbf Z_2$ parity under which the SM fields are even and the new fields are odd, or an equivalent continuous symmetry.
Finally, unless otherwise stated, we will usually assume the interactions in the scalar potential\,---\,such as the unavoidable quartic couplings between our new scalars $\Phi_X$ and the Higgs field $H$ of the form $\Phi_X^\dag \Phi_X H^\dag H$\,---\,be small enough to have only subdominant effects on the phenomenology of our models.
 
\begin{table}[t!]
\centering
\renewcommand{\arraystretch}{1.1}
\begin{tabular}{ | c | c  c  c |  }
\hline
Label & $\Phi _Q$ & $\Phi_L$ & $\Psi$ \\
\hline
$\mathcal{F}_\text{IA;\,-1}$
& $({\bf 3},{\bf 2},7/6)$ & $({\bf 1},{\bf 2},1/2)^\star$ & $({\bf 1},{\bf 1},-1)$ \\
\rowcolor{cyan!40}
$\mathcal{F}_\text{IA;\,0}$
& $({\bf 3},{\bf 2},1/6)$ & $({\bf 1},{\bf 2},-1/2)^\star$ & $({\bf 1},{\bf 1},0)^\star$ \\
\hline
\rowcolor{cyan!40}
$\mathcal{F}_\text{IB;\,-1/3}$
& $({\bf 1},{\bf 2},1/2)^\star$ & $({\bf \bar 3},{\bf 2},-1/6)$ & $({\bf 3},{\bf 1},-1/3)$ \\
$\mathcal{F}_\text{IB;\,2/3}$
& $({\bf 1},{\bf 2},-1/2)^\star$ & $({\bf \bar 3},{\bf 2},-7/6)$
& $({\bf 3},{\bf 1},2/3)$ \\
\hline
$\mathcal{F}_\text{IIA}$
& $({\bf 3},{\bf 1},2/3)$ & $({\bf 1},{\bf 1},0)^\star$ & $({\bf 1},{\bf 2},-1/2)$ \\
\hline
$\mathcal{F}_\text{IIB}$  
& $({\bf 1},{\bf 1},0)^\star$ & $({\bf \bar 3},{\bf 1},-2/3)$ & $({\bf 3},{\bf 2},1/6)$ \\
\hline
$\mathcal{F}_\text{IIIA;\,-3/2}$  
& $({\bf 3},{\bf 3},5/3)$ & $({\bf 1},{\bf 3},1)^\star$ & $({\bf 1},{\bf 2},-3/2)$ \\
\rowcolor{cyan!40}
$\mathcal{F}_\text{IIIA;\,-1/2}$
& $({\bf 3},{\bf 3},2/3)$ & $({\bf 1},{\bf 3},0)^\star$ & $({\bf 1},{\bf 2},-1/2)$ \\
$\mathcal{F}_\text{IIIA;\,1/2}$
& $({\bf 3},{\bf 3},-1/3)$ & $({\bf 1},{\bf 3},-1)^\star$ & $({\bf 1},{\bf 2},1/2)$ \\
\hline
$\mathcal{F}_\text{IIIB;\,-5/6}$
& $({\bf 1},{\bf 3},1)^\star$ & $({\bf \bar 3},{\bf 3},1/3)$ & $({\bf 3},{\bf 2},-5/6)$ \\
$\mathcal{F}_\text{IIIB;\,1/6}$
& $({\bf 1},{\bf 3},0)^\star$ & $({\bf \bar 3},{\bf 3},-2/3)$ & $({\bf 3},{\bf 2},1/6)$ \\
$\mathcal{F}_\text{IIIB;\,7/6}$
& $({\bf 1},{\bf 3},-1)^\star$ & $({\bf \bar 3},{\bf 3},-5/3)$ & $({\bf 3},{\bf 2},7/6)$ \\
\hline
$\mathcal{F}_\text{IVA;\,-1}$  
& $({\bf 3},{\bf 2},7/6)$ & $({\bf 1},{\bf 2},1/2)^\star$ & $({\bf 1},{\bf 3},-1)$ \\
$\mathcal{F}_\text{IVA;\,0}$  
& $({\bf 3},{\bf 2},1/6)$ & $({\bf 1},{\bf 2},-1/2)^\star$ & $({\bf 1},{\bf 3},0)^\star$ \\
\hline
$\mathcal{F}_\text{IVB;\,-1/3}$  
& $({\bf 1},{\bf 2},1/2)^\star$ & $({\bf \bar 3},{\bf 2},-1/6)$ & $({\bf 3},{\bf 3},-1/3)$ \\
$\mathcal{F}_\text{IVB;\,2/3}$  
& $({\bf 1},{\bf 2},-1/2)^\star$ & $({\bf \bar 3},{\bf 2},-7/6)$
& $({\bf 3},{\bf 3},2/3)$ \\
\hline
$\mathcal{F}_\text{VA}$  
& $({\bf 3},{\bf 3},2/3)$ & $({\bf 1},{\bf 1},0)^\star$ & $({\bf 1},{\bf 2},-1/2)$ \\
\hline
$\mathcal{F}_\text{VB;\,-5/6}$  
& $({\bf 1},{\bf 3},1)^\star$ & $({\bf \bar 3},{\bf 1},1/3)$ & $({\bf 3},{\bf 2},-5/6)$ \\
$\mathcal{F}_\text{VB;\,1/6}$  
& $({\bf 1},{\bf 3},0)^\star$ & $({\bf \bar 3},{\bf 1},-2/3)$ & $({\bf 3},{\bf 2},1/6)$ \\
$\mathcal{F}_\text{VB;\,7/6}$  
& $({\bf 1},{\bf 3},-1)^\star$ & $({\bf \bar 3},{\bf 1},-5/3)$ & $({\bf 3},{\bf 2},7/6)$ \\
\hline
$\mathcal{F}_\text{VIA;\,-3/2}$  
& $({\bf 3},{\bf 1},5/3)$ & $({\bf 1},{\bf 3},1)^\star$ & $({\bf 1},{\bf 2},-3/2)$ \\
$\mathcal{F}_\text{VIA;\,-1/2}$  
& $({\bf 3},{\bf 1},2/3)$ & $({\bf 1},{\bf 3},0)^\star$ & $({\bf 1},{\bf 2},-1/2)$ \\
$\mathcal{F}_\text{VIA;\,1/2}$  
& $({\bf 3},{\bf 1},-1/3)$ & $({\bf 1},{\bf 3},-1)^\star$ & $({\bf 1},{\bf 2},1/2)$ \\
\hline
$\mathcal{F}_\text{VIB}$  
& $({\bf 1},{\bf 1},0)^\star$ & $({\bf \bar 3},{\bf 3},-2/3)$ & $({\bf 3},{\bf 2},1/6)$ \\
\hline
\end{tabular}
\caption{Models with a fermion flavour mediator (Class $\mathcal{F}$). The fields are denoted by their transformation properties under, respectively, ($SU(3)_c$, $SU(2)_L$, $U(1)_Y$). We highlight in cyan the models that we study in detail in Section~\ref{sec:results}. \label{tab:fmodels}}
\end{table}
 
\begin{table}[!th]
\centering
\renewcommand{\arraystretch}{1.1}
\begin{tabular}{ | c | c  c  c |  }
\hline
Label & $\Psi _Q$ & $\Psi_L$ & $\Phi$ \\
\hline
 \rowcolor{cyan!40}
 $\mathcal{S}_\text{IA}$  
 & $({\bf 3},{\bf 2},1/6)$ & $({\bf 1},{\bf 2},-1/2)$ & $({\bf 1},{\bf 1},0)^\star$ \\
\hline
$\mathcal{S}_\text{IIA;\,-1/2}$  
& $({\bf 3},{\bf 1},2/3)$ & $({\bf 1},{\bf 1},0)^\star$ & $({\bf 1},{\bf 2},-1/2)^\star$ \\
$\mathcal{S}_\text{IIA;\,1/2}$  
  & $({\bf 3},{\bf 1},-1/3)$ & $({\bf 1},{\bf 1},-1)$ & $({\bf 1},{\bf 2},1/2)^\star$ \\
\hline
\rowcolor{cyan!40}
$\mathcal{S}_\text{IIB}$  
& $({\bf 1},{\bf 1},0)^\star$ & $({\bf \bar 3},{\bf 1},-2/3)$ & $({\bf 3},{\bf 2},1/6)$ \\
\hline
\rowcolor{cyan!40}
$\mathcal{S}_\text{IIIA;\,-1/2}$  
&  $({\bf 3},{\bf 3},2/3)$ & $({\bf 1},{\bf 3},0)^\star$ & $({\bf 1},{\bf 2},-1/2)^\star$ \\
$\mathcal{S}_\text{IIIA;\,1/2}$  
& $({\bf 3},{\bf 3},-1/3)$ & $({\bf 1},{\bf 3},-1)$ & $({\bf 1},{\bf 2},1/2)^\star$ \\
\hline
$\mathcal{S}_\text{IIIB}$  
&  $({\bf 1},{\bf 3},0)^\star$ & $({\bf \bar 3},{\bf 3},-2/3)$ & $({\bf 3},{\bf 2},1/6)$ \\
\hline
$\mathcal{S}_\text{IVA;\,-1}$  
& $({\bf 3},{\bf 2},7/6)$ & $({\bf 1},{\bf 2},1/2)$ & $({\bf 1},{\bf 3},-1)^\star$ \\
$\mathcal{S}_\text{IVA;\,0}$  
& $({\bf 3},{\bf 2},1/6)$ & $({\bf 1},{\bf 2},-1/2)$ & $({\bf 1},{\bf 3},0)^\star$ \\
$\mathcal{S}_\text{IVA;\,1}$  
& $({\bf 3},{\bf 2},-5/6)$ & $({\bf 1},{\bf 2},-3/2)$ & $({\bf 1},{\bf 3},1)^\star$ \\
\hline
$\mathcal{S}_\text{VA;\,-1/2}$  
& $({\bf 3},{\bf 3},2/3)$ & $({\bf 1},{\bf 1},0)^\star$ & $({\bf 1},{\bf 2},-1/2)^\star$ \\
$\mathcal{S}_\text{VA;\,1/2}$  
& $({\bf 3},{\bf 3},-1/3)$ & $({\bf 1},{\bf 1},-1)$ & $({\bf 1},{\bf 2},1/2)^\star$ \\
\hline
$\mathcal{S}_\text{VB}$  
& $({\bf 1},{\bf 3},0)^\star$ & $({\bf \bar 3},{\bf 1},-2/3)$ & $({\bf 3},{\bf 2},1/6)$ \\
\hline
$\mathcal{S}_\text{VIA;\,-1/2}$  
& $({\bf 3},{\bf 1},2/3)$ & $({\bf 1},{\bf 3},0)^\star$ & $({\bf 1},{\bf 2},-1/2)^\star$ \\
$\mathcal{S}_\text{VIA;\,1/2}$  
& $({\bf 3},{\bf 1},-1/3)$ & $({\bf 1},{\bf 3},-1)$ & $({\bf 1},{\bf 2},1/2)^\star$ \\
\hline
$\mathcal{S}_\text{VIB}$  
& $({\bf 1},{\bf 1},0)^\star$ & $({\bf \bar 3},{\bf 3},-2/3)$ & $({\bf 3},{\bf 2},1/6)$ \\
\hline
\end{tabular}
\caption{Same as Table~\ref{tab:fmodels} for models with a scalar flavour mediator (Class $\mathcal{S}$). \label{tab:smodels}}
\end{table}
 
The possible gauge quantum numbers of the extra fields follow from the requirement of gauge invariance of the above Lagrangians and the additional condition that at least one component is uncoloured and electrically neutral, so to provide a viable DM candidate. 
Considering that the involved SM fields only are quark doublets $Q$, whose quantum numbers under $SU(3)_c\otimes SU(2)_L \otimes U(1)_Y$ are $({\mathbf 3},{\mathbf 2},1/6)$, and lepton doublets $L$, $({\mathbf 1},{\mathbf 2},-1/2)$, the new fields have to belong to the representations of $SU(3)_c$ and $SU(2)_L$ displayed in Table~\ref{tab:reps}.
Notice that, while all the possible representations of $SU(3)_c$ are listed in the table~\ref{tab:reps} (as combinations involving larger representations would not feature any colour singlet), in the case of $SU(2)_L$ only representations with dimension $d\le 3$ are displayed.
The hypercharge assignment is in general not unique but, as mentioned above, it is restricted by the requirement that at least one state is neutral, i.e.~$Q = T_3+ Y = 0$. Setting the hypercharge of the flavour mediator ($\Psi$ or $\Phi$) as a free parameter $X$, the hypercharge of other fields are then  derived from gauge invariance as shown in the last line of Table~\ref{tab:reps}.

The resulting combinations of quantum numbers are shown in Tables~\ref{tab:fmodels} and~\ref{tab:smodels} that collect the possible models with, respectively, a fermion and a scalar flavour mediator. 
The models have been labelled according to the spin of the flavour mediator ($\mathcal F$ or $\mathcal S$), the combination of $SU(2)_L$ and $SU(2)_c$ representations of the fields as given in Table~\ref{tab:reps} and\,---\,for categories containing more than one model\,---\,the hypercharge of the flavour mediator.
The tables include only models featuring at least a viable DM candidate, i.e.~an electrically neutral stable state. The representation to which this state belongs have been marked with~$^\star$. Notice that the DM candidate can belong to any of the three NP fields for both classes of models.

DM candidates with non-zero hypercharge are severely constrained by the direct detection experiments, as consequence of the coherently enhanced Spin Independent interactions with nuclei mediated by the $Z$ boson.
To keep  the  particle content of our models minimal, we only consider fermion DM candidates with $Y=0$.\footnote{
Scenarios with Dirac fermion DM and $Y\neq 0$  would be still viable if the field content of the model is extended beyond our minimality criterion, such that the DM field mixes
with an additional Majorana fermion  making the lightest state Majorana, see e.g.~Ref.~\cite{Cirelli:2005uq,Hisano:2011cs}. This is the well-known case of the supersymmetric Higgsinos, $SU(2)_L$ doublet fermions mixing with a Majorana singlet (the Bino) and a Majorana triplet (the Wino).} 
Instead, if DM belongs to a scalar multiplet one can evade the direct detection bounds by introducing a suitable  mass splitting between CP-odd and CP-even components through couplings of the scalar potential. Since this avoids the dangerous DM coupling to the $Z$ without introducing more fields, we include in Tables~\ref{tab:fmodels} and~\ref{tab:smodels} also solutions with DM belonging to scalar multiplets with $Y\neq 0$. 
Let us mention that some of the models included in the classification shown in Tables~\ref{tab:fmodels} and~\ref{tab:smodels} were previously studied in the literature, namely 
$\mathcal{F}_\text{IA;\,0}$~\cite{Cerdeno:2019vpd},
$\mathcal{F}_\text{IIA}$~\cite{Huang:2020ris}, $\mathcal{S}_\text{IA}$~\cite{Kawamura:2017ecz,Grinstein:2018fgb}, $\mathcal{S}_\text{IIA;\,-1/2}$~\cite{Cline:2017qqu} and
$\mathcal{S}_\text{IIA;\,1/2}$~\cite{Barman:2018jhz}.
 
In the following sections, we will discuss in details the different constraints (from flavour anomalies, LHC, relic density and direct detection) for the above-defined class of models.


\section{Strategy}
\label{sec:strategy}
In this section we present the strategy that we are going to employ for each model under scrutiny. We take into account constraints coming from flavour physics, from the recasting of direct searches at the LHC and from DM searches. All these constraints are considered in a systematic and comprehensive way in order to assess whether, for a given model, a region of the parameter space where all bounds are evaded exists, or the model is excluded by the combination of all the constraints. We remind that the parameter space of each model is fully determined by three couplings plus three mass parameters, namely
\beq
\label{eq:parameters}
\{\Gamma^Q_b\,, \Gamma^Q_s\,, \Gamma_\mu^L\} \quad\text{plus} \quad \{M_\Psi\,, M_{\Phi_Q}\,, M_{\Phi_L}\}\quad \text{or plus} \quad \{ M_\Phi\,, M_{\Psi_Q}\,, M_{\Psi_L}\}
\eeq
for  models with a fermion or a scalar flavour mediator, respectively.

\subsection{Fit to flavour anomalies}
\label{sec:fit}
The first step of our strategy consists in considering the constraints coming from flavour physics. In particular, we take into account data coming from the ratios $R_K$~\cite{Aaij:2019wad} and $R_{K^*}$~\cite{Aaij:2017vbb,Abdesselam:2019wac}\,---\,which are tests of LFU violation (LFUV)\,---\,from the angular analyses of the semi-leptonic decays $B\to K^{(*)}\ell^+\ell^-$~\cite{Aaij:2015oid,Aaij:2015dea,Aaij:2016flj,Wehle:2016yoi,Aaboud:2018krd,Khachatryan:2015isa,Sirunyan:2017dhj,Aaij:2020nrf,Aaij:2020ruw} and $B_s \to \phi \, \ell^+ \ell^-$~\cite{Aaij:2015esa}, from the branching ratios of the fully-leptonic decays $B_s\to \ell^+\ell^-$~\cite{Aaij:2017vad,Chatrchyan:2013bka,Aaboud:2018mst,Amhis:2016xyh} and from the branching ratios of the radiative decays $B \to K^* \gamma$~\cite{Amhis:2016xyh}, $B \to \phi \gamma$~\cite{Aaij:2012ita} and $B\to X_s\gamma$~\cite{Amhis:2016xyh}. In order to systematically consider all the above experimental data, we directly rely on the constraints on the NP contributions to the Wilson coefficients (WC) of the operators in Eq.~\eqref{eq:Heff} obtained by global fits performed on this full data set, see e.g.~Refs.~\cite{DAmico:2017mtc,Ciuchini:2019usw,Alguero:2019ptt,Alok:2019ufo,Datta:2019zca,Aebischer:2019mlg,Kowalska:2019ley,Ciuchini:2020gvn,Hurth:2020ehu}. As reference value, we will employ the $2\sigma$-range obtained in the fit in Ref.~\cite{Ciuchini:2020gvn}:
\beq
\label{eq:c9mc10}
-0.72 \leq \delta C^9_\mu = -\delta C^{10}_\mu \leq -0.36\,, \quad \mathrm{(at\, 2\sigma)}\,.
\eeq
Moreover, we also take into account the latest SM  results~\cite{DiLuzio:2019jyq,DiLuzio:2017fdq} on the mass difference of the neutral $B_s$ mesons $\Delta M_s$~\cite{Amhis:2016xyh}: 
\beq
\label{eq:dms}
\frac{\Delta M_s^{\rm SM}}{\Delta M_s^{\rm exp}} = 1.04^{+0.04}_{-0.07} \,.
\eeq
In a similar fashion to what we did for $b\to s \ell\ell$ transitions, we can interpret the experimental data in terms of a bound to NP effects to a WC: introducing the effective $\Delta B=2$ Hamiltonian
\beq
\mathcal{H}_{\rm eff}^{B \bar B} \supset C^{B\bar B} (\overline{s}\gamma_\mu P_L b)(\overline{s}\gamma_\mu P_L b)\,,
\eeq
we can observe that the relation between the NP contribution to $C^{B\bar B}$ and $\Delta M_s$ reads
\beq
\label{eq:cBB}
\frac{C^{B\bar B}}{C^{B\bar B} + \delta C^{B\bar B}} = \frac{\Delta M_s^{\rm SM}}{\Delta M_s^{\rm exp}} = 1.04^{+0.04}_{-0.07}\,.
\eeq
In order to cast these constraints on the models described in the previous section, we remind that for each of the two classes of models defined in Eqs.~\eqref{eq:Linta}-\eqref{eq:Lintb} it is possible to write~\cite{Arnan:2016cpy}:
\begin{equation}
\begin{aligned}
\left(\delta C^9_\mu\right)_\mathcal{F} = -\left(\delta C^{10}_\mu\right)_\mathcal{F} &= \frac{\sqrt2}{4 G_F V_{tb}V_{ts}^*} \frac{\Gamma_Q |\Gamma_\mu^L|^2}{32\pi \alpha_{\rm EM} M_\Psi^2} \left( \eta F\left( {x_Q,x_L}\right) + 2 \chi^M\eta^M G\left( {x_Q,x_L}\right)  \right)\,,\\
\left(\delta C^9_\mu\right)_\mathcal{S} = -\left(\delta C^{10}_\mu\right)_\mathcal{S} &= -\frac{\sqrt2}{4 G_F V_{tb}V_{ts}^*}\frac{\Gamma_Q |\Gamma_\mu^L|^2}{32\pi \alpha_{\rm EM} M_\Phi^2} \left( \eta - \chi^M\eta^M\right) F\left( {y_Q,y_L} \right)\,,
\end{aligned}
\label{eq:C9_I,II}
\end{equation} 
and
\begin{equation}
\begin{aligned}
\left(\delta C^{B\bar B}\right)_\mathcal{F} &= \frac{\Gamma_Q^2}{128\pi^2 M_\Psi^2} \left( \eta_{BB} F\left( {x_Q,x_L}\right) + 2 \chi^M\eta^M G\left( {x_Q,x_L}\right)  \right)\,,\\
\left(\delta C^{B\bar B}\right)_\mathcal{S} &= \frac{\Gamma_Q^2}{128\pi^2 M_\Phi^2} \left( \eta_{BB} - \chi^M\eta^M\right) F\left( {y_Q,y_L} \right)\,,
\end{aligned}
\label{eq:CBB_I,II}
\end{equation}
where we have defined $\Gamma_Q \equiv \Gamma^Q_b \Gamma^{Q\ast}_s$ and we have introduced the compact notation $x_Q= M_{\Phi_Q}^2/M_\Psi^2$, $x_L=M_{\Phi_L}^2/M_\Psi^2$ and $y_Q=M_{\Psi_Q}^2/M_\Phi^2$, $y_L=M_{\Psi_L}^2/M_\Phi^2$, respectively. The $SU(2)_L$-factors $\eta$, $\eta^M$ and $\eta_{BB}$ are tabulated in Table~\ref{tab:eta}, while the $SU(3)_c$-factor $\chi^M$ is equal to 1 only if $\Psi$ ($\Phi$) is a Majorana fermion (real scalar) in representation A of Table~\ref{tab:reps}, vanishing otherwise. 
\begin{table}[t]
\centering
\renewcommand{\arraystretch}{1.2}
\begin{tabular}{ | c | c  c  c  c  c  c |  }
\hline
${SU\left( 2 \right)_L}$ & I & II & III & IV & V & VI \\
\hline
$\eta$ & 1 & 1 & 5/16 & 5/16 & 1/4 & 1/4\\
$\eta^M$ & 1 & 0 & 0 & 1/16 & 0 & 0\\
$\eta_{BB}$ & 1 & 1 & 5/16 & 5/16 & 5/16 & 1\\
\hline  
\end{tabular}
\caption{Table of the $SU(2)_L$-factors entering the Wilson coefficients in Eqs.~\eqref{eq:C9_I,II},\eqref{eq:CBB_I,II}.
\label{tab:eta}}
\end{table}
Moreover, we have introduced the following loop functions
\begin{equation}
\begin{aligned}
F(x,y) &= 
\frac{1}{(1-x)(1-y)} + \frac{x^2 \log{x}}{(1-x)^2(x-y)} +\frac{y^2  \log{y}}{(1-y)^2(y-x)} \,,\\
G(x,y) &= \frac{1}{(1-x)(1-y)} + \frac{x \log{x}}{(1-x)^2(x-y)} +\frac{y \log{y}}{(1-y)^2(y-x)} \,.
\end{aligned}
\label{eq:FG}
\end{equation}

The constraints in Eqs.~\eqref{eq:c9mc10} and \eqref{eq:cBB} are exploited through a combined fit to the relevant set of parameters in Eq.~\eqref{eq:parameters} using the \texttt{HEPfit} package~\cite{deBlas:2019okz}. For each parameter, we adopt a flat prior with the following ranges:
\begin{itemize}
\item For the lepton coupling, we allow it to be in the range $\vert \Gamma_\mu^L \vert \in [0,4]$. Regarding the quark coupling $\Gamma_Q$ we first notice from Eq.~\eqref{eq:C9_I,II} that, since the loop functions assume positive values for the mass regimes under scrutiny, it is the only free parameter capable to affect the sign of the Wilson coefficient; hence, in order to obtain the desired sign for $\delta C^9_\mu$ according to Eq.~\eqref{eq:c9mc10} and remembering that $\text{Re}(V_{ts})<0$, we allow it to be in the range $\Gamma_Q \in [0,2]$ for models in class $\mathcal{F}$, and to be in the range $\Gamma_Q \in [-2,0]$ for models in class $\mathcal{S}$; this choice will not affect the result for $\delta C^{B\bar B}$, since the coupling is squared in Eq.~\eqref{eq:CBB_I,II};
\item For the DM candidate mass, we require for it to be lighter than the other NP fields;
\item For the remaining NP masses, we let them vary up to 5 TeV.
\end{itemize}
\begin{figure}[!t]
\centering
\includegraphics[width=0.78\textwidth]{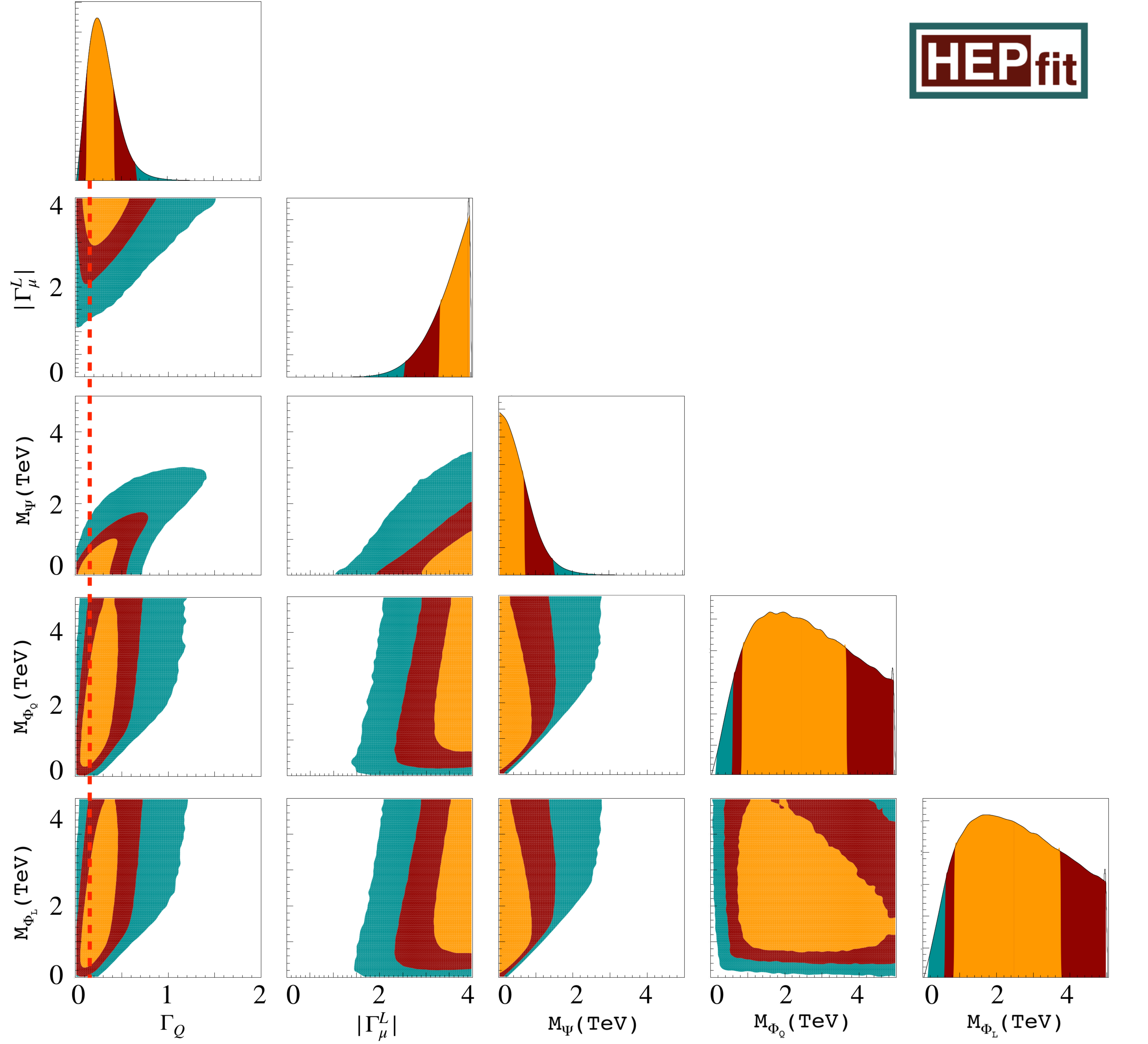}
\caption{
Results of the flavour fit for the parameters $\Gamma_Q \equiv \Gamma^Q_b \Gamma^{Q\ast}_s$, $\vert\Gamma_\mu^L\vert$, $M_\Psi$, $M_{\Phi_Q}$ and $M_{\Phi_L}$ in the scenario $\mathcal{F}_\text{IA;\,0}$ with Majorana DM. Both the 1D distribution for each of the 5 parameters and the 2D correlations between each of them are shown (with 1$\sigma$, 2$\sigma$ and 3$\sigma$ regions in orange, red and blue respectively). The red dashed line correspond to the benchmark value $\tilde\Gamma_Q = 0.15$.
\label{fig:F_IA_M_flavour}}
\end{figure}
The outcome of the fit is then summarised in terms of posterior probability distribution functions (\emph{p.d.f.}) for each parameter, together with correlation plots between each pair of them.

As an example, we show in Figure~\ref{fig:F_IA_M_flavour} the outcome of the fit to the model $\mathcal{F}_\text{IA;\,0}$, where the DM candidate is the Majorana fermion $\Psi$. The diagonal of the triangle plot contains the 1D \emph{p.d.f.}s for the 5 parameters constrained by the model, while in the other panels we give the combined 2D \emph{p.d.f.}s showing the correlations among each couple of parameters. Let us now discuss the information we can extract from this result, since it will also be useful for the other models we consider in Section~\ref{sec:results}. Starting from the couplings 1D \emph{p.d.f.}s, we observe that $\Gamma_Q$ is constrained to small values: this is a byproduct of the inclusion of $\Delta M_s$ in the fit, which disfavours higher values for this coupling. However, this implies that $\vert\Gamma_\mu^L\vert$ is pushed to values at the boundary of the perturbative region in order to satisfactory address the $B$-anomalies. Concerning the masses 1D \emph{p.d.f.}s, we first notice that the DM mass is allowed to grow only up to a few TeV, due to the requirement that it has to be the lightest state of the NP sector. On the other hand, no upper bound can be inferred on the masses of the other NP fields, which are found to be unconstrained in the whole mass range under scrutiny. Moving on to the analysis of the combined 2D \emph{p.d.f.}s, we focus our attention on the first column of panels of Figure~\ref{fig:F_IA_M_flavour}, namely the ones showing the correlations among $\Gamma_Q$ and the other NP parameters included in the fit, where we have highlighted with a dashed red line the value $\Gamma_Q \sim 0.15$. From these panels it is indeed possible to infer that such value, sitting in the 1D \emph{p.d.f.} of $\Gamma_Q$ in the 1$\sigma$ region and close to its mode, is an interesting benchmark point when looking also at the other parameters. Indeed, from the 2D \emph{p.d.f.} describing the correlation among $\Gamma_Q$ and $\vert\Gamma_\mu^L\vert$, we observe that this choice for the quark coupling implies that the 1$\sigma$ and 2$\sigma$ regions reach the lowest allowed values for the lepton coupling, corresponding to $\vert\Gamma_\mu^L\vert$ = 3 and 2, respectively, with the 3$\sigma$ bound reaching $\vert\Gamma_\mu^L\vert$ = 1. Hence, the benchmark value $\Gamma_Q \sim 0.15$ is the one that better justifies the benchmark assignments $\vert\Gamma_\mu^L\vert = 1,2,3$, while a different value for the quark coupling would induce bigger values for the leptonic one. Finally, from the other 2D \emph{p.d.f.}s we observe that in correspondence to this benchmark value the 1$\sigma$ (2$\sigma$) upper bound for $M_{\rm DM}$ is around 700 (1100) GeV, while for the other two fields the whole mass range is allowed at the 2$\sigma$ level.

An analogous behaviour for the posterior \emph{p.d.f.}s has been found in all the models analysed in Section~\ref{sec:results}, with a value for $\vert\Gamma_Q\vert \sim 0.15$ (positive or negative, depending on whether the model belongs to class $\mathcal{F}$ or $\mathcal{S}$ respectively) such that the allowed 1$\sigma$ and 2$\sigma$ bounds on $\vert\Gamma_\mu^L\vert$ are the lowest, the 1$\sigma$ and 2$\sigma$ upper bounds for the DM mass are around 700 GeV and 1100 GeV respectively, and no upper bound is found at the 2$\sigma$ level on the remaining masses. Therefore, in the following analysis we will fix the value for the quark coupling to the benchmark value $\vert\tilde\Gamma_Q\vert = 0.15$, and we will consider 3 different benchmark values for the leptonic coupling $\vert\Gamma_\mu^L\vert = 1,2,3$, corresponding to the 3$\sigma$, 2$\sigma$ and 1$\sigma$ lower bounds, respectively. Moreover, we observe that the fit always allows to fix the mass of one the two heavy NP partners of the DM candidate to be equal to 700 GeV (1100 GeV), value corresponding to the 1$\sigma$ (2$\sigma$) upper bound for the DM (which being the lightest NP particle obviously sets a lower bound on the rest of the spectrum).

Let us finish this subsection with a comment on another observable featuring a long-standing deviation from the SM prediction: the muon anomalous magnetic moment, $(g-2)_\mu$. 
Notice that the subset of our NP fields coupling to muons indeed contributes to the dipole operator responsible for $(g-2)_\mu$. Such a contribution is, however, chirally suppressed (our fields do not couple to right-handed muons), thus too small to account for the current $\sim 3.5\,\sigma$ discrepancy between SM prediction and experimental measurement~\cite{Aoyama:2020ynm}, with the possible exception of tuned regions of the paramater space, as systematically discussed in~\cite{Calibbi:2018rzv}. A successful fit of the $(g-2)_\mu$ discrepancy would require the introduction of extra fields coupling to the SM Higgs field and thus providing a chiral enhancement to the muon dipole.
Examples of such ``next-to-minimal'' dark matter scenarios naturally accounting for the $(g-2)_\mu$ discrepancy are given in~\cite{Kowalska:2017iqv,Calibbi:2018rzv} as well as, in combination with the $B$-physics anomalies, in~\cite{Calibbi:2019bay,Darme:2020hpo}.

\subsection{LHC constraints}
\label{sec:lhc}
The second step of our strategy involves studying the bounds following from direct searches at the LHC. We will consider here as an example, without any loss of generality, the case where the flavour mediator is a scalar field (being the conclusions in the case of a  fermion mediator identical). DM scenarios and LHC searches can be separated in two categories, depending on whether the DM candidate is the fermion field coupling to SM leptons ($\Psi_L$), or one of the two other possibilities ($\Psi_Q$ or $\Phi$).

\noindent\paragraph{$\bullet$ Case 1.} 
If the DM candidate is either the field $\Psi_Q$ or $\Phi$, the main production channel at the LHC will be via QCD-mediated processes, such as
\beq\label{eq:qq_decaychannel}
pp \to \Psi_Q \Psi_Q \to  qq^\prime +  \slashed{E}_T\,,
\qquad \text{or} \qquad
pp \to \Phi\Phi \to qq^\prime+  \slashed{E}_T\,,
\eeq 
with DM being the field $\Phi$ or $\Psi_Q$, respectively, and appearing as missing energy in the detector. Here $q,q^\prime=s,c,b,t$, that is, each of the quarks can be either a light quark, producing a jet, a bottom quark, producing a $b$-jet, or a top quark, producing a $b$-jet plus the products of the decay of the $W$ boson.\footnote{Here and below, we are neglecting the production of $u$ quarks, which are CKM-suppressed.} Notice the similarity of this setup to simplified supersymmetric (SUSY) models with squarks decaying into neutralino DM. As a consequence, in order to produce exclusion plots in the $(M_{\Psi_Q},M_\Phi)$ plane, it is possible to recast limits from LHC squark searches involving missing energy plus 2-6 jets, allowing also the presence of $b$-jets. The latest of such analyses
has been performed by CMS~\cite{Sirunyan:2019ctn}. Let us stress once again that this recasting can be applied no matter which of the two NP fields in the couple $\{\Psi_Q,\Phi\}$ is the DM candidate, since the final states and hence the LHC signature is the same for both cases, as shown in Eq.~\eqref{eq:qq_decaychannel}.

In order to set bounds on $M_{\Psi_L}$,  we now have to distinguish whether the DM candidate is the scalar $\Phi$ or the fermion $\Psi_Q$ coupling to SM quarks. In the former case, one can constrain $M_{\Psi_L}$ with a procedure analogous to the one outlined above: a pair of $\Psi_L$ fields will be mainly produced by electroweak Drell-Yann, and they will subsequently decay into muons and DM, i.e.~missing energy. Hence, this channel can be described as 
\beq\label{eq:mumu_decaychannel}
pp \to \Psi_L\Psi_L \to \mu^+\mu^- +  \slashed{E}_T\,.
\eeq 
Therefore, similarly to what described above, in order to produce exclusion plots in the plane $(M_{\Psi_L},M_\Phi)$ it is possible to recast limits from LHC slepton searches involving missing energy plus a muon pair. The latest of such analyses has been performed by ATLAS~\cite{Aad:2019vnb}. Moreover, there are also searches considering soft leptons performed by both ATLAS~\cite{Aad:2019qnd} and CMS~\cite{Sirunyan:2018iwl}. 
In the second case (DM in $\Psi_Q$), on the other hand, $\Psi_L$ does not couple directly to the DM candidate, hence the above signature cannot be used. The main decay channel of the $\Psi_L$ fields will consist instead in the cascade production of two muons plus two NP scalars $\Phi$, which will further decay into quarks and DM candidates, i.e.~missing energy. Hence, in this last scenario the signature will be 
\beq\label{eq:mumuqq_decaychannel}
pp \to \Psi_L\Psi_L \to \mu^+\mu^-+\,\Phi\Phi \to  \mu^+\mu^-+ qq^\prime +  \slashed{E}_T\,.
\eeq 
Therefore, in order to produce exclusion plots in the plane $(M_{\Psi_L},M_{\Psi_Q})$, it is necessary to recast limits from combined LHC searches for stops cascade decaying into sleptons/charginos, involving missing energy plus 2 muons and 2-6 jets (including $b$-jets). The latest of such analyses can be found in~\cite{Sirunyan:2020tyy,ATLAS:2020dav}.

\noindent
\paragraph{$\bullet$ Case 2.}
If the DM candidate is the field $\Psi_L$, analogous considerations to the ones reported above Eq.~\eqref{eq:mumu_decaychannel}, can be applied, with the signature reading now 
\beq
\label{eq:mumu_decaychannel2}
pp \to \Phi\Phi \to \mu^+\mu^- +  \slashed{E}_T\,.
\eeq 
One can therefore use again the searches from Refs.~\cite{Aad:2019vnb,Aad:2019qnd,Sirunyan:2018iwl} to constrain the plane $(M_\Phi,M_{\Psi_L})$. Further, bounds on $M_{\Psi_Q}$ can be obtained through the cascade decay
\beq
\label{eq:mumuqq_decaychannel2}
pp \to \Psi_Q\Psi_Q \to qq^\prime +\,\Phi\Phi \to  qq^\prime +  \mu^+\mu^- + \slashed{E}_T\,.
\eeq 
Hence, exclusion plots in the plane $(M_{\Psi_Q},M_{\Psi_L})$ can be obtained again by means of the analyses in~\cite{Sirunyan:2020tyy,ATLAS:2020dav}.
\smallskip

While some combination of the above production and decay modes will appear in all models under study, there is a further collider signature that is  possible if DM is part of an $SU(2)_L$ multiplet, where charged states will be also present. Due to electroweak radiative
corrections the charged states are typically $\mathcal{O}(100)$~MeV heavier than the neutral DM state~\cite{Cirelli:2005uq,Yamada:2009ve,Ibe:2012sx}.
Because of such a small mass splitting, the charged states\,---\,that can be pair produced through electroweak Drell-Yann at the LHC\,---\,are long-lived, that is, they can travel a macroscopic distance (typically a few cm) in the detector before decaying (through the exchange of an off-shell $W$ boson) into DM and a very soft and thus undetectable pion. The resulting signature is a so-called ``disappearing track'' observable in the inner tracker of the detector. Searches for this kind of events have been performed by both ATLAS~\cite{Aaboud:2017mpt} and CMS~\cite{Sirunyan:2020pjd}. 

In order to exploit the LHC searches discussed above\footnote{The following procedure applies to the ``prompt'' signatures of Eqs.~\eqref{eq:qq_decaychannel}-\eqref{eq:mumu_decaychannel2}, while we will assess the bounds from disappearing tracks from existing literature, as explained in the following section for the models where these searches are relevant.} we have implemented, as first step, the models  in \texttt{Feynrules~2.3}~\cite{Alloul:2013bka} in order to generate UFO files, which are subsequently passed to \texttt{MadGraph5\_aMC@NLO~2.7}\cite{Alwall:2014hca} where the matrix elements are calculated and a set of 50k events is generated. The partonic events are showered using \texttt{Pythia 8}~\cite{Sjostrand:2014zea}, the detector effects are simulated by means of \texttt{DELPHES 3}~\cite{deFavereau:2013fsa} and the result is eventually passed to \texttt{CheckMATE~2}~\cite{Dercks:2016npn}, which compares the simulated signal with the experimental searches at the LHC and determines whether the model point is excluded at the $90\%$ confidence level. For the LHC searches selected above that are not yet implemented in the current version of \texttt{CheckMATE}, we employ the \texttt{AnalysisManager} framework in order to define them ourselves~\cite{Kim:2015wza}.

As said above, such procedure determines whether a model is excluded at the $90\%$ confidence level for a fixed value of the 6 parameters $\bar\Gamma^Q_b\,, \bar\Gamma^Q_s\,, \bar\Gamma_\mu^L\,, \bar M_\Phi\,, \bar M_{\Psi_Q}$ and $\bar M_{\Psi_L}$. This means that, in order to produce an exclusion plot in the $(M_{\Psi_Q},M_\Phi)$ plane (or in the $(M_{\Psi_L},M_\Phi)$ one, if $\Psi_L$ is the DM candidate), we have to fix the values of the 3 couplings and of the remaining mass to some set values. The chosen values for these 4 parameters will be guided by the benchmark values inferred from the flavour fit performed in the previous section. Regarding the lepton coupling, we can directly adopt one of the 3 benchmark values inferred from the flavour fit, i.e.~$|\Gamma_\mu^L| = 1,2,3$. However, the same cannot be done for the quark couplings since the flavour fit gave a benchmark value for the product $\vert\Gamma^Q_b \Gamma^{Q\ast}_s\vert$, i.e. $\vert\tilde\Gamma_Q\vert = 0.15$. On the other hand, LHC constraints are sensitive to the individual values to the two quark couplings and not only to their product. Hence, in the following we will inspect for each model 2 different benchmark cases:
\beq\label{eq:casesAB}
i)\quad\vert\bar\Gamma^Q_b\vert = 1\,,~\vert\bar\Gamma^Q_s\vert = \vert\tilde\Gamma_Q\vert\,, \qquad \qquad  
ii)\quad\vert\bar\Gamma^Q_b\vert = \vert\bar\Gamma^Q_s\vert = \sqrt{\vert\tilde\Gamma_Q\vert}\,.
\eeq

We are now left to fix the value of one of the 3 masses. In the case of the DM being $\Psi_L$, we will fix the value of the mass of $\Psi_Q$, which is a coloured particle not directly coupling to the DM candidate. The produced signature will be the cascade decay shown in Eq.~\eqref{eq:mumuqq_decaychannel2}, and hence we will have to rely on the recasting from Refs.~\cite{Sirunyan:2020tyy,ATLAS:2020dav}. We infer that (unless $M_{\Psi_Q}-M_{\Psi_L} \lesssim 50$ GeV) we have to consider as a benchmark value $M_{\Psi_Q} = 1400$ GeV. This is indeed the lowest allowed value for this parameter such that no lower bound is induced on the DM mass. In a similar fashion, if the DM candidate is $\Psi_Q$ we will fix the mass of the coloured field $\Psi_L$. Once again, given the signature of Eq.~\eqref{eq:mumuqq_decaychannel} we employ the recasting from Refs.~\cite{Sirunyan:2020tyy,ATLAS:2020dav} and obtain also for this field the benchmark value $M_{\Psi_L} = 1400$ GeV. Finally, if the flavour mediator $\Phi$ is the DM particle, the particle whose mass we will fix is yet again $\Psi_L$, but this field is now a colour singlet directly coupling to the DM candidate. We can therefore directly employ the recasting from Refs.~\cite{Aad:2019vnb,Aad:2019qnd,Sirunyan:2018iwl}, and observe that we can set the mass of $\Psi_L$ to the one of the 2 benchmark values inferred from the flavour fit of the previous section, i.e.~$M_{\Psi_L}$=700 GeV or 1100 GeV. Indeed, the highest excluded value from the recasting of such results is always found to be at most $\mathcal{O}(600)$~GeV.

Summarizing, given the above benchmark values for 4 parameters, it is possible to combine either in the $(M_{\Psi_Q},M_\Phi)$ plane or in the $(M_{\Psi_L},M_\Phi)$ one (according to whether the DM candidate is $\Psi_Q$ or $\Phi$, in the former case, or $\Psi_L$, in the latter). Indeed, the WC $\delta C^9_\mu = -\delta C^{10}_\mu$ is simply a function of the 2 masses, once all the other parameters are fixed to benchmark values. Following the procedure outlined above it is therefore possible to visualise on the same plane both the region that can account for the flavour anomalies at the 2$\sigma$ level as in Eq.~\eqref{eq:c9mc10}, and the one excluded at 90\% CL by our recasting of direct searches at LHC.


\subsection{Constraints from DM phenomenology}
\label{sec:DM}
As a last step, the flavour and LHC constraints  can be complemented with the ones from DM phenomenology. Concerning the latter, the models considered here belong to the category of the so-called $t$-channel portals \cite{Bai:2013iqa,Bai:2014osa,Arcadi:2017kky,Mohan:2019zrk,Arina:2020udz,Arina:2020tuw} (a ``flavoured'' variant of this kind of setup has been also considered here \cite{Agrawal:2011ze,Agrawal:2014aoa,Blanke:2017tnb,Blanke:2020bsf,Liu:2021crr}). Our assumption that a good fit of the $B$-anomalies is achieved introduces, however, some relevant variation in the phenomenology of this kind of models, especially for what concerns direct detection.

As will be shown in the following, being some model parameters fixed by the fit of flavour observables, the latter constraints can be easily visualised into two-dimensional mass-mass plots to compare the corresponding viable regions with those fulfilling the requirements from both LHC and flavour physics.

\subsubsection{Relic density}
\label{sec:relic}
The DM relic abundance has been measured with great precision by the Planck experiment \cite{Aghanim:2018eyx} and it is represented by the parameter $\Omega_{\rm DM}h^2$ whose value is:
\beq\label{eq:DMOmega}
\Omega_{\rm DM} h^2 = 0.1199 \pm 0.0022\,.
\eeq
As will be evident in the next sections, for values of the couplings compatible with flavour anomalies, the DM is capable of reaching thermal equilibrium in the Early Universe and, hence, can achieve its final relic density through the freeze-out mechanism. In such a case, the DM abundance is the solution of a Boltzmann equation which can be written as \cite{Gondolo:1990dk,Edsjo:1997bg}:
\begin{equation}
\label{eq:relic_thermal}
\Omega_{\rm DM}h^2 \approx 8.76 \times 10^{-11}\,{\mbox{GeV}}^{-2}{\left[\int_{T_{\rm f.o.}}^{T_0} g_{*}^{1/2} \langle \sigma v \rangle_{\rm eff}\frac{dT}{M_{\rm DM}}\right]}^{-1}\,,
\end{equation}
where $\langle \sigma v \rangle_{\rm eff}$ is the effective  thermally-averaged DM annihilation cross-section including coannihilations and $g_*$ is the effective number of relativistic
degrees of freedom. The integral is computed between the freeze-out temperature $T_{\rm f.o.}$ (we remind that for WIMP $T_{\rm f.o.} \sim \frac{M_{\rm DM}}{20}-\frac{M_{\rm DM}}{30}$) and the current temperature of the universe $T_0$. 
Defining $M$ the field that is $t$-channel exchanged in DM pair annihilations and also contributes to co-annihilation processes, the DM effective annihilation cross-section can be written as \cite{Bai:2013iqa}:
\begin{align}
\langle \sigma v \rangle_{\rm eff}~=~ &\frac{1}{2}\langle \sigma v \rangle_{\rm DM\, DM}\frac{g_{\rm DM}^2}{g_{\rm eff}^2}+\langle \sigma v \rangle_{\rm DM\, M}\frac{g_{\rm DM} g_{\rm M}}{g^2_{\rm eff}}{\left(1+\Delta\right)}^{3/2} \exp\left[-x \Delta \right]+ \\
& \frac{1}{2}\langle \sigma v \rangle_{\rm M^{\dagger}M}\frac{g_{\rm M}^2}{g_{\rm eff}^2}{\left(1+\Delta\right)}^3 \exp\left[-2 x \Delta \right]\,,
\end{align}
where $x=M_{\rm DM}/T$. $\langle \sigma v \rangle_{\rm DM\,DM}$ describes DM pair annihilation processes into SM fermions mediated by $t$-channel exchange of the field $M$. Given the assumptions stated in the previous sections, the possible final states are $\mu^+ \mu^{-}$, and $\bar q q^\prime$ where $q,q^\prime=s,c,b,t$.\footnote{Up quarks can be as well annihilation final states. Their contribution to the DM relic density is, however, negligible because of the CKM suppression of the couplings.} $\langle \sigma v \rangle_{\rm DM\, M}$ represents coannihilation processes with a DM and a $M$ particle in the initial states while $\langle \sigma v \rangle_{\rm M^\dagger M}$ describes the contribution of $M$ pair annihilation processes to the DM effective annihilation cross-section provided that the mass splitting between DM and $M$ is sufficiently small. 
Notice that the expression above is valid for complex scalar and Dirac fermion DM. In the case of real scalar and Majorana DM we have a slightly different expression:
\begin{align}
\langle \sigma v \rangle_{\rm eff} 
~=~ & \langle \sigma v \rangle_{\rm DM \,DM}\frac{g_{\rm DM}^2}{g_{\rm eff}^2}+\langle \sigma v \rangle_{\rm DM\, M}\frac{g_{\rm DM} g_{\rm M}}{g^2_{\rm eff}}{\left(1+\Delta\right)}^{3/2} \exp\left[-x \Delta \right] +
\nonumber \\
& \left(\langle \sigma v \rangle_{\rm M^{\dagger}M}+\langle \sigma v \rangle_{\rm M\,M}\right)\frac{g_{\rm M}^2}{g_{\rm eff}^2}{\left(1+\Delta\right)}^3 \exp\left[-2 x \Delta \right]\,.
\end{align}
In the above equations $\Delta={(M_{\rm M}-M_{\rm DM})}/{M_{\rm DM}}$ is the relative DM/mediator mass splitting while:
\begin{equation}
g_{\rm eff}=g_{\rm DM}+g_{\rm M}{\left(1+\Delta\right)}^{3/2}\exp\left[-x \Delta \right]\,,
\end{equation}
where $g_{\rm M}$ and $g_{\rm DM}$ are the internal degrees of freedom of the mediator and of the DM.

In our analysis, the DM relic density, including coannihilations, have been computed with great numerical precision through the package micrOMEGAs \cite{Belanger:2015nma}. To clarify our results we provide nevertheless analytical expressions of the DM annihilation cross-section into fermion pairs, the most relevant in the regions of parameter space favored by $B$-anomalies (see below), at the leading order in the conventional velocity expansion (as further simplification we have neglected the masses of the final state fermions)\cite{Bai:2013iqa,Bai:2014osa,Giacchino:2015hvk,Arcadi:2017kky}:
\begin{equation}
\begin{aligned}
\langle \sigma v \rangle_{{\rm DM}\, {\rm DM}}^{\text{Complex}}&=\sum_f N_c\frac{\lambda_f^4 M_{\Phi_{\rm DM}}^2 v^2}{48 \pi {\left(M_{\Phi_{\rm DM}}^2+M_{F_f}^2\right)}^2},
& \langle \sigma v \rangle_{{\rm DM}\, {\rm DM}}^{\text{Dirac}}&=\sum_f N_c\frac{\lambda_f^4 M_{\Psi_{\rm DM}}^2}{32 \pi {\left(M_{\Psi_{\rm DM}}^2+M_{S_f}^2\right)}^4},\\
\langle \sigma v \rangle_{{\rm DM}\, {\rm DM}}^{\text{Real}}&=\sum_f N_c \frac{\lambda_f^4 M_{\Phi_{\rm DM}}^6 v^4}{60 \pi {\left(M_{\Phi_{\rm DM}}^2+M_{F_f}^2\right)}^4},
& \langle \sigma v \rangle_{{\rm DM}\,{\rm DM}}^{\text{Majorana}}&=\sum_f N_c\frac{\lambda_f^4 M_{\Psi_{\rm DM}}^2 \left(M_{\Psi_{\rm DM}}^4+M_{S_f}^4\right) v^2}{48 \pi {\left(M_{\Psi_{\rm DM}}^2+M_{S_f}^2\right)}^4},
\end{aligned}
\end{equation}
where $N_c=3(1)$ in case of colour charged (colour singlet) final state fermions. The sums run over the kinematically accessible final states, depending on the value of the DM mass.

The four expressions refer, as indicated, to real scalar, complex scalar, Dirac fermion and Majorana fermion DM. Scalar and fermionic DM candidates have been generically called $\Phi_{\rm DM}$ and $\Psi_{\rm DM}$, respectively, while the states exchanged in the $t$-channel Feynman diagrams and interacting with the fermion $f$ have been called $S_f$ and $F_f$. Finally $\lambda_{f}$ correspond to suitable assignments of $\Gamma_{\mu,s,b}^{L,Q}$ according to the final states.\footnote{Notice that colour charged NP field might interact with different quark generations. The expression of the annihilation cross-section might slightly change in this case.} 
While the four cross-sections have a very similar mass dependence in the limit in which DM  is much lighter than the NP field exchanged in the $t$-channel, they feature a very different velocity dependence. We notice indeed that the annihilation cross-section is $s$-wave dominated, in the case of Dirac fermion, $p$-wave suppressed in the case of complex scalar DM and Majorana fermion, and even further ($d$-wave) suppressed in the case of real scalar DM. Given that $v^2 \sim 0.1$, we expect that, while Dirac DM will easily comply with the requirement of the correct DM relic density, real scalar DM will, instead, by typically overabundant in light of its very suppressed annihilation cross-section, unless the latter will be enhanced e.g. by coannihilations.

Annihilations into SM fermion pairs and coannihilations mediated by $M$ represent the main contribution to the DM relic density in the case the DM belongs to an $SU(2)_L$ singlet. In the case DM belongs to an $SU(2)_L$ multiplet, it can also annihilate, through gauge interactions, into $W$ and $Z$ boson pairs. The latter annihilation processes easily become the dominant contribution to DM pair annihilations since the corresponding annihilation rate is not suppressed by the mass of the field $M$. For DM masses above the TeV, such cross-section is further increased by the so called Sommerfeld enhancement \cite{Iengo:2009ni,Feng:2010zp,Hryczuk:2011vi,Beneke:2014hja} as well as by bound state formation \cite{Mitridate:2017izz}. Additional coannihilation processes, due to other components of the DM multiplet, are present as well. 

As will be clear from the following analysis, imposing the correct relic density, Eq.~\eqref{eq:DMOmega}, translates into a very strong constraint, only marginally compatible with the flavour anomalies and other phenomenological bounds. For this reason we will just apply, through Eq.~\eqref{eq:relic_thermal},  an overclosure bound $\Omega_{\rm DM}h^2 \leq 0.12$.
However, while requiring that thermal DM production does not exceed the observed relic density, we will also assume that our DM candidate always accounts for 100\% of DM, including in the regions of the parameter space where it would be underabundant, as a consequence of some (unspecified) non-thermal DM production mechanism, 
see e.g.~\cite{Arcadi:2011ev,Drees:2017iod}.

\subsubsection{Direct detection}
\label{sec:DD}
The scattering of the DM with nucleons and nuclei, which is at the base of direct detection (DD), is typically described through effective four-field operators coupling pairs of DM particles with SM quark or gluon pairs. For all the models considered in this work, the strongest constraints come from Spin Independent (SI) interactions. For our analysis we have adopted the world leading limits given by the XENON1T collaboration~\cite{Aprile:2018dbl}. How effective are the resulting constraints depends on the spin of the DM and, in the case of scalar DM, on whether the field is real or complex, while in the case of fermionic DM, on its Dirac rather than Majorana nature. In the following illustrative discussion, we focus for simplicity on the case in which the DM is an $SU(2)_L$ singlet.

In the case of complex scalar DM,  the effective Lagrangian for DD reads:
\begin{align}
\label{Scalar:leff}    
\mathcal{L}_\text{eff}^{{\rm Scalar},q} = & \sum_{q=u,d} c^q
\left( \Phi_{\rm DM}^\dagger i\overset{\leftrightarrow}{\partial_\mu} \Phi_{\rm DM}\right) \bar q \gamma^\mu q + \sum_{q=u,d,s} d^q M_q \Phi_{\rm DM}^\dagger \Phi_{\rm DM}\, \bar q q+ d^g \frac{\alpha_s}{\pi}\Phi_{\rm DM}^\dagger \Phi_{\rm DM} \, G^{a\mu \nu}G^a_{\mu \nu}  \\ 
& + \sum_{q=u,d,s}g_1^q \frac{ \Phi_{\rm DM} \left(i \partial^\mu\right)\left(i \partial^\nu \right) \Phi_{\rm DM}\, \mathcal{O}^{q}_{\mu \nu}}{M_{\Phi_{\rm DM}}^2} + g_1^g\frac{ \Phi_{\rm DM} \left(i \partial^\mu\right)\left(i \partial^\nu \right) \Phi_{\rm DM}\, \mathcal{O}^{g}_{\mu \nu}}{M_{\Phi_{\rm DM}}^2} \,,
\end{align}
where $\mathcal{O}^{q}_{\mu \nu}$ and $\mathcal{O}^{g}_{\mu \nu}$ are the twist-2 operators:
\begin{align}
\mathcal{O}^q_{\mu \nu}=\bar q i\left(\dfrac{D_\mu \gamma_\nu+D_\nu \gamma_\mu}{2}-\frac{1}{4}g_{\mu \nu}\slashed{D}\right)q\,,\quad\quad
\mathcal{O}^g_{\mu \nu}=G_\mu^{a\rho} G^a_{\nu \rho}-\frac{1}{4}g_{\mu \nu}G^a_{\rho \sigma}G^{a\rho \sigma}\,.
\end{align}

As we can see, the effective Lagrangian considers just interactions with light quarks ($q=u,d,s$) and gluons. This is because the typical energy scale for DM scattering processes is of the order of 1 GeV and, hence, heavy quark flavours, $c,b,t$, are integrated out.

\begin{figure}
\begin{center}
\subfloat{\includegraphics[scale=0.8]{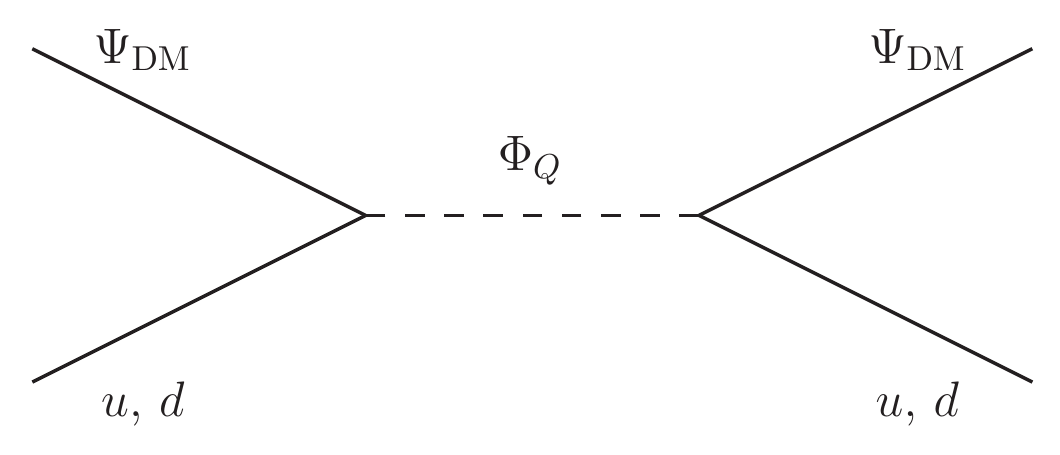}}
\hfill
\subfloat{\includegraphics[scale=0.8]{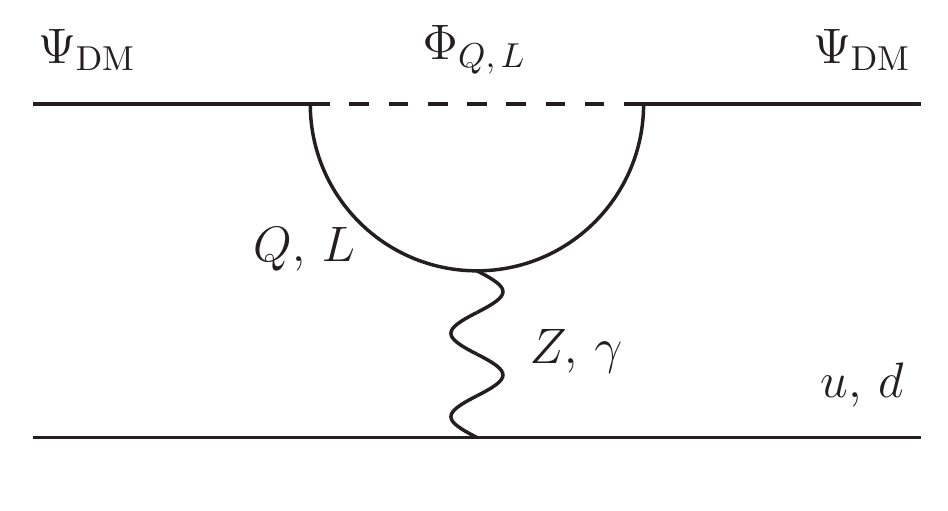}}\\
\subfloat{\includegraphics[scale=0.8]{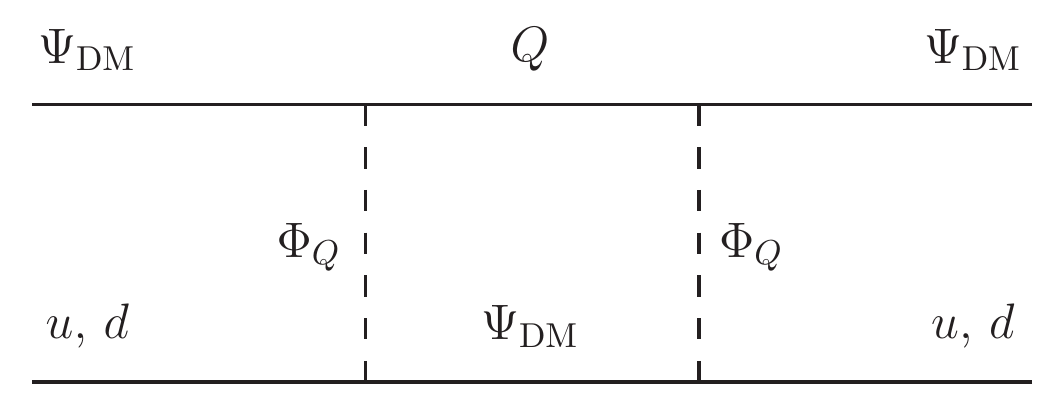}}
\hfill
\subfloat{\includegraphics[scale=0.8]{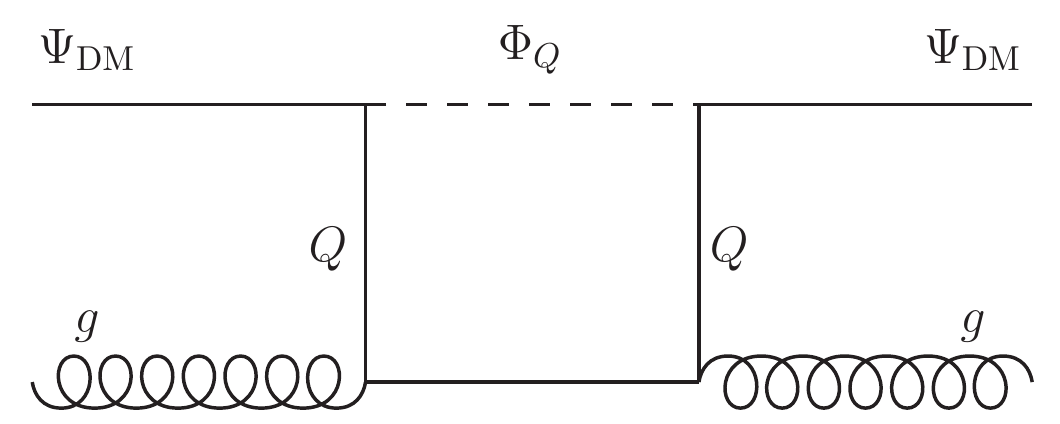}}
\end{center}
\caption{
Representative diagrams contributing to the DM-nucleon cross section relevant for direct detection experiments. For illustration, we only show some diagrams for the case of DM belonging to the fermionic flavour mediator $\Psi$. Analogous diagrams are present in the other cases (with the exception of models with DM belonging to $\Psi_L/\Phi_L$ where there is no tree-level diagram).
\label{diag}
 }
\end{figure}

The coefficient $c^q$ in \eq{Scalar:leff} can be further decomposed as $c^q=c_{\rm tree}^q+c_Z^q+c_\gamma^q+c_{\rm box}^q$. 
Illustrative diagrams associated to these different contributions are shown in Figure~\ref{diag}. Notice that the figure actually displays the case of fermionic DM, discussed in the following. However topologically analogous diagrams can be  also obtained for scalar DM, since as we will discuss below, the two cases share many similarities.  

$c_{\rm tree}^q$  is the tree-level induced contribution from diagrams with $s$-channel exchange of colour charged scalar NP field. Since such tree level contribution describes the interactions of vector currents, the corresponding coefficient for the nucleons is just a linear combination of the contributions of the valence quarks, namely:
\begin{equation}
c^p_{\rm tree}=2 c^u_{\rm tree}+c^d_{\rm tree},\,\,\,\,\,c^n_{\rm tree}=c^u_{\rm tree}+2 c^d_{\rm tree}.
\end{equation}
In the models considered in this work the $c^{N=p,n}_{\rm tree}$ coefficients are generated only be the CKM mixing. Having chosen a basis in which the down-type quark mass matrix is flavour diagonal we have:
\begin{equation}
c_{\rm tree}^u=\frac{(\Gamma_s^{Q}V_{us}+\Gamma_b^Q V_{ub})^2}{ 4(M_{\Phi_{\rm DM}}^2-M_{F_f}^2)}, \,\,\,\,\,c_{\rm tree}^d=0\,.
\end{equation}
Given this result, one cannot neglect a priori contributions from loop-level induced interactions. Indeed, the $c_{Z,\gamma}^q$ coefficients are generated by penguin-like diagrams as the ones shown in the second panel of Figure~\ref{diag}, charged under the SM EW group, with SM $\gamma$ and $Z$ bosons while $c_{\rm box}^q$ is the coefficient associated to box diagrams analogous to the one present in the third panel of Figure~\ref{diag}. The remaining operators in Eq.~(\ref{Scalar:leff}) arise, again at the loop level, from QCD interactions of the colour charged new fermions, possibly present in the theory~\cite{Hisano:2015bma}. Full analytical expressions for the $c_{Z,\gamma}^q,\,c_{\rm box}^q,\,d^{q,g},\,g_1^{q,g}$ can be found e.g.~in Refs.~\cite{Bhattacharya:2015xha,Hisano:2015bma,Kawamura:2017ecz}. As will be discussed in the following, the strongest limits will arise from $\gamma,Z$ penguins. We  report here the relevant simplified expressions, as given in \cite{Kawamura:2017ecz}, in the limit $M_{\Phi_{DM}} \ll M_{F_f}$. For what concerns the $\gamma$ penguin we have:
\begin{equation}
c^q_\gamma \approx \frac{e^2 Q_q}{96 \pi^2}\sum_{f=s,c,b,t,\mu}\frac{Q_f \lambda_f^2}{M_{F_f}^2}N_c  \left[3+\log\left(\frac{M_f^2}{M_{F_f}^2}\right)\right]\,.
\end{equation}
The coefficient associated to the $Z$-penguin, in the same approximation, instead reads:\footnote{In these expression, as well as the corresponding ones for fermionic DM, we have omitted for simplicity $O(1)$ factors that depend on the gauge quantum numbers of the field $F_f$ ($S_f$ for fermionic DM).}
\begin{equation}
c^q_Z \approx v^q\frac{G_F}{\sqrt{2}}\sum_{f=s,c,b,t,\mu} \frac{\lambda_f^2 T_3^f N_c}{16\pi^2}\frac{M_f^2}{M_{F_l}^2}\left(\frac{3}{2}+\log\left(\frac{M_f^2}{M_{F_f}^2}\right)\right)\,,
\end{equation}
where
\begin{equation}
v^u=\frac{8}{3}\sin^2 \theta_W-1,\,\,\,\,\,v^d=-\frac{4}{3}\sin^2 \theta_W+1\,.
\end{equation}
Among these two contributions, photon penguins give typically the dominant contribution with the exception of the case in which the effective coupling of the DM with top quarks is sizable,  as a consequence of the enhancement proportional to the square of the top mass in $c^q_Z$.  

The effective Lagrangian in Eq.~(\ref{Scalar:leff}) gives rise to the following scattering cross-section for the DM over nucleons (for illustration we focus on the case of the proton):
\begin{equation}
\sigma_{\Phi_{\rm DM}}^{{\rm SI},\, p}= \frac{\mu_{p}^2}{\pi}\, \frac{\left[Z f_p +(A-Z)f_n\right]^2}{A^2}\,,
\end{equation}
where $\mu_{p} = M_{\Phi_{\rm DM}} M_p/(M_{\Phi_{\rm DM}}+M_p)$ is the DM/proton reduced mass.  The extra factor depending on $A,Z$, being respectively the mass and atomic number of the detector material, allows a consistent comparison with experimental limits which assume equal coupling of the DM with protons and neutrons \cite{Feng:2013vod}.
$f_{p,n}$ represent, in fact, the effective coupling of the DM with protons and neutrons and read:
\begin{align}
\label{eq:fpfn_cs}
f_p= &c_{\rm tree}^p+c_Z^p+c_\gamma^p+c_{\rm box}^p+M_p\sum_{q=u,d,s}\left(f_q^p d_q  +\frac{3}{4}g_1^q \left(q(2)+\bar{q}(2)\right)\right)+ \frac{3}{4}M_p\sum_{q=c,b,t}g_1^g G(2) -\frac{8}{9}f_{TG}f_G\,, \\
f_n= & c_{\rm tree}^n+c_Z^n+c_{\rm box}^n +\frac{3}{4}M_n\sum_{q=u,d,s}\left(q(2)+\bar{q}(2)\right)g_1^q+
\frac{3}{4}M_n\sum_{q=c,b,t}g_1^g G(2)
-\frac{8}{9}f_{TG}f_G\,,
\end{align}
where  $c_{i}^p=2 c_i^u+c_i^d$, $c_i^n=c_i^u+2 c_i^d$ with $i={\rm tree},\, Z,\,\gamma,\,{\rm box},$ and  $f_G=d_c^g+d_b^g+d_t^g$. The parameters $f_q^{N=n,p},\,f_{TG},\,q(2),$ and $G(2)$ are nucleon form factors defined as:
\begin{align}
& \langle N | m_q \bar q q|N \rangle= M_N f_q^N,\,\,\,\,\,f_{TG}=1-\sum_{q=u,d,s}f_q^N \,,\\
& \langle N |\mathcal{O}^q_{\mu \nu} |N \rangle=\frac{1}{M_N}\left(p_\mu p_\nu-\frac{1}{4}M_N^2 g_{\mu \nu}\right) \left(\bar q(2)+q(2)\right)\,, \\
& \langle N |\mathcal{O}^g_{\mu \nu} |N \rangle=\frac{1}{M_N}\left(p_\mu p_\nu-\frac{1}{4}M_N^2 g_{\mu \nu}\right) G(2)\,.
\end{align}
For our analysis we have used the default values implemented into the micrOMEGAs package~\cite{Belanger:2015nma}.

It is important to remark that, contrary to the other coefficients, including the ones generated at the tree-level, $c_{\gamma,Z}^p$ coefficients can be present in models in which the DM is coupled only to NP states charged under the electroweak gauge groups but that are colour singlets.

If DM is a real scalar, the $\Phi_{\rm DM}^\dagger i\overset{\leftrightarrow}{\partial_\mu} \Phi_{\rm DM}$ operator identically vanishes. Hence, the DM direct detection cross-section is expected to be suppressed.

Moving to fermionic DM, which we just call $\Psi_{\rm DM}$, the effective Lagrangian for the Dirac case is:
\begin{align}
\label{eq:Lfermion}
\mathcal{L}_\text{eff}^{\text{Dirac},q}&= \sum_{q=u,d} c^q \bar\Psi_{\rm DM} \gamma_\mu \Psi_{\rm DM} \bar q \gamma^\mu q
+ \sum_{q=u,d,s} d^q M_q\bar \Psi_{\rm DM} \Psi_{\rm DM} \bar q q 
+\sum_{q=c,b,t} d_q^g \frac{\alpha_s}{\pi} \bar \Psi_{\rm DM} \Psi_{\rm DM} G^{a\mu \nu}G^a_{\mu \nu} \\
&+ \sum_{q=u,d,s} \left(g_{1}^{q} \frac{\bar \Psi_{\rm DM} i \partial^\mu \gamma^\nu \Psi_{\rm DM} \mathcal{O}^q_{\mu \nu} }{M_{\Psi_{\rm DM}}}+ g_{2}^{q} \frac{\bar \Psi_{\rm DM} \left( i  \partial^\mu \right)\left(i \partial^\nu \right) \Psi_{\rm DM} \mathcal{O}^q_{\mu \nu} }{M_{\Psi_{\rm DM}}^2}\right)\\
& + \sum_{q=c,b,t}\left( g_{1}^{g,q} \frac{\bar \Psi_{\rm DM} i \partial^\mu \gamma^\nu \Psi_{\rm DM} \mathcal{O}^g_{\mu \nu} }{M_{\Psi_{\rm DM}}}+ g_{2}^{g,q} \frac{\bar \Psi_{\rm DM} \left( i  \partial^\mu \right)\left(i \partial^\nu \right) \Psi_{\rm DM} \mathcal{O}^g_{\mu \nu} }{M_{\Psi_{\rm DM}}^2}\right)\,.
\end{align}
Again, the coefficient $c^q$ is a combination of CKM-suppressed tree-level and loop-induced contributions: 
\begin{equation}
c^q=c_{\rm tree}^q+c_Z^q+c_H^q+c_{\rm box}^q\,,
\end{equation}
where, in contrast to the case of scalar DM, a contribution from Higgs penguin diagrams is present as well. The operator $\bar \Psi_{\rm DM} \gamma_\mu \Psi_{\rm DM} \bar q \gamma^\mu q$ behaves, with respect to direct detection, in an analogous way as $\left(\phi^\dagger_{\rm DM} i\overset{\leftrightarrow}{\partial_\mu} \phi_{\rm DM}\right) \bar q \gamma^\mu q$. So we have again that the coefficients $c^{p,n}$ at the nucleon level are linear combinations of the coefficients associated to up and down quarks. The coefficients $d^{q,g}$ can be decomposed into QCD contributions, which we label $d^{q,g}_{\rm QCD}$, analogous to the ones discussed for scalar DM, and a contribution from Higgs penguin diagrams, labelled as $d_{H}^{q,g}$. The coefficients $g_1^{g,q}$ finally come from QCD interactions mediated by quarks/gluons and possible NP colour-charged states. Effective interactions mediated by the photon are present as well. The latter are described, this time, by the following Lagrangian:
\begin{equation}
\label{eq:Lfermionphoton}
\mathcal{L}_{\rm eff}^{\text{Dirac}, \gamma}=\frac{\tilde{b}_\Psi}{2}\bar \Psi_{\rm DM} \sigma^{\mu \nu}\Psi_{\rm DM} F_{\mu \nu}+b_\Psi \bar \Psi_{\rm DM} \gamma^\mu \Psi_{\rm DM} \partial^\nu F_{\mu \nu}\,,
\end{equation}
with the two terms dubbed, respectively, magnetic dipole moment and charge radius operators.

Concerning the relative contribution of the different coefficients, as already discussed in Ref.~\cite{Ibarra:2015fqa}, the situation is analogous to the case of complex scalar DM. The dominant contribution is typically associated to the charged radius and dipole operators, whose coefficient can be approximately written as:
\begin{align}
& b_\psi \approx -\frac{e}{48\pi M_{\Psi_\text{DM}}^2}\sum_{f=s,c,b,t,\mu}N_c Q_f \lambda_f^2 \frac{M_{\Psi_\text{DM}}^2}{M_{S_f}^2} \left(1-2\log\left(\frac{M_f^2}{M_{S_f}^2}\right)\right)\,, \\
& \tilde{b}_\psi \approx -\frac{e}{8\pi M_{\Psi_\text{DM}}}\sum_{f=s,c,b,t,\mu}N_c Q_f \lambda_f^2 \frac{M_{\Psi_\text{DM}}^2 M_f^2}{M_{S_f}^2} \left(2+\log\left(\frac{M_f^2}{M_{S_f}^2}\right)\right) \,.
\end{align}
Again, in case of sizable couplings of the DM with the top quark, the latter terms are overcome by $c_Z^q$ which can be approximately written as:
\begin{equation}
c_Z^q=v^q\sum_{f=s,c,b,t,\mu} \frac{G_F}{\sqrt{2}}\frac{N_c \lambda_f^2}{32\pi^2}\frac{M_f^2}{M_{S_f}^2}\left(1+\log\left(\frac{M_f^2}{M_{S_f}^2}\right)\right).
\end{equation}
Given the presence of  dipole operators, direct detection phenomenology is not fully caught by the scattering cross-section over nucleons but, on the contrary, one has to rely on the DM scattering rate over nuclei: 
\begin{align}
\label{eq:DM_scattering_rate}
& \frac{d\sigma}{dE_R}=
\left(\frac{M_T}{2\pi v^2}|f^T|^2 + \alpha_{\rm em}\tilde{b}_\Psi^2 Z^2 \left(\frac{1}{E_R}-\frac{M_T}{2 \mu_{T }^2 v^2}\right)\right)|F_{\rm SI}(E_R)|^2 
+\tilde{b}_\Psi^2 \frac{\mu_T^2 M_T}{\pi v^2}\frac{J_T+1}{3 J_T}|F_{\rm D}(E_R)|^2,
\end{align}
where
\begin{equation}
f^T=Z f_p+(A-Z) f_n,
\end{equation}
and $\mu_T=M_{\Psi_{\rm DM}} M_T/(M_{\Psi_{\rm DM}}+M_T)$ with $M_T$ being the mass of target nucleus. $F_{SI}$ is the conventional SI nuclear form factor~\cite{Lewin:1995rx}, while $F_{\rm D}$ is the form factor associated to dipole scattering~\cite{Banks:2010eh}. Experimental limits have been obtained, in this case, with the procedure illustrated in Ref.~\cite{Hisano:2018bpz}.

In the case of Dirac DM the coefficients $f_p$ and $f_n$ are written as:
\begin{align}
\label{eq:fpfn}
f_p & = c_{\rm tree}^p+c_Z^p+c_{\rm box}^p-e b_\Psi-\frac{e \tilde{b}_\Psi}{2 M_\Psi}
+M_p\sum_{q=u,d,s} \left( f_q^p d_q  +\frac{3}{4}\left(q(2)+\bar{q}(2)\right)\left(g_1^q+g_2^q\right)\right) \\
&+\frac{3}{4}M_p\sum_{q=c,b,t}G(2)\left(g_1^{g,q}+g_2^{g,q}\right)    -\frac{8}{9}f_{TG}f_G\,,\\
f_n & = c_{\rm tree}^n+c_Z^n+c_H^n+c_{\rm box}^n  
+\frac{3}{4}M_n\sum_{q=u,d,s}\left(q(2)+\bar{q}(2)\right)\left(g_1^q+g_2^q\right)\\
&+\frac{3}{4}M_n\sum_{q=c,b,t}G(2)\left(g_1^{g,q}+g_2^{g,q}\right)
-\frac{8}{9}f_{TG}f_G\,.
\end{align}

Changing the nature of the DM, this time from Dirac to Majorana, leads to markedly different case. Indeed, the $\bar \Psi \gamma^\mu \Psi$ and $\bar \Psi \sigma^{\mu \nu} \Psi$ operators are identically null and the DD phenomenology is again fully captured by the SI scattering cross-section of the DM on protons:
\begin{equation}
\sigma_{\Psi_{\rm DM}}^{{\rm SI},\, p}=4 \frac{\mu_{p}^2}{\pi}\,  \frac{\left[Z f_p +(A-Z)f_n\right]^2}{A^2}\,,
\end{equation}
where $f_{p,n}$ are defined analogously to Eq.~\eqref{eq:fpfn}.

In the case in which the DM belongs to an $SU(2)_L$ multiplet, no new operators are generated in the Lagrangians in Eqs.~\eqref{Scalar:leff} and \eqref{eq:Lfermion}. The coefficients of the effective operators get, however, additional contributions from loop diagrams in which $Z,W$ bosons are exchanged. The case of Majorana DM has been discussed extensively e.g.~in Refs.~\cite{Hisano:2010fy,Hisano:2011cs} while the case of real DM has been considered, to a lower extent e.g.~in Ref.~\cite{Hisano:2015bma}. To our knowledge no analogous computations are available for complex scalar and Dirac fermionic DM. 

\subsubsection{Indirect detection}
As well known, indirect detection (ID) for WIMPs, relies on the search of the products of residual annihilation processes for DM occurring at present times. Similarly to the case of direct detection it is convenient to distinguish the cases in which the DM is an $SU(2)_L$ singlet or not. In the former case, the main annihilation channels to consider are the ones into SM fermion pair final states. These lead mostly to continuous $\gamma$-ray signals which, for the ranges of DM masses considered in this work, can be probed by telescopes such as Fermi-LAT~\cite{Ahnen:2016qkx,Hoof:2018hyn}.
The impact of the resulting constraints is highly model-dependent though. Indeed, having in mind the velocity expansion:
\begin{equation}
\langle \sigma v \rangle \approx a+b v^2\,,
\end{equation}
we have that only for $s$-wave dominated annihilation cross-section, i.e.~$a \neq 0$, the values of the cross-section at thermal freeze-out and and present times are comparable, so that eventual ID limits are effective. On the contrary, $p$-wave, i.e. the $b$ coefficient is the leading contribution, dominated cross-section are affected by ID to a negligible extent. Notice as well that coannihilations are also mostly effective at thermal freeze-out while their rates are, instead, exponentially suppressed at present times. Given this, among the models presented in this work, only scenarios with Dirac fermionic DM can be probed by indirect detection.\footnote{Notice that also for Majorana DM $a \neq 0$. However this term is helicity suppressed and, hence, typically subdominant with respect to the$p$-wave contribution.}

Summarising, in the following section we will apply the strategy here described to several models of interest, in order to study whether such models allow for a region of the parameter space where all the constraints here described are evaded, or the model is excluded by the combination of all the constraints.


\section{Results and discussion}
\label{sec:results}
In this section, we analyse some of the models listed in Tables~\ref{tab:fmodels} and~\ref{tab:smodels}, following the strategy illustrated in the previous section. Such models have been highlighted in cyan in the tables. Our selection covers a broad variety of cases, including both scalar and fermion flavour mediators, as well as both scalar and fermion DM. Furthermore we will separately discuss, where appropriate, both real and complex scalar DM as well as both Dirac and Majorana nature for fermionic DM. Finally, notice that our selection comprises models where DM is a pure SM singlets as well as cases where it belongs to $SU(2)_L$ multiplets.

We will write for each model the Lagrangian responsible for the phenomenology we are interested in, according to the quantum numbers of the NP particles, and determine the regions of parameter space for which the $B$-anomalies are accounted for. We will then combine this requirement with the constraints from collider searches of the NP particles as well as from DM phenomenology, in particular relic density and direct detection. As discussed in Section~\ref{sec:relic}, we will assume that in the regions of the parameter space where thermal DM production is insufficient some non-thermal mechanism is at work such that our DM candidate always account for 100\% of the observed DM abundance.

%
\subsection{\texorpdfstring{$\mathcal{F}_\text{IA;\,0}$}{FIA}, Dirac singlet  DM}
\label{sec:FIA_D}
We start considering the model $\mathcal{F}_\text{IA;\,0}$ with singlet Dirac DM. This case, which is among the simplest in Tables~\ref{tab:fmodels} and~\ref{tab:smodels}, has never been studied before and, as we will see, is subject to strong constraints. It is a good example to illustrate how bounds from different sources can altogether exclude a model.  
The Lagrangian of this model reads:
\begin{eqnarray}
{\cal L}_{\rm int} =  {\Gamma^Q_i\bar Q_i}{P_L}{\Psi}{\Phi_Q} + \Gamma^L_i \bar L_i {P_R}{\Psi}{\Phi_L} +
{\rm{h}}{\rm{.c.}}\,,
\label{eq:L_IIA_D}
\end{eqnarray}
with the fields $\Phi_Q$, $\Phi_L$ and $\Psi$ carrying respectively the following $SU(3)_c\otimes SU(2)_L\otimes U(1)_Y$ quantum numbers: $({\bf 3},{\bf 2},1/6)$, $({\bf 1},{\bf 2},-1/2)$ and $({\bf 1},{\bf 1},0)$. As mentioned above, the DM candidate is the Dirac field $\Psi_\text{DM}=\Psi$, which also plays the role of the flavour mediator in the diagram in Fig.~\ref{fig:boxes}. This scenario has been selected for our analysis since it features the highest degree of correlation. Indeed, notice that the DM field couples to both the NP fields $\Phi_Q$ and $\Phi_L$, which are charged under the SM gauge group. Consequently, all the three couplings $\Gamma_s^Q$,  $\Gamma_b^Q$, $\Gamma_\mu^L$, entering $\delta C^{9,10}_\mu$ are, as well, contributing to the DM annihilation and scattering rates. For this reason we will also investigate, in the following, similar models with Majorana fermion and scalar DM. 

\begin{figure}[!t]
\centering
\subfloat{\includegraphics[width=0.33\linewidth]{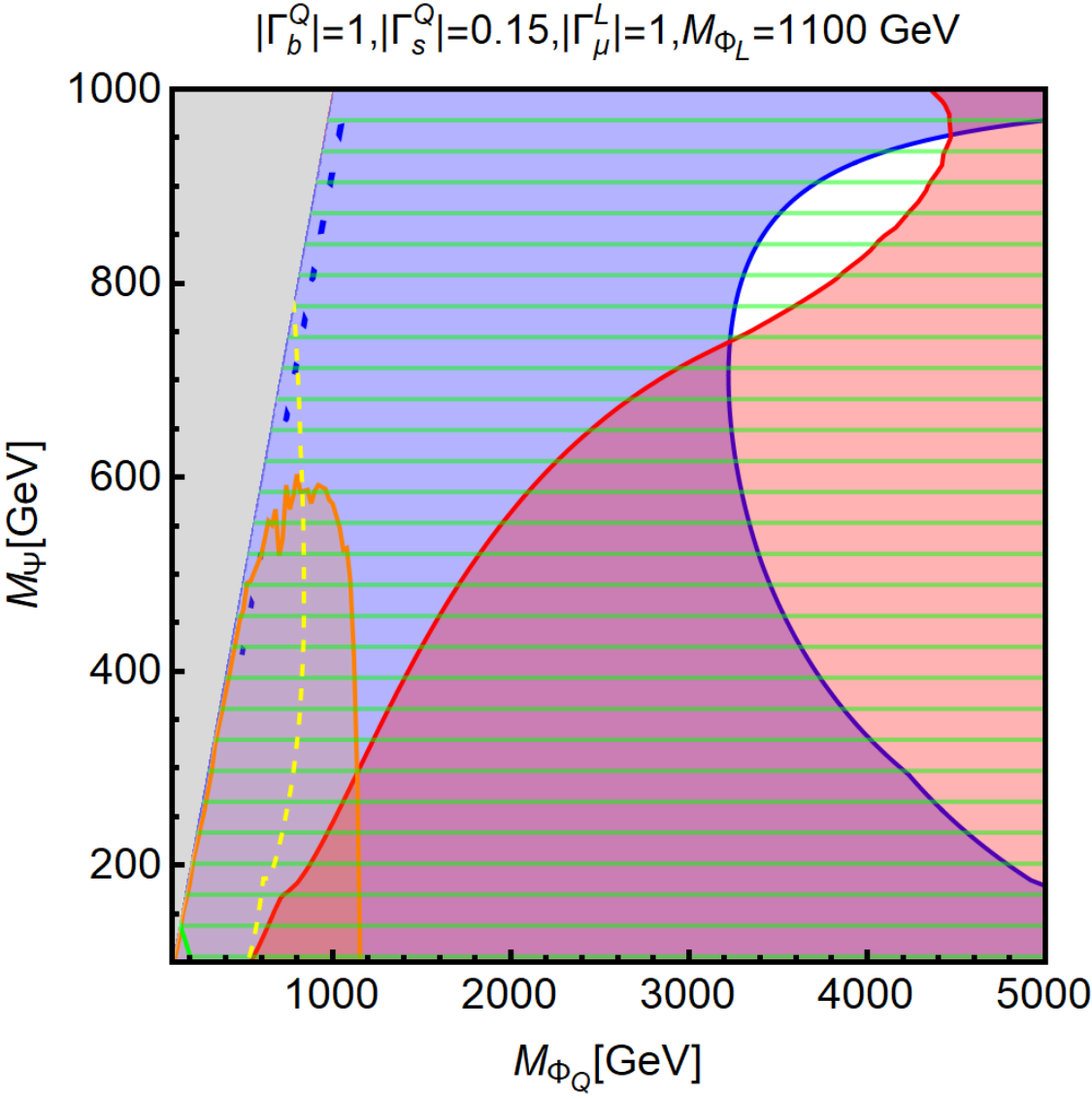}}
\subfloat{\includegraphics[width=0.33\linewidth]{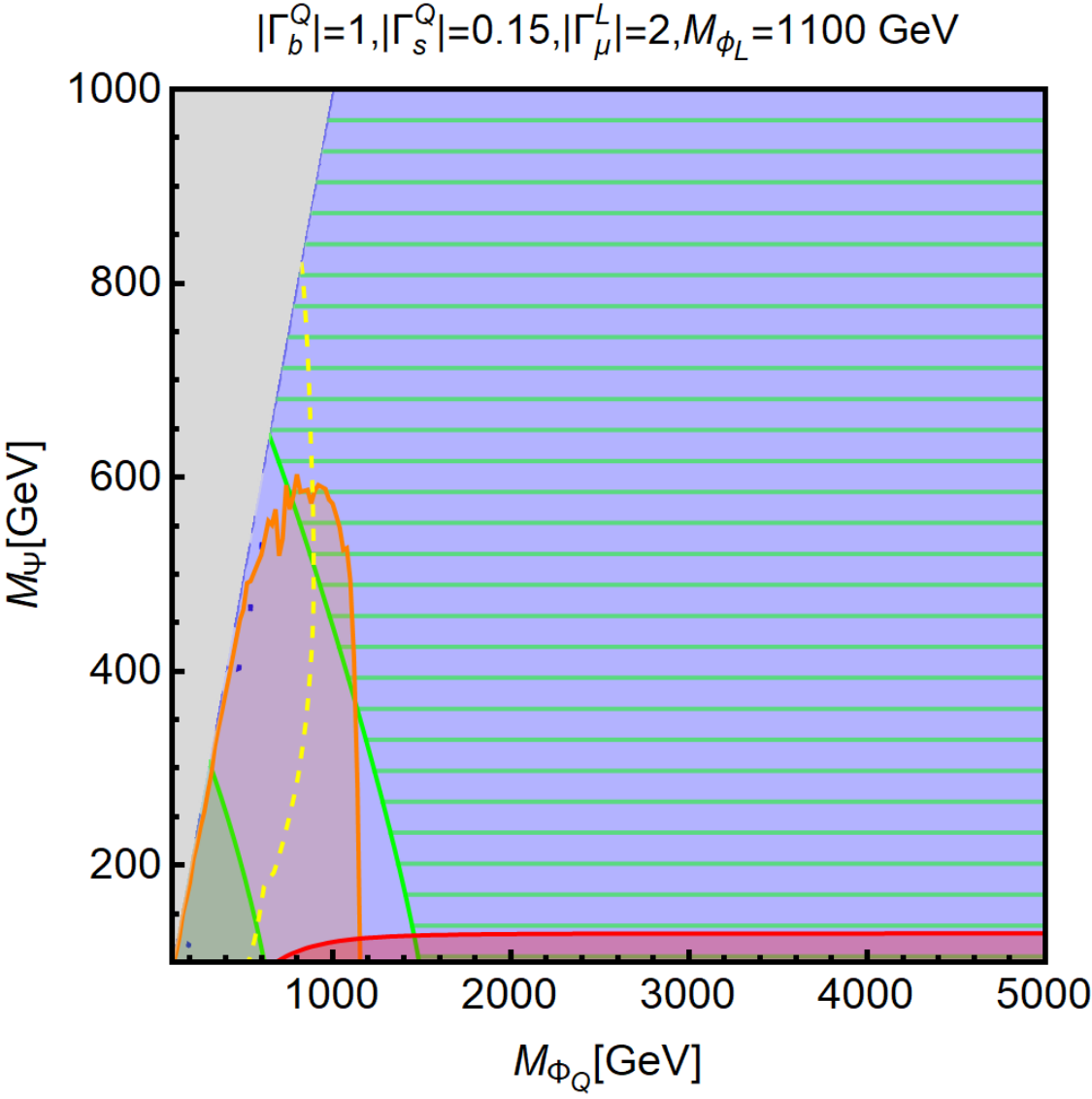}}
\subfloat{\includegraphics[width=0.33\linewidth]{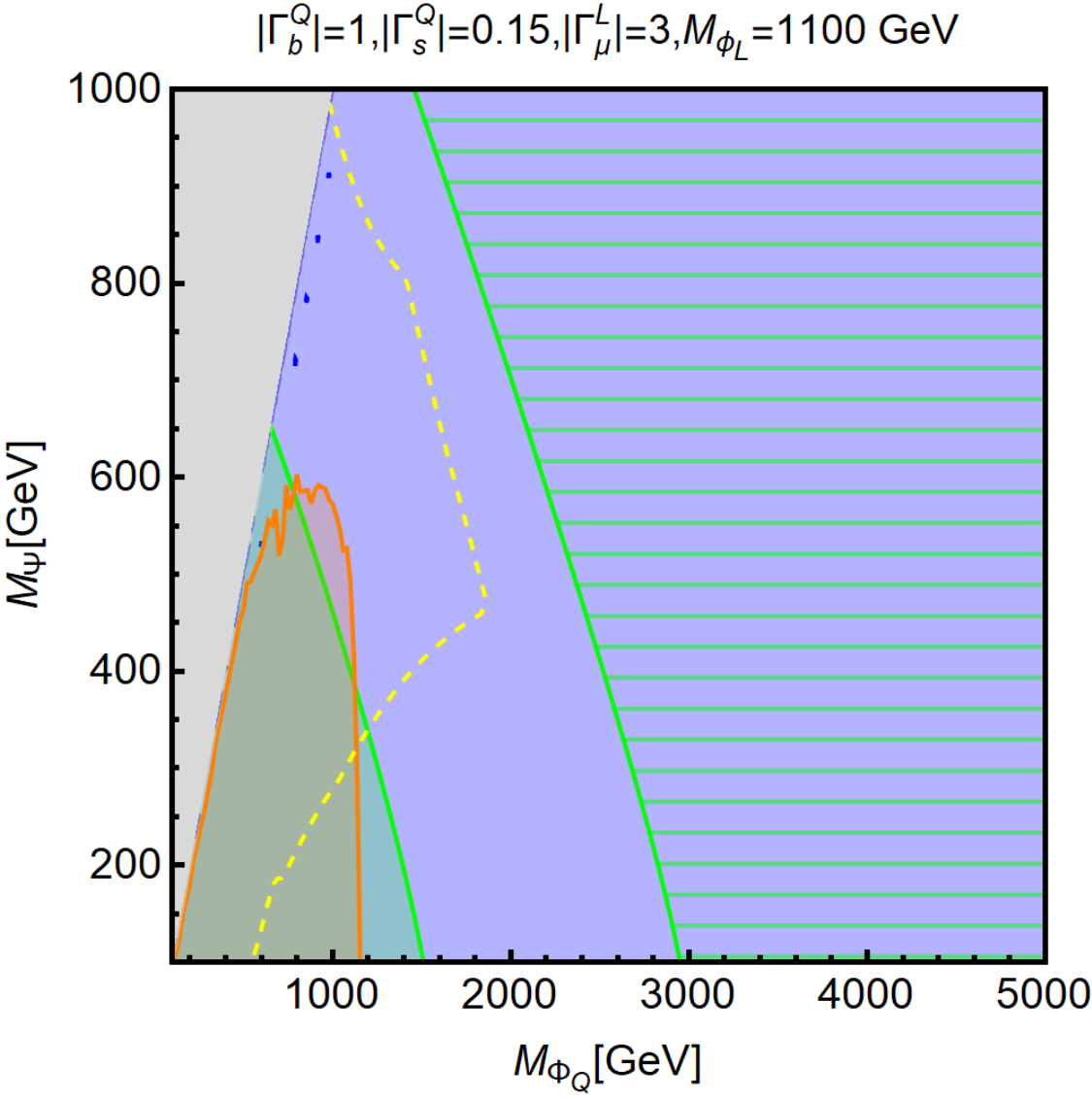}}\\
\subfloat{\includegraphics[width=0.33\linewidth]{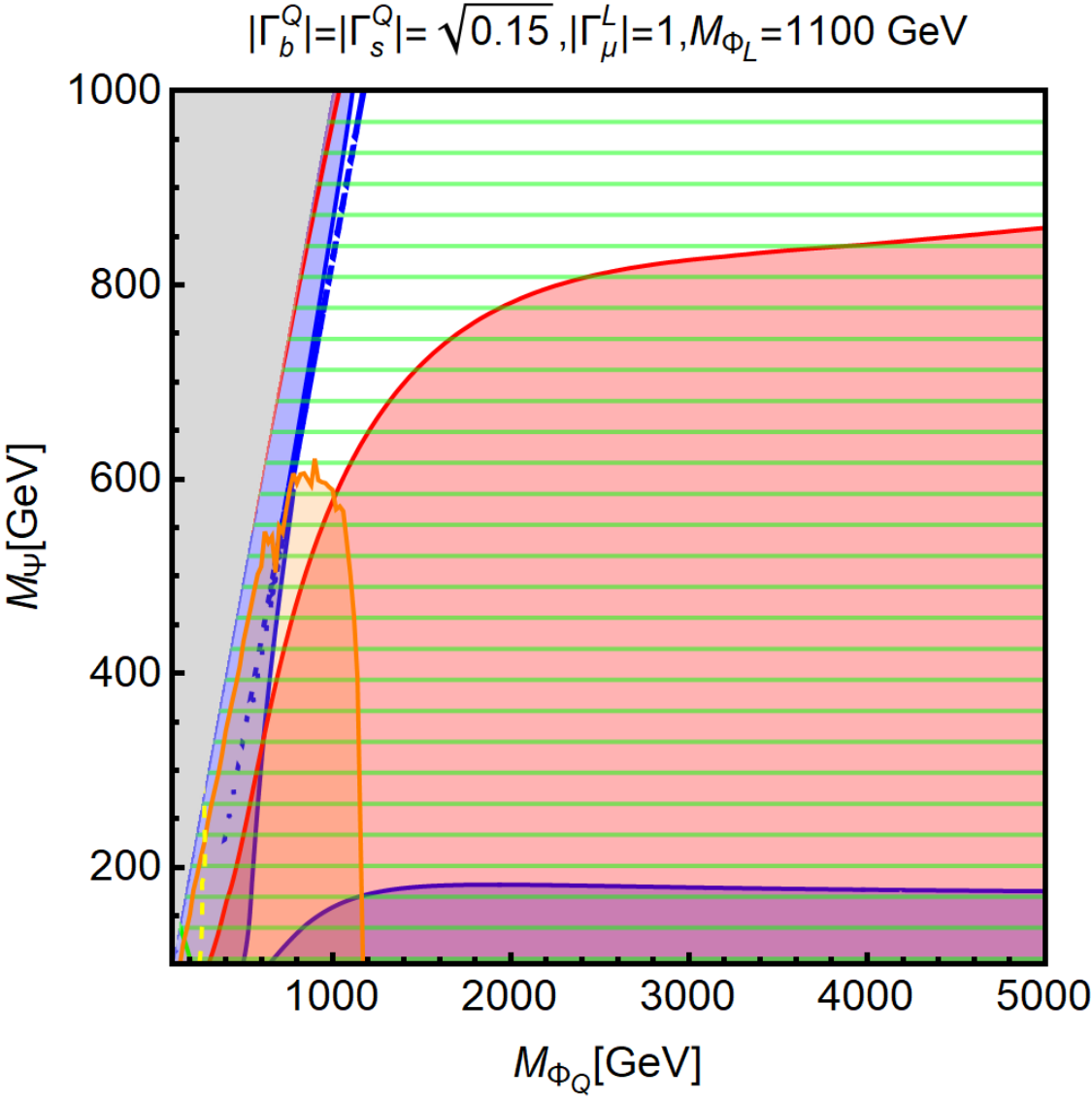}}
\subfloat{\includegraphics[width=0.33\linewidth]{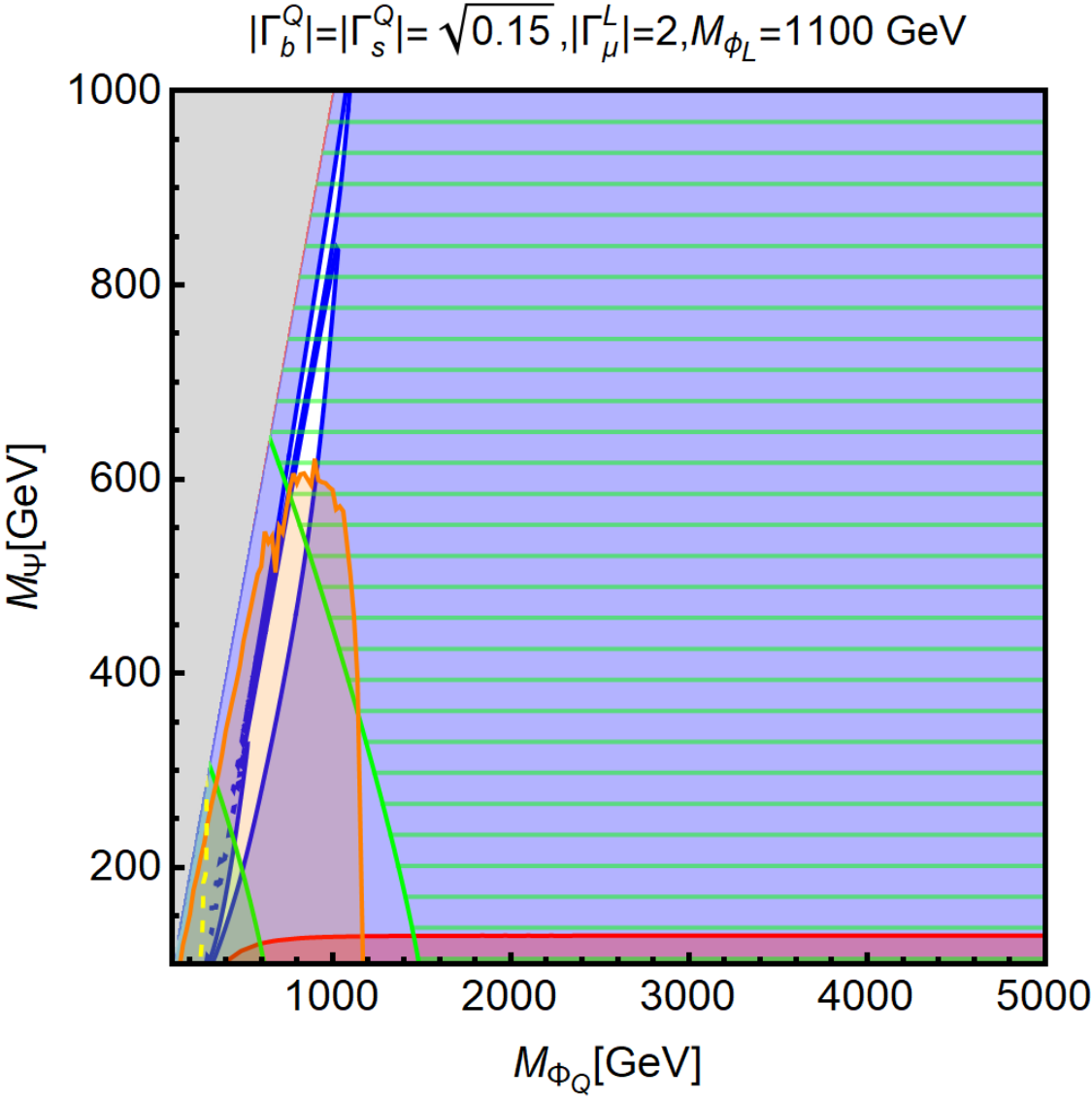}}
\subfloat{\includegraphics[width=0.33\linewidth]{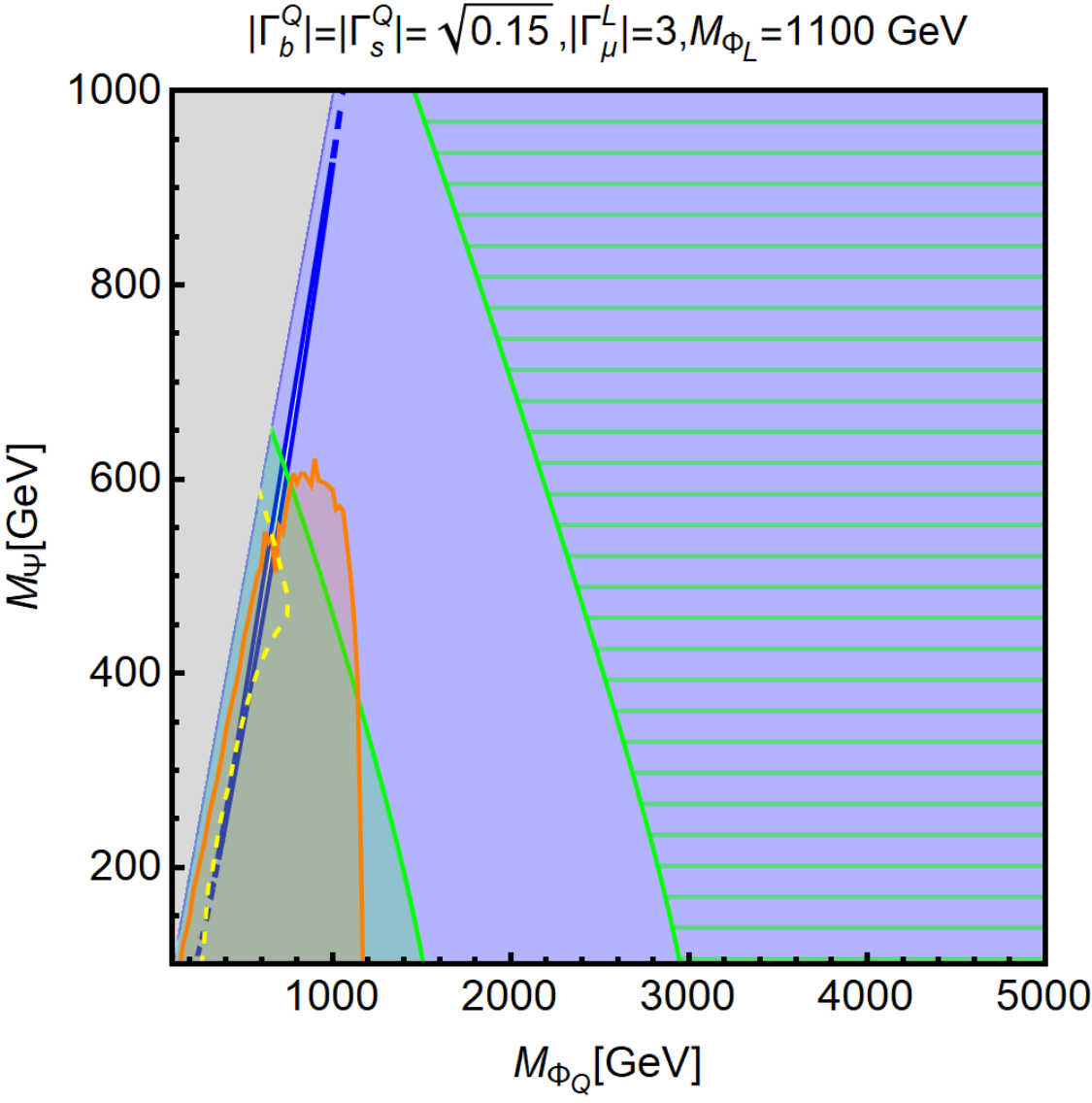}}
\caption{Summary of the constraints for the model $\mathcal{F}_\text{IA;\,0}$ with Dirac dark matter. The first row corresponds to the benchmark case \emph{i}), with $|\Gamma_b^Q|=1$, $|\Gamma_s^Q|=0.15$, while in the second row we show the benchmark case \emph{ii}), with $|\Gamma_b^Q|=|\Gamma_s^Q|=\sqrt{0.15}$. For both cases we set $M_{\Phi_L}=1100\,\mbox{GeV}$ and we show the three assignments $|\Gamma_\mu^L|=1,2,3$. Regions outside green contours are not compatible with the $B$-anomalies, with the solid green regions being strictly ruled out while the hatched green ones corresponding to contributions to $\delta C^{9,10}_\mu$ not significantly deviating from the SM predictions, hence not strictly ruled out as the anomalies are still awaiting full experimental confirmation. The orange region represents the exclusion from direct searches at the LHC (see the text for details). Red regions are excluded because they correspond to overproduction of DM compared to the observed relic density. Blue regions are excluded by DD, while regions at the left of the yellow dashed contour are excluded by ID.
}
\label{fig:F_IA_D}
\end{figure} 

The results of our analysis for this model are presented in Figure~\ref{fig:F_IA_D}, displayed in the $(M_\Psi,M_{\Phi_Q})$ two-dimensional plane. The two rows in the figure correspond to the two different assignments of the couplings ($|\Gamma_s^Q|$, $|\Gamma_b^Q|$), dubbed \emph{i}) ($|\Gamma_s^Q|=1$, $|\Gamma_b^Q|=0.15$) and \emph{ii}) ($|\Gamma_s^Q|=|\Gamma_b^Q|=\sqrt{0.15}$), which proved to provide an equally good fit of the $B$-anomalies, cf.~Section~\ref{sec:fit}. For each coupling configuration we show only one of our benchmark values for $M_{\Phi_L}$, namely 1100~GeV, and three values of $|\Gamma_\mu^L|$. 

In each plot, the region compatible with the flavour anomalies is the one enclosed within the two green contours. As can be seen from the different filling styles, the regions outside these bands should be interpreted differently. Indeed the regions on the left of the green contours (filled in green) are ruled out, since they correspond to the case in which $\delta C^{9,10}_\mu$ exceed the experimental limits. On the right of the contours, on the contrary, NP contributions to $C^{9,10}_\mu$ are increasingly suppressed so that the these observables do not deviate, to a statistically relevant extent, with respect to the SM expectation. While current flavour anomalies are not reproduced in the latter parameter regions, we cannot strictly regard them as ruled out as the anomalies are still awaiting full experimental confirmation. These regions are denoted by a green horizontal hatching.

The orange region represents the exclusion from LHC searches for the signatures with jets and/or muons and missing energy  described in Section~\ref{sec:lhc}.
For this model, we show our recasting of the bound 
from Ref.~\cite{Sirunyan:2019ctn} on $pp \to \Phi_Q \Phi_Q \to qq^\prime + \Psi_\text{DM} \Psi_\text{DM}$.

Moving to DM phenomenology, the constraints from DM relic density are represented as red regions. As already mentioned, throughout our study we will just require that the value of $\Omega_{\rm DM}h^2$, determined by applying the conventional thermal freeze-out paradigm, does not exceed the experimental determination, namely $\Omega_{\rm DM}h^2 \le 0.12$. In each plot the region of parameter space which does not fulfill this constraint has been marked in red. Being the DM a SM singlet, its relic density is determined, with the exception of the coannihilation region, by annihilations into SM fermion pairs. As the value of $\Gamma_\mu^L$ increases, the region with overabundant DM progressively reduces and, for $\Gamma_\mu^L = 3$ the DM is always underabundant within the whole range of $M_\Psi$ and $M_{\Phi_Q}$ shown in the plot. The blue region corresponds, instead, to the case in which the DM interactions with nuclei, as given by Eq.~\eqref{eq:DM_scattering_rate}, exceed the constraints from XENON1T. Finally, being the DM a Dirac fermion, constraints from indirect detection should be taken also into account. The regions of parameter space at the left side of the dashed yellow contour are excluded by the latter type of searches.

\begin{figure}[!t]
\centering
\subfloat{\includegraphics[width=0.33\linewidth]{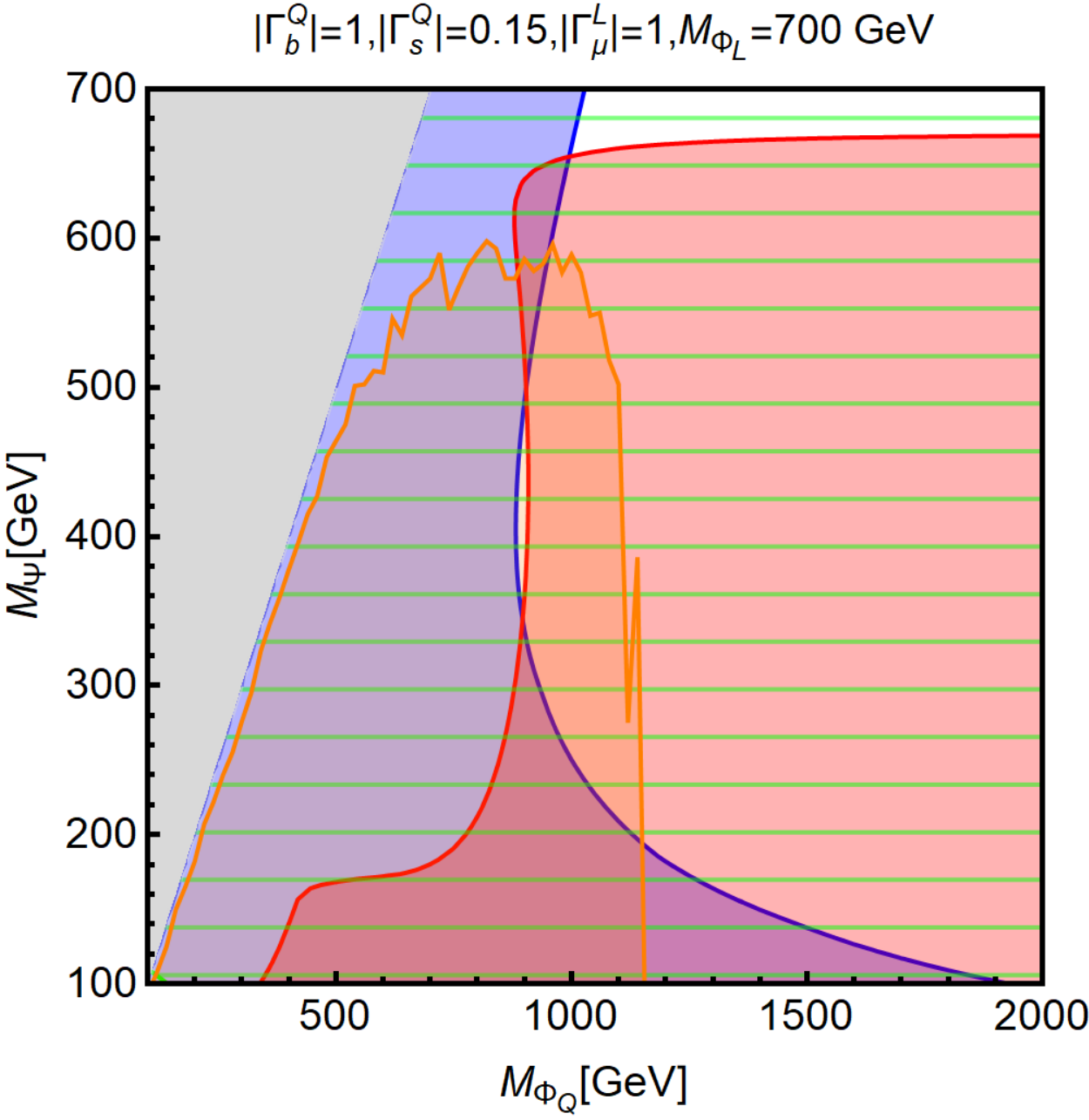}}
\subfloat{\includegraphics[width=0.33\linewidth]{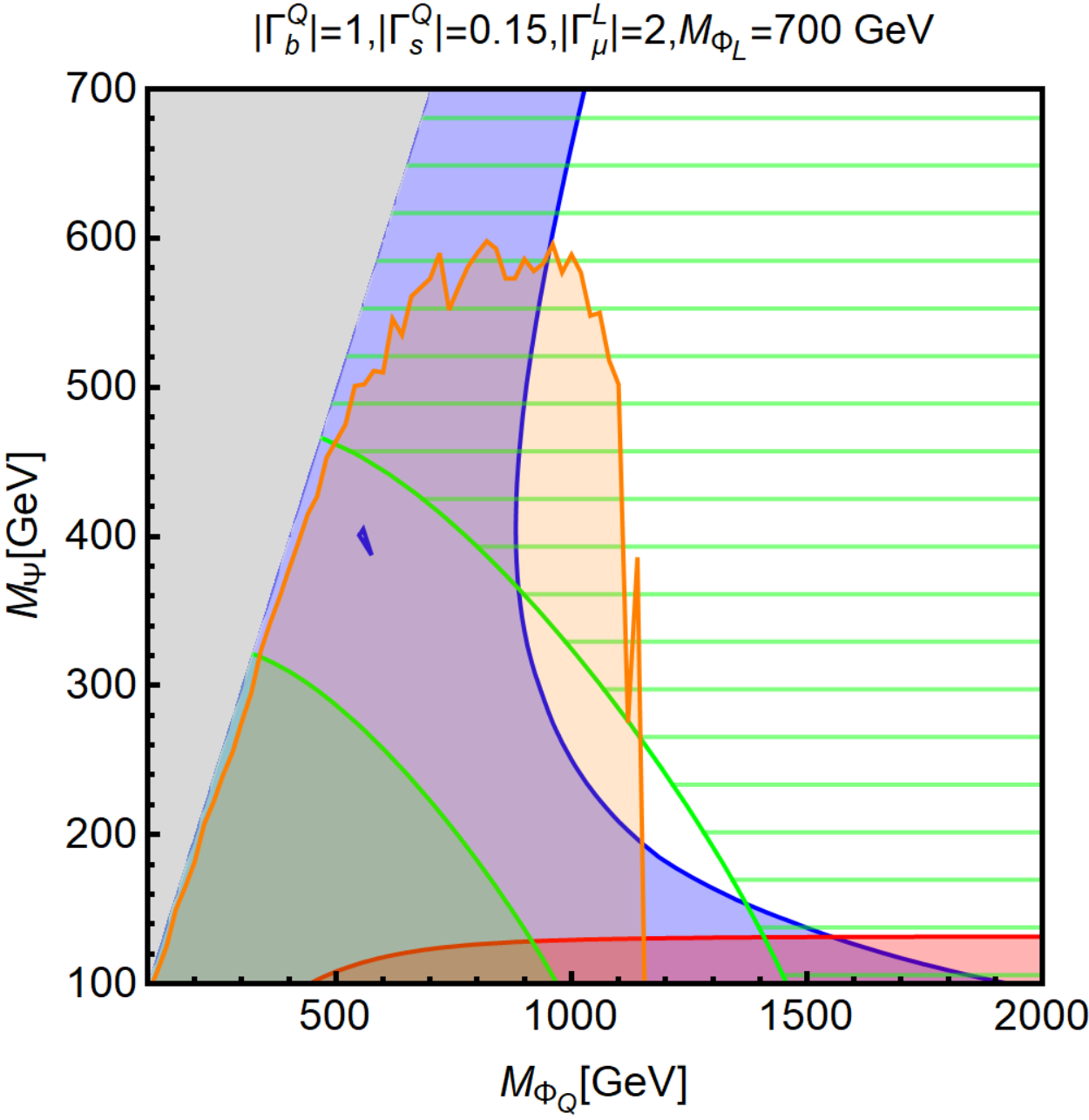}}
\subfloat{\includegraphics[width=0.33\linewidth]{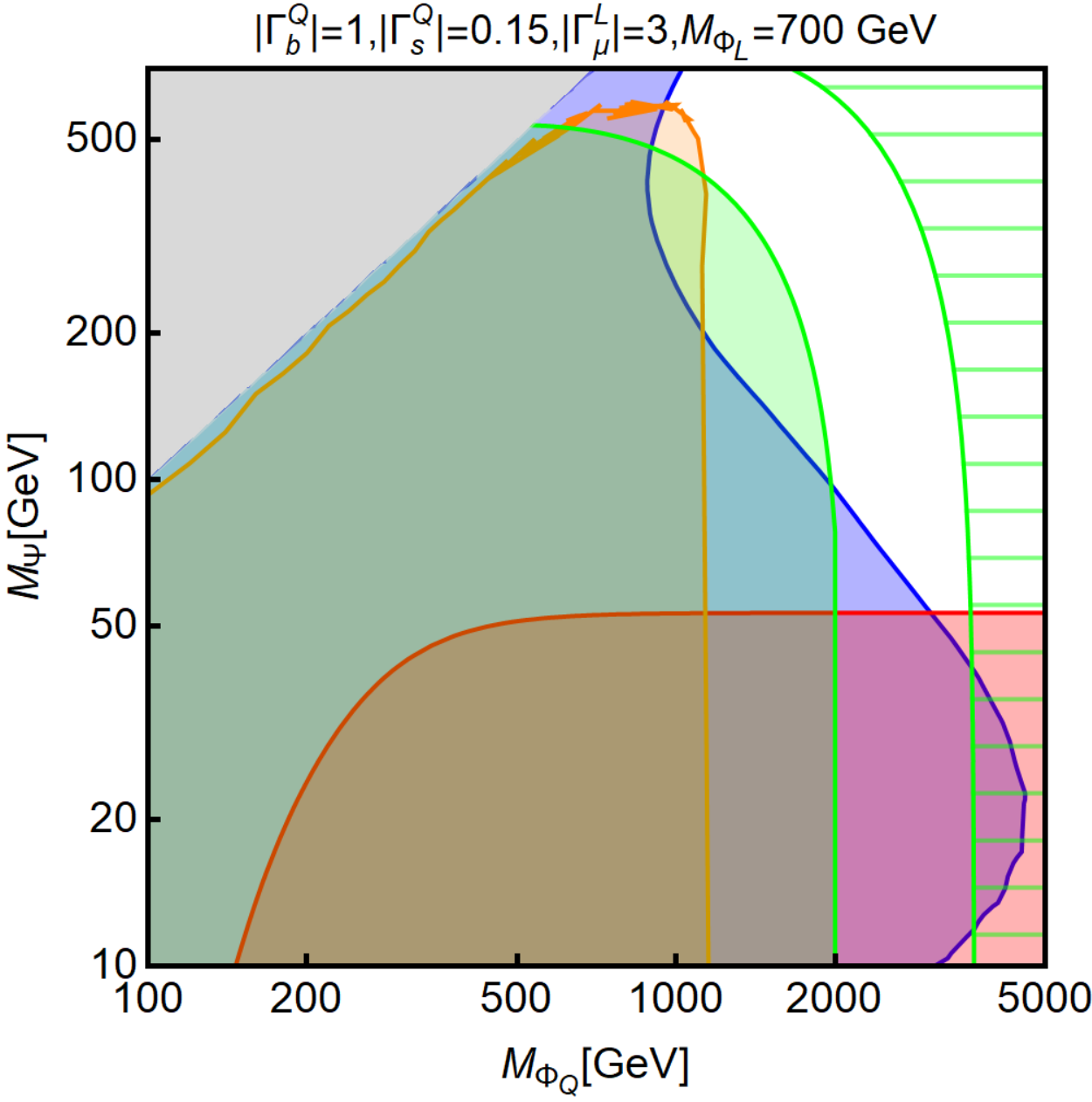}}\\
\subfloat{\includegraphics[width=0.33\linewidth]{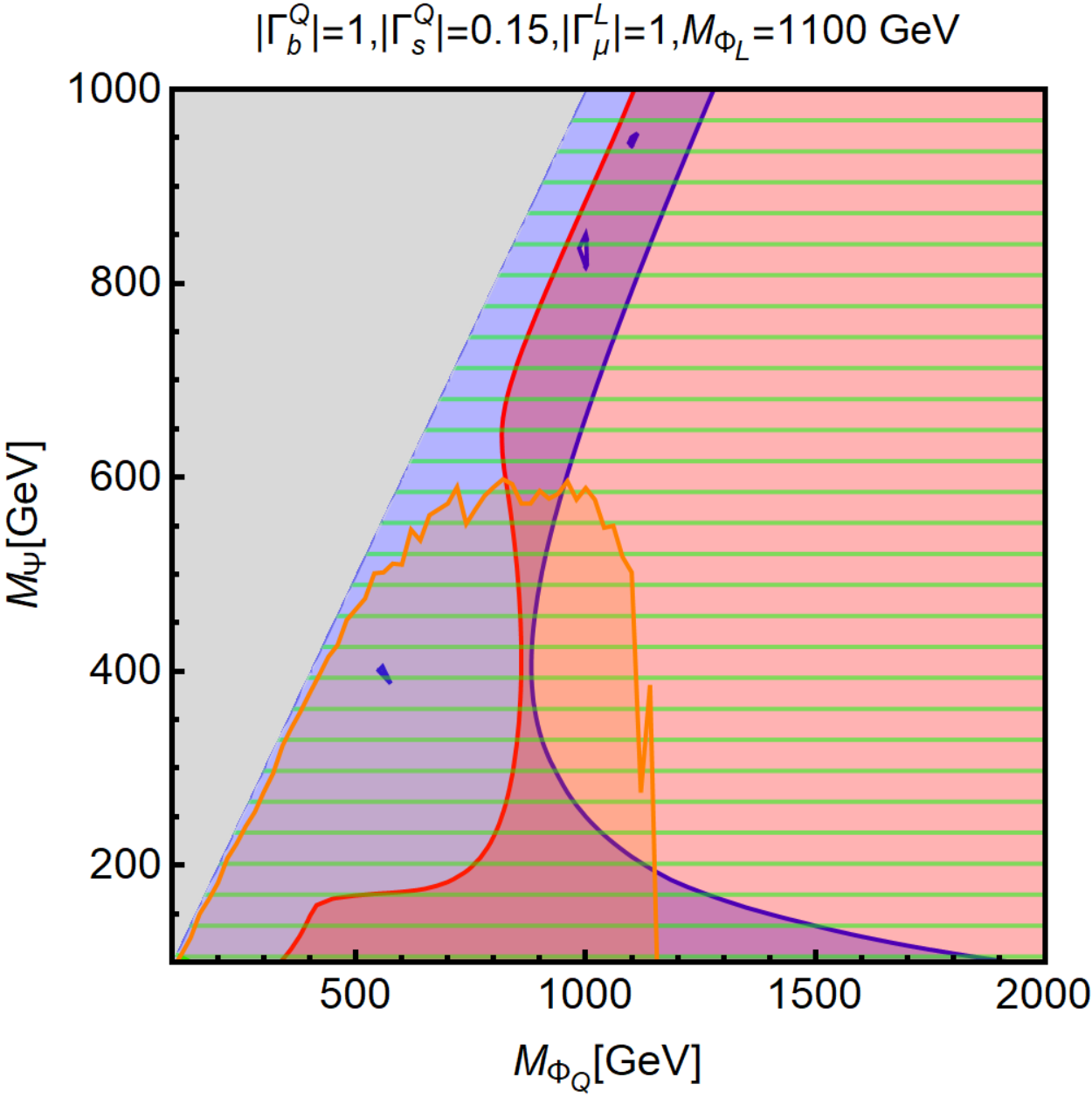}}
\subfloat{\includegraphics[width=0.33\linewidth]{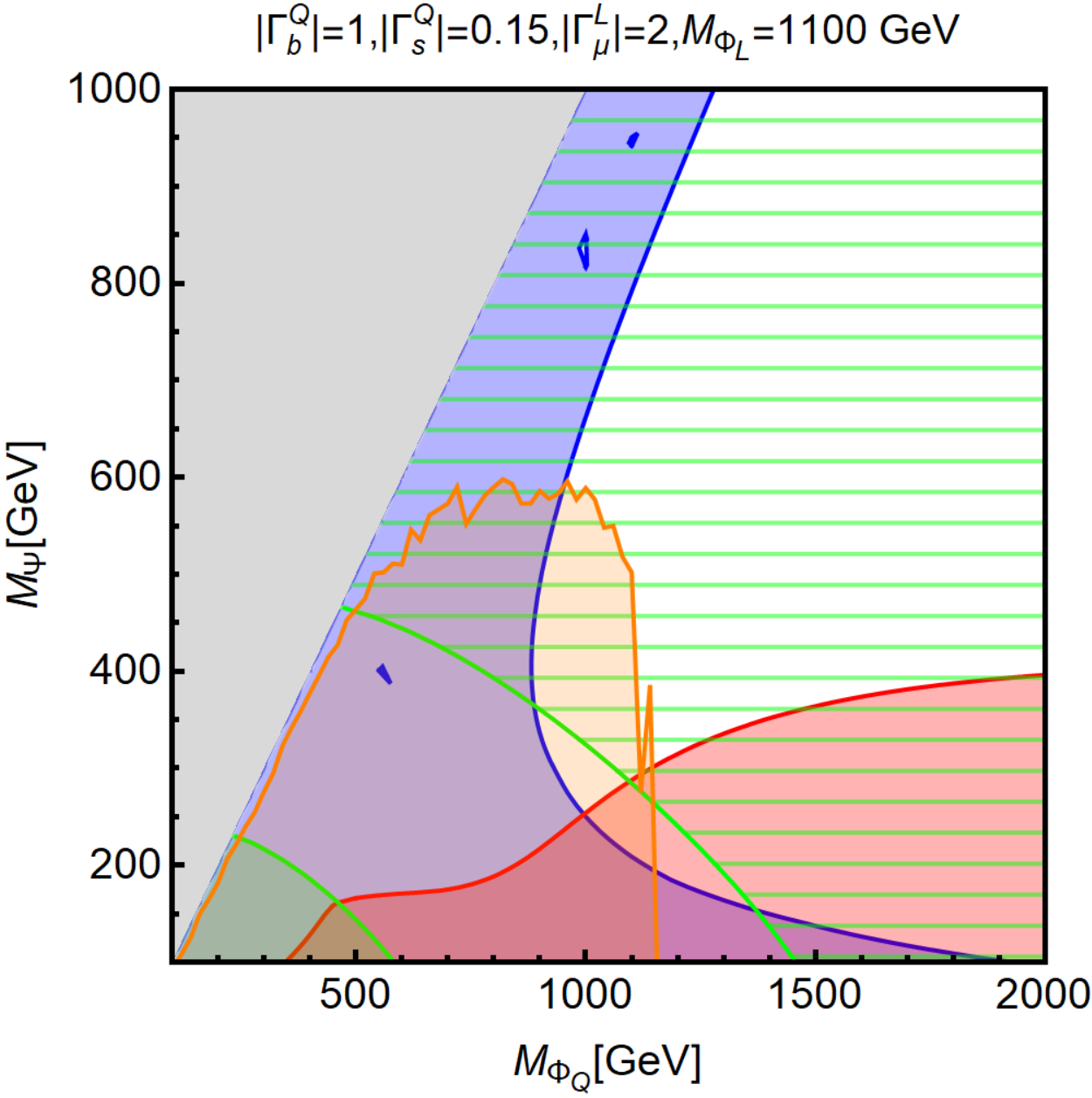}}
\subfloat{\includegraphics[width=0.33\linewidth]{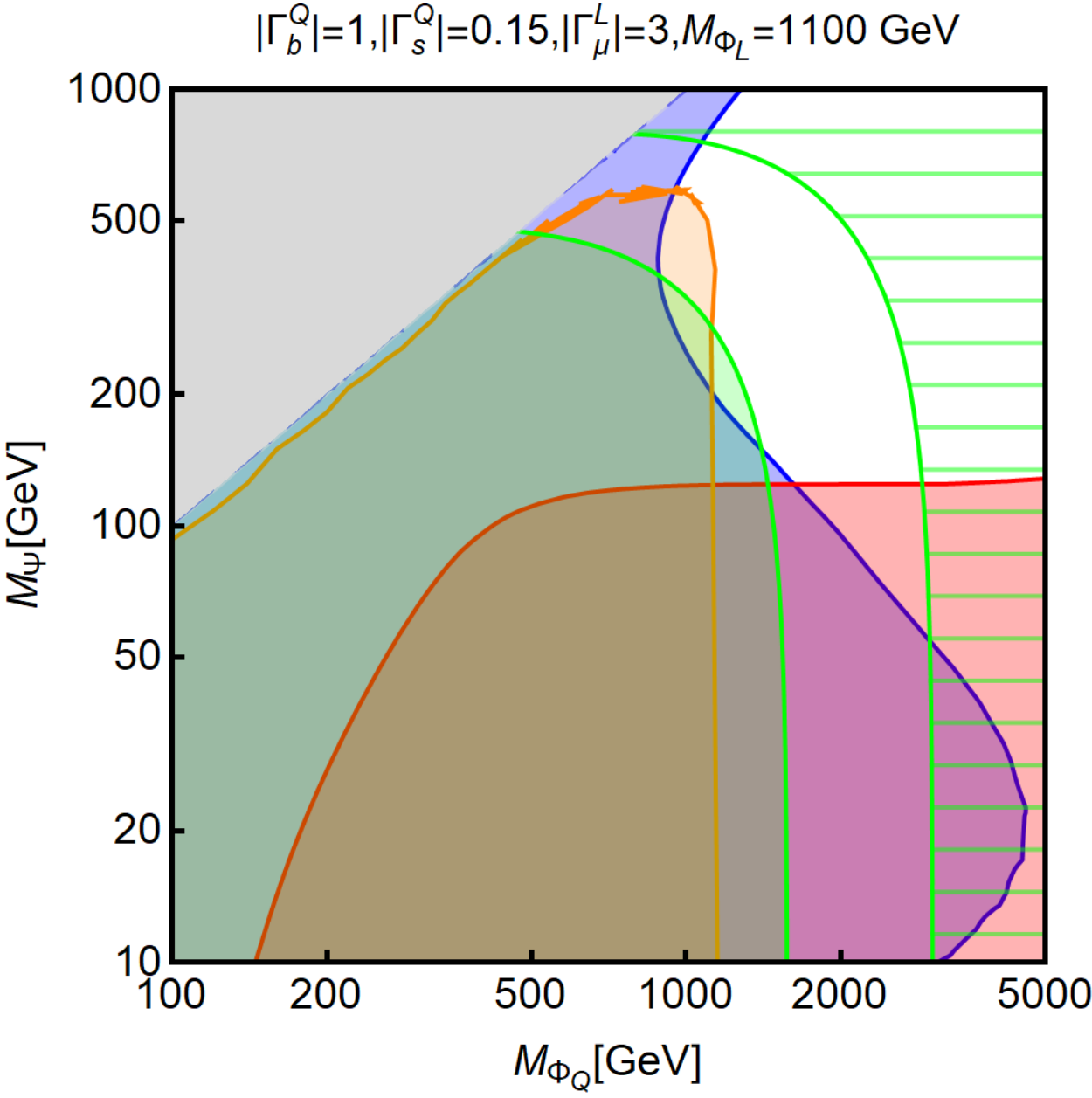}}
\caption{Summary of the constraints for the model $\mathcal{F}_\text{IA;\,0}$ with Majorana dark matter, with $|\Gamma_b^Q|=1$, $|\Gamma_s^Q|=0.15$, and the three assignments $|\Gamma_\mu^L|=1,2,3$. The upper (lower) row corresponds to $M_{\Phi_L}=700~(1100)\,\mbox{GeV}$. The colour scheme is as defined in the caption of Figure~\ref{fig:F_IA_D}.}
\label{fig:F_IA_M_A_all}
\end{figure}

As evident, in all the plots the region compatible with the flavour anomalies falls at least into one of the experimental exclusions. Among them, the strongest by far comes from direct detection, which excludes the whole range of masses considered in the different plots, besides the case $\Gamma_\mu^L=1$: with such an assignment, only a region with $M_{\Phi_Q}=4-5\,\mbox{TeV}$ survives in case \emph{i}), while a broader area is allowed in case \emph{ii}). Nevertheless, in these the NP contribution $\delta C^9_\mu = -\delta C^{10}_\mu$ is too small to account for the observed anomalies.
The direct detection bound extends even beyond multi-TeV masses for the $\Phi_Q$ field because it is actually saturated by the charge radius and magnetic dipole operators in Eq.~\eqref{eq:DM_scattering_rate} which are dominated by the contribution of the colour singlet field $M_{\Phi_L}$, whose mass and coupling $\Gamma_\mu^L$ were kept fixed in the analysis. The case $M_{\Phi_L}=1.1\,\mbox{TeV}$ is hence ruled out by DM DD regardless the assignment of the other parameters. For the same reasons we have shown no plot for $M_{\Phi_L}=700\,\mbox{GeV}$ since, in such a case, also the case $\Gamma_\mu^L=1$ would be completely ruled out by direct detection.


\begin{figure}[!t]
\centering
\subfloat{\includegraphics[width=0.33\linewidth]{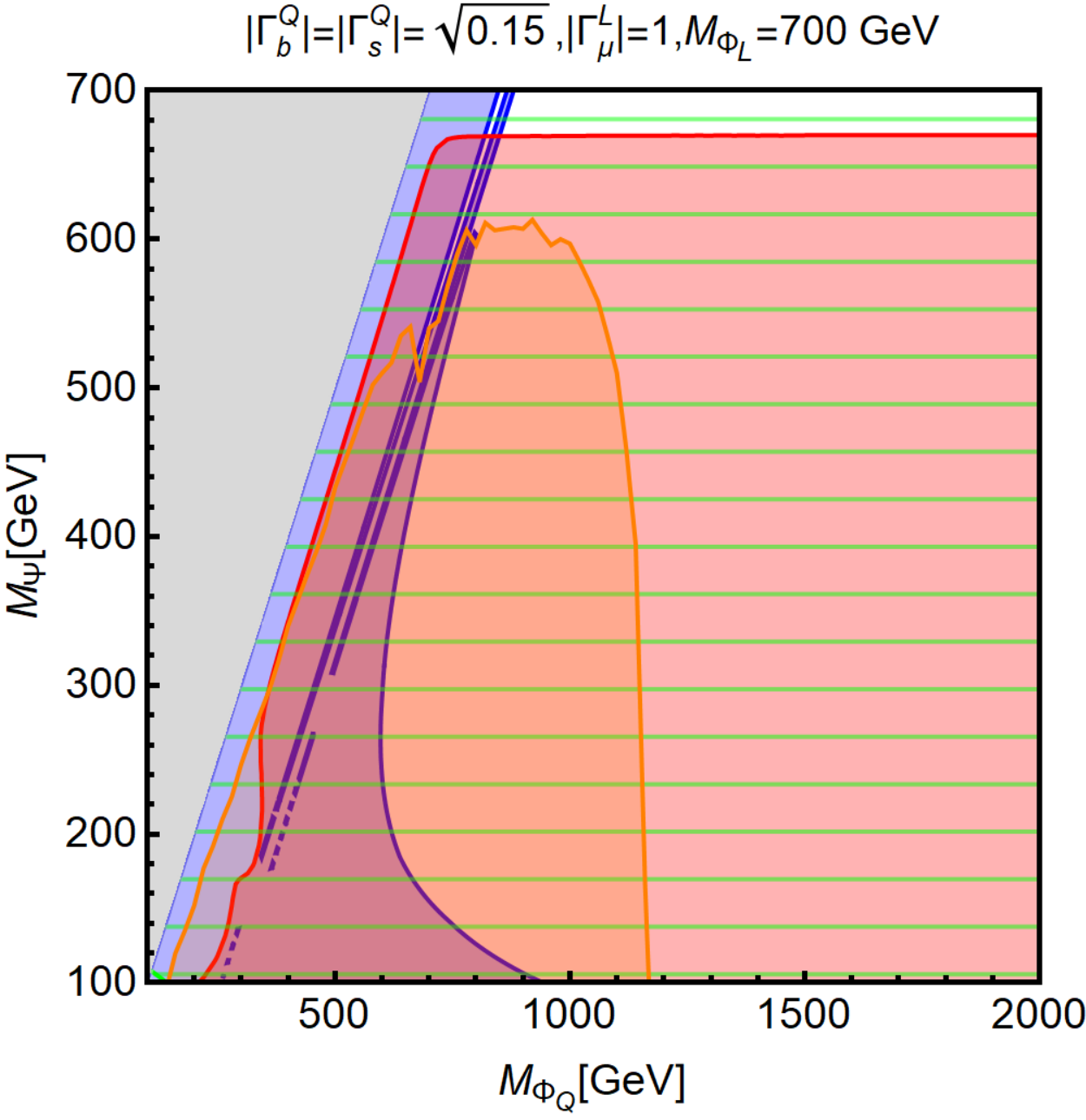}}
\subfloat{\includegraphics[width=0.33\linewidth]{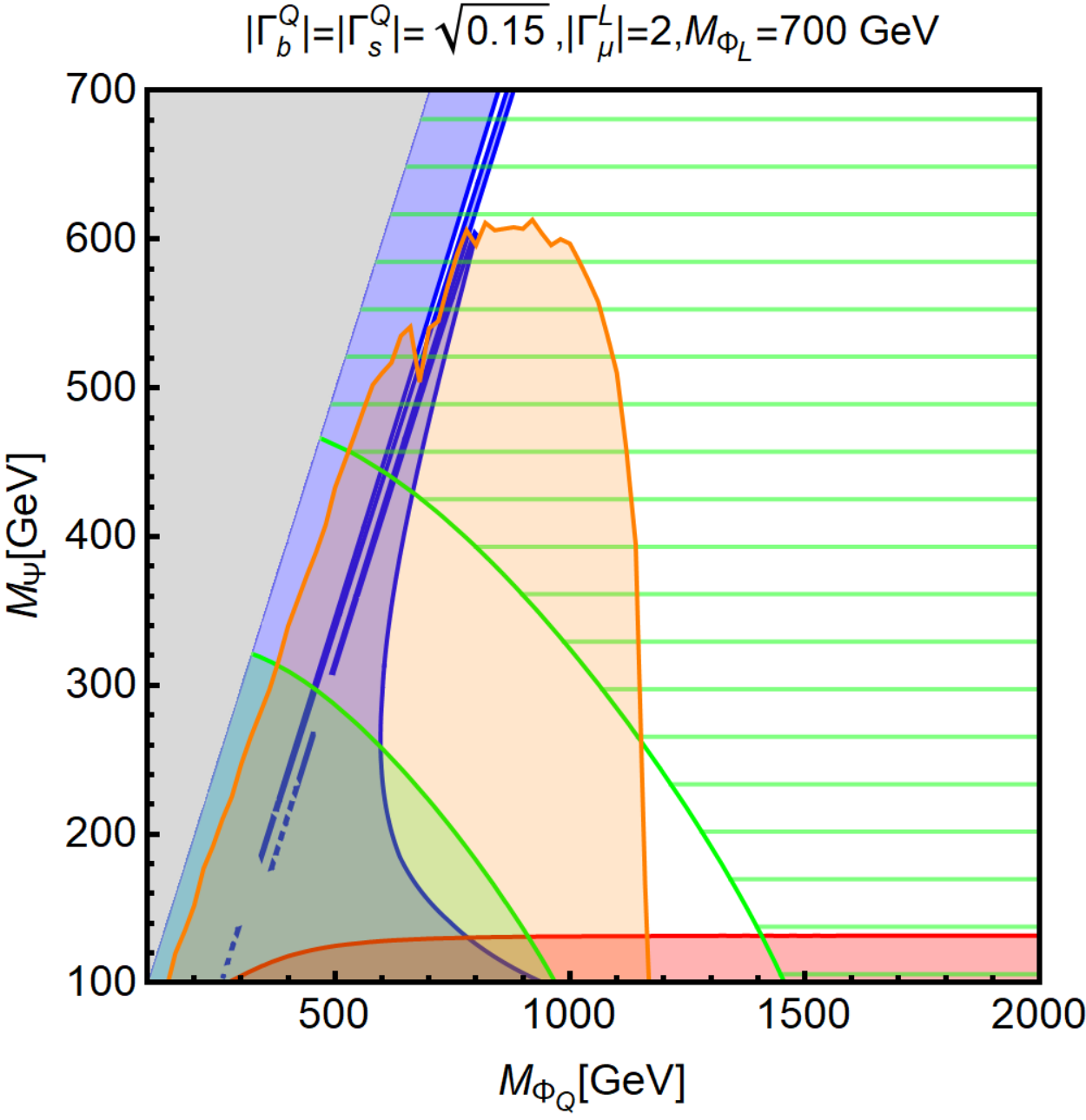}}
\subfloat{\includegraphics[width=0.33\linewidth]{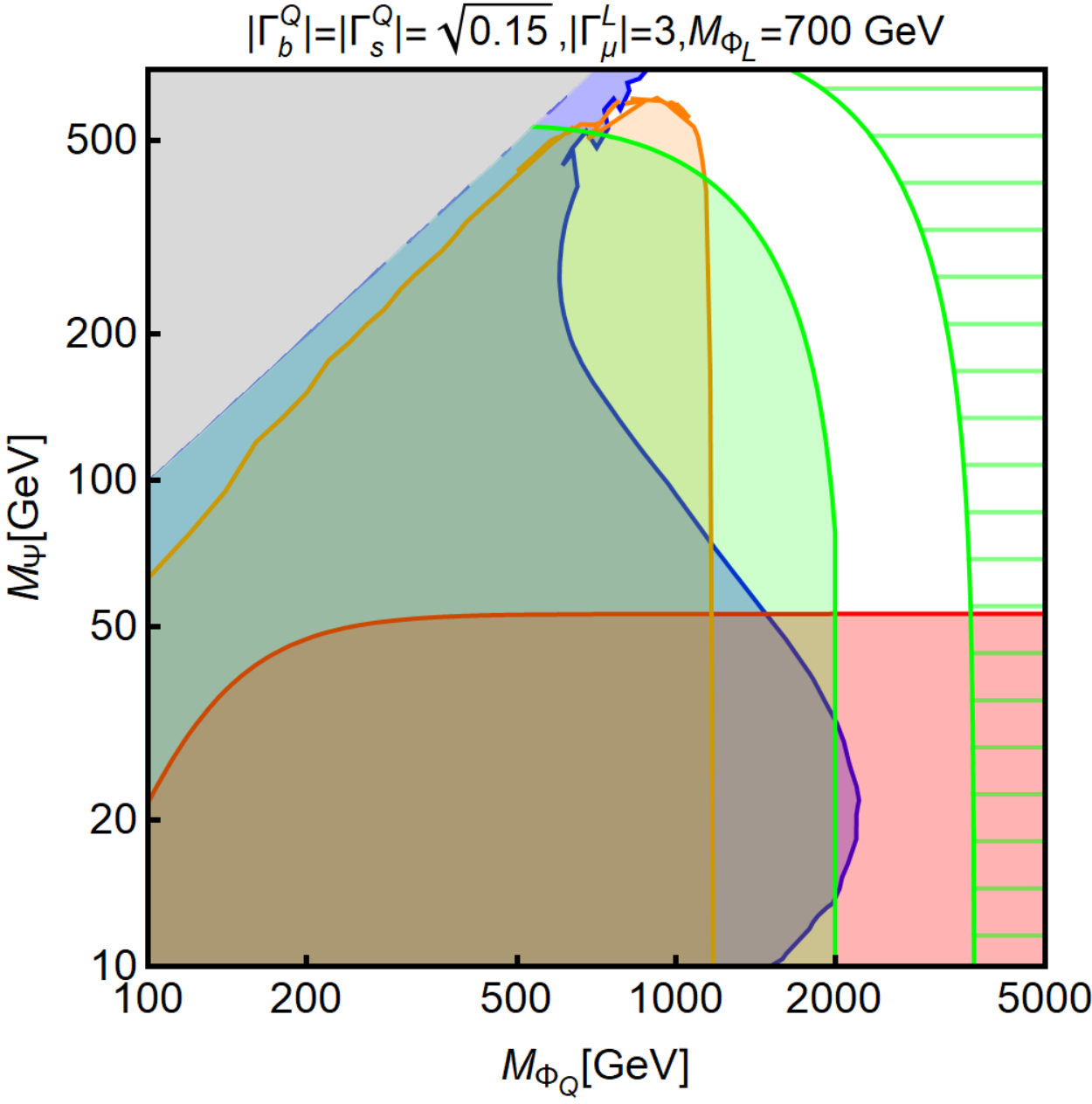}}\\
\subfloat{\includegraphics[width=0.33\linewidth]{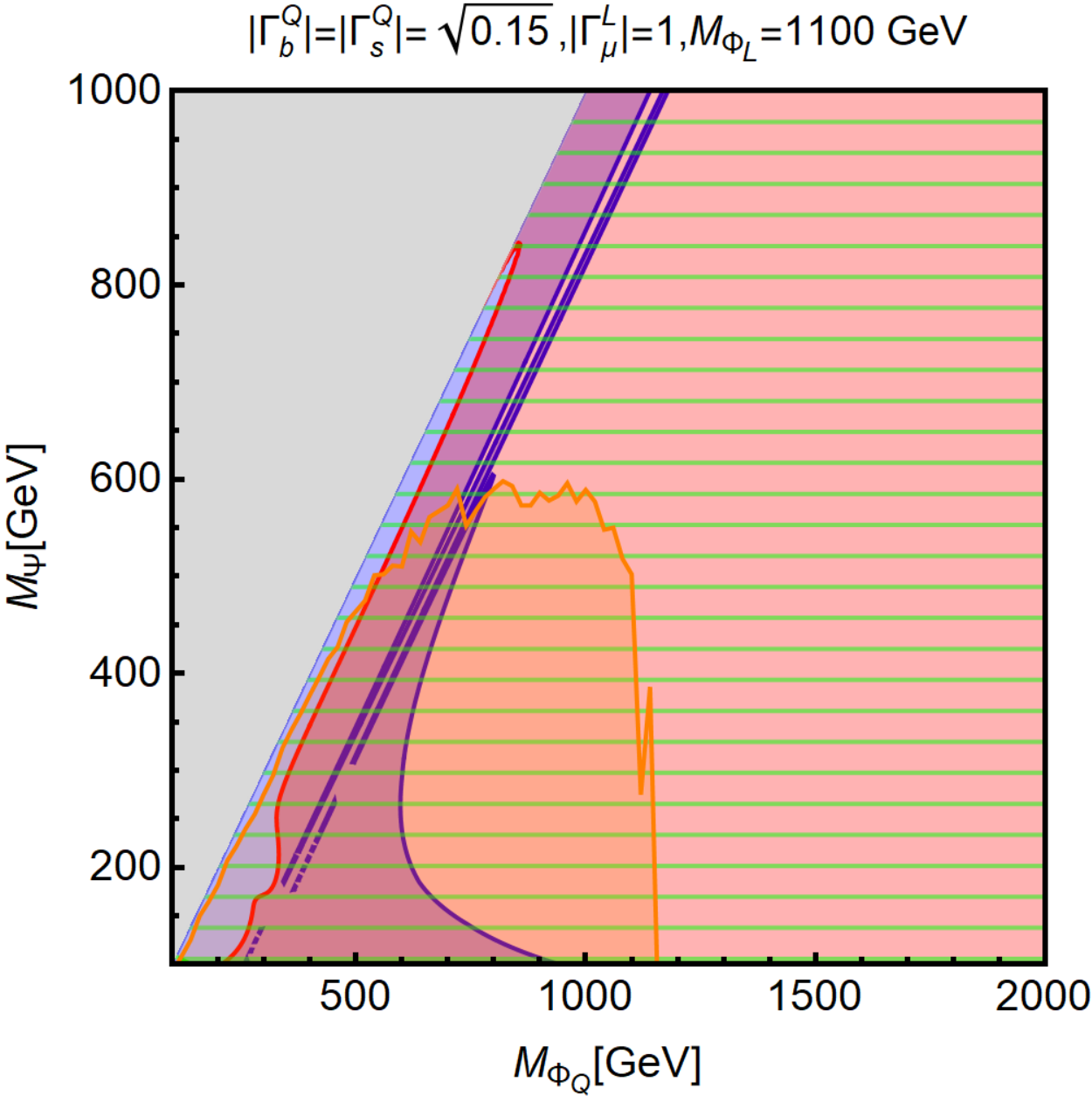}}
\subfloat{\includegraphics[width=0.33\linewidth]{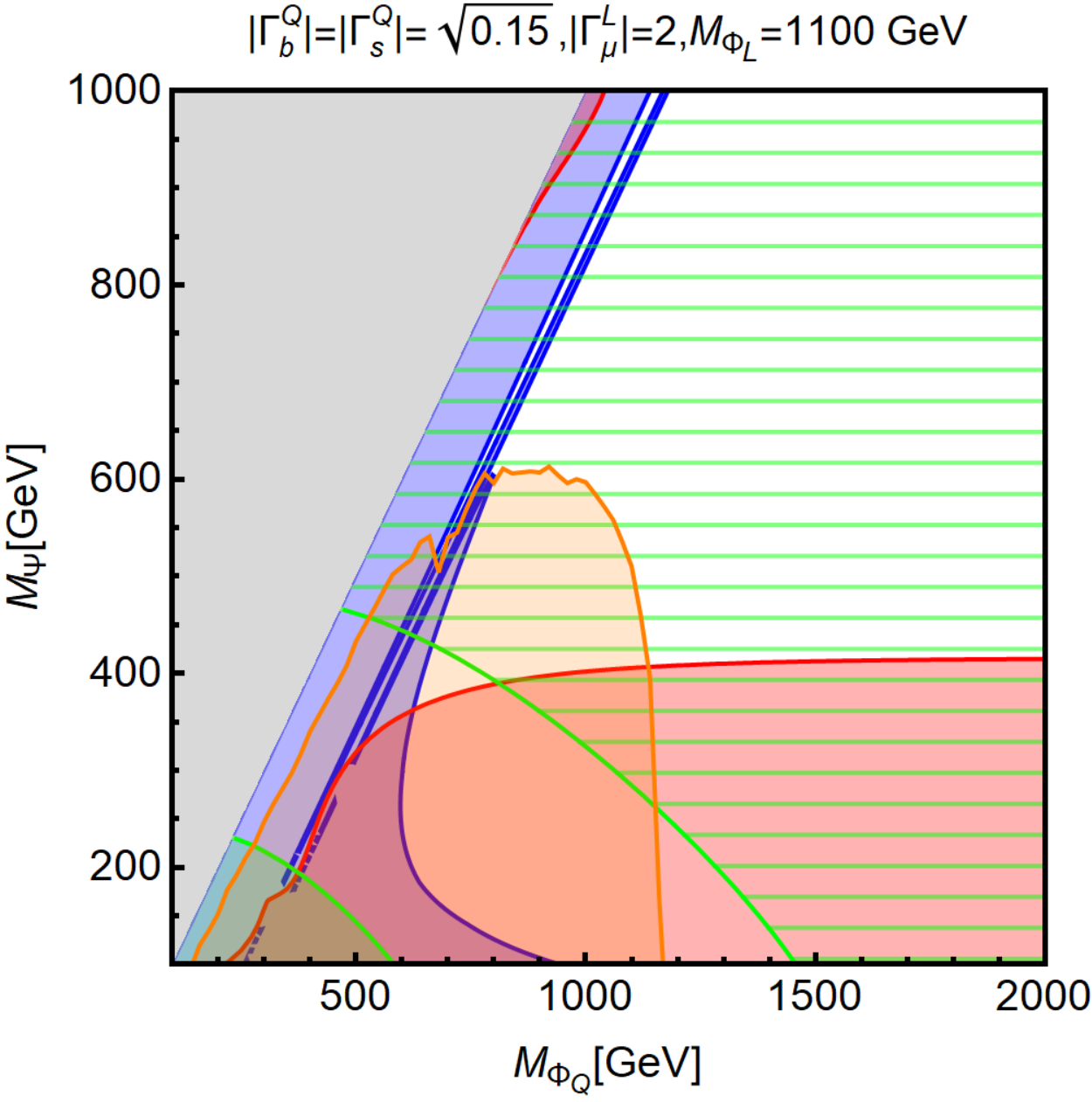}}
\subfloat{\includegraphics[width=0.33\linewidth]{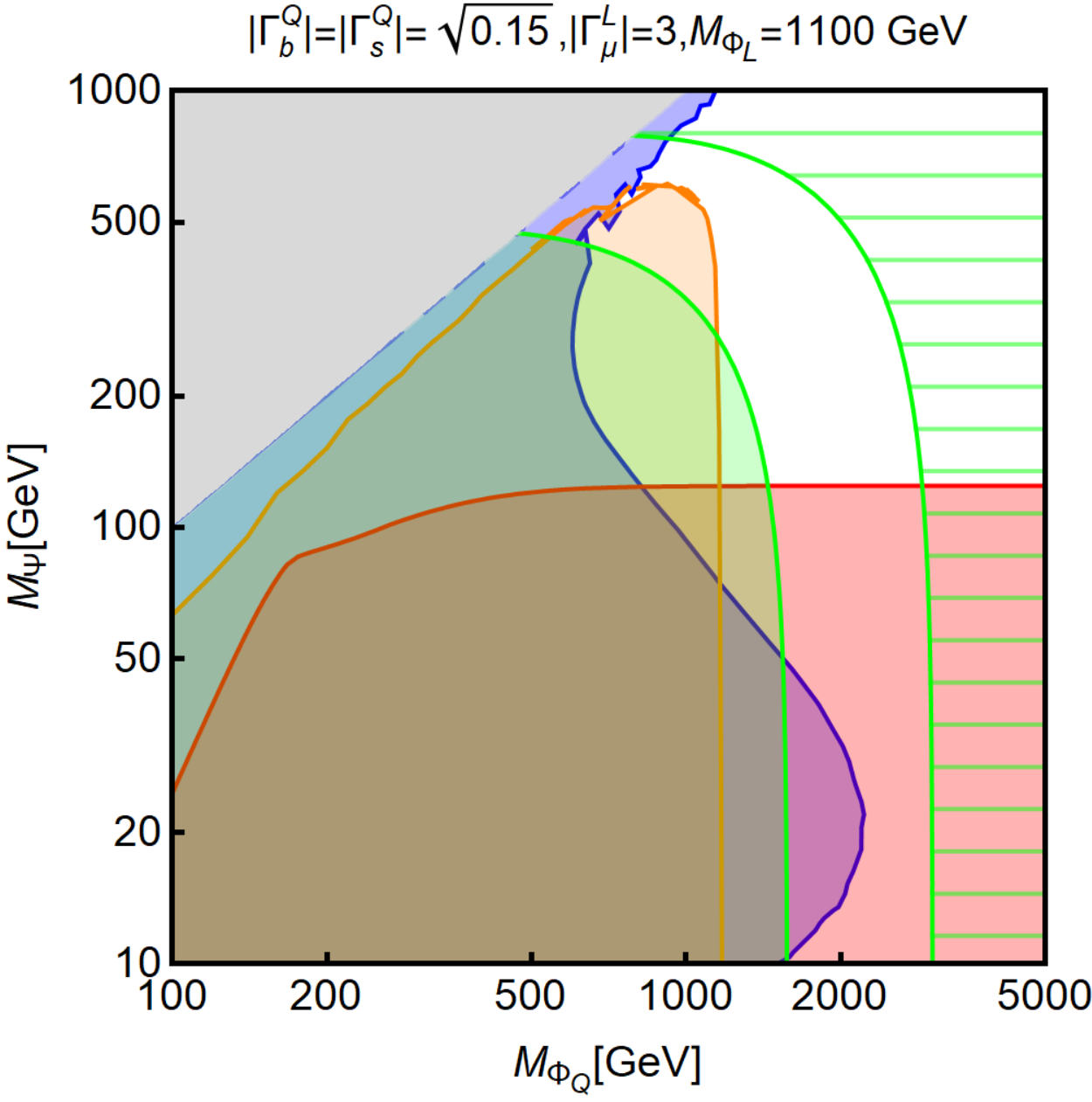}}
\caption{Summary of the constraints for the model $\mathcal{F}_\text{IA;\,0}$ with a Majorana dark matter, with $|\Gamma_b^Q|=|\Gamma_s^Q|=\sqrt{0.15}$, and the three assignments $|\Gamma_\mu^L|=1,2,3$. The upper (lower) row corresponds to $M_{\Phi_L}=700~(1100)\,\mbox{GeV}$.  
The colour scheme is as defined in the caption of Figure~\ref{fig:F_IA_D}.}
\label{fig:F_IA_M_B_all}
\end{figure}

\subsection{\texorpdfstring{$\mathcal{F}_\text{IA;\,0}$}{FIA},  Majorana singlet DM}
\label{sec:FIA_M}
In this subsection we study the same model discussed in the previous one, defined by the Lagrangian in Eq.~\eqref{eq:L_IIA_D}. The only difference is the nature of the DM field $\Psi$, now corresponding to a Majorana fermion. As pointed out above, this kind of scenario is particularly interesting since it features the highest degree of correlations between $B$-anomalies and the other phenomenological observables considered in the present study. A model analogous to $\mathcal{F}_\text{IA;\,0}$ with Majorana DM has been already studied in Ref.~\cite{Cerdeno:2019vpd}. The analysis in the latter reference differs from the present work in the fact that they strictly imposed the requirement of the correct relic density and used it to fix $M_{\Phi_L}$ as a function of the other parameters, in particular the coupling $\Gamma_\mu^L$. Furthermore, the latter parameter has been allowed to reach the perturbativity bound $\Gamma_\mu^L=\sqrt{4\pi}$. As already mentioned, we adopt here a more conservative approach for what concerns the DM relic density, keeping $M_{\Phi_L}$ and $\Gamma_\mu^L$ as free parameters in the fit of flavour observables and considering $\Gamma_\mu^L \leq 3$ in our phenomenological study. Given these different assumptions our findings slightly differ from the ones reported in Ref.~\cite{Cerdeno:2019vpd}.

We repeat the analysis whose procedure has been illustrated in detail in the previous section. The  results are shown in Figures~\ref{fig:F_IA_M_A_all} and~\ref{fig:F_IA_M_B_all}. As we can see, we find a very different picture with respect to the results of the previous case as shown in Figure~\ref{fig:F_IA_D}. This is a consequence of the different nature of the fermionic DM candidate. One can indeed find regions of the parameter space compatible with the flavour anomalies and not in tension with other  experimental bounds, even for the lighter benchmark value $M_{\phi_L}=700\,\mbox{GeV}$. For $\Gamma_\mu^L \gtrsim 2$ it is also possible to have, albeit in a narrow window of the parameter space, a good fit of the flavour anomalies and, simultaneously, saturate the observed DM relic density ($\Omega_\text{DM}h^2\simeq 0.12$) with a standard thermal WIMP. This occurs for DM masses between approximately 50 and 150~GeV.      

This different outcome compared to the Dirac case is mostly due to the fact that, for Majorana DM, the effective operators accounting for DM interactions with nucleons mediated by the $Z$ boson and the photon identically vanish. As a consequence, the strength of the interactions between DM and nuclei is strongly reduced and, hence, DD bounds are significantly weaker. Furthermore, the annihilation cross-section of Majorana DM is $p$-wave suppressed, and thus ID constraints are not present. 
Although this is a successful scenario according to our criteria, we remark again that, in order to have compatibility between the fit of the flavour anomaly and the other constraints, rather large values of the coupling to muons $|\Gamma_\mu^L|$ are required: $|\Gamma_\mu^L| \gtrsim 2$ for $M_{\Phi_L} \gtrsim 700\,\mbox{GeV}$ and $|\Gamma_\mu^L| \gtrsim 3$ for $M_{\Phi_L} \gtrsim 1100\,\mbox{GeV}$. 


\subsection{\texorpdfstring{$\mathcal{F}_\text{IB;\,-1/3}$}{FIB},  Real-scalar doublet DM}
\begin{figure}[!t]
\centering
\subfloat{\includegraphics[width=0.33\linewidth]{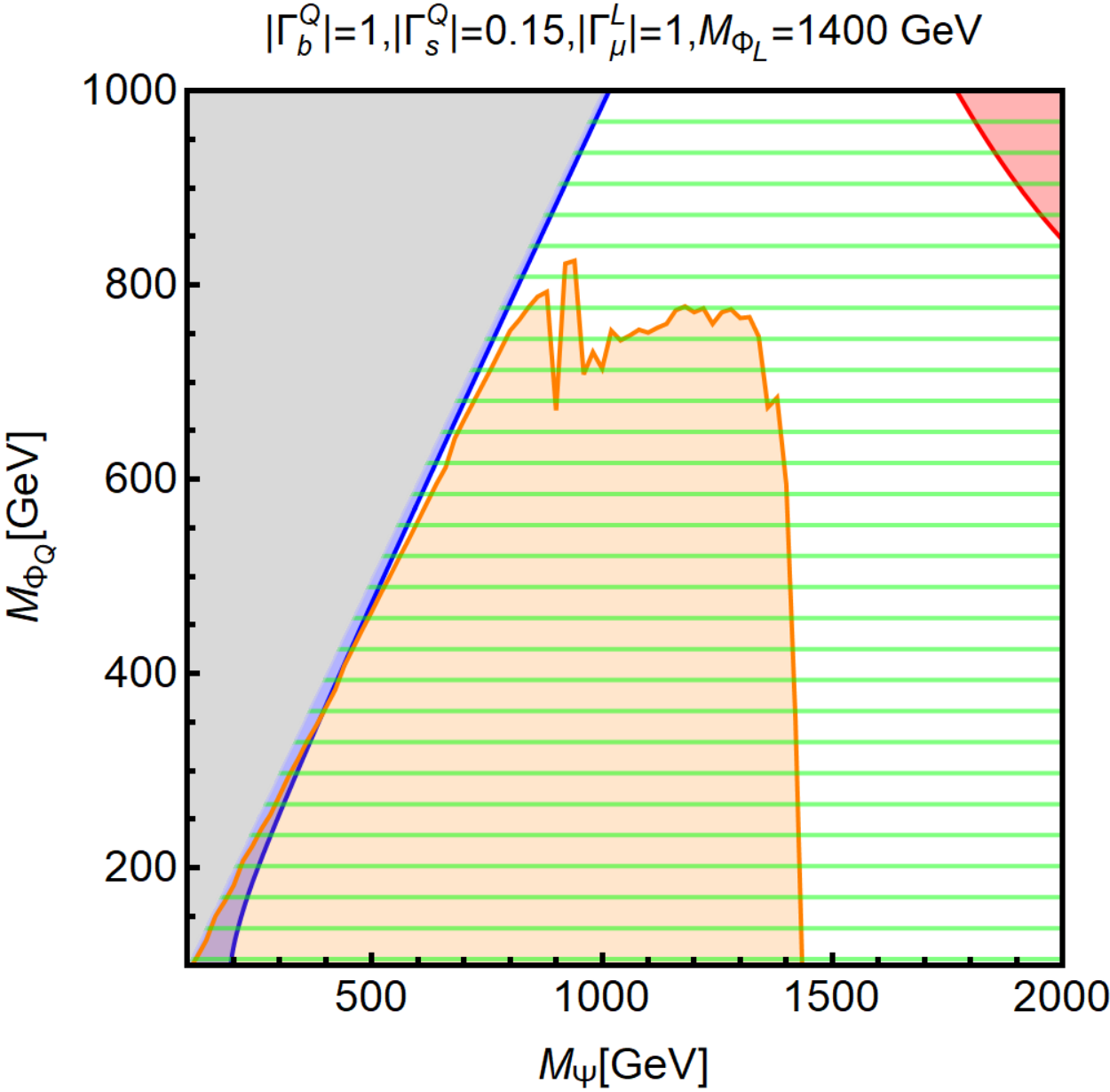}}
\subfloat{\includegraphics[width=0.33\linewidth]{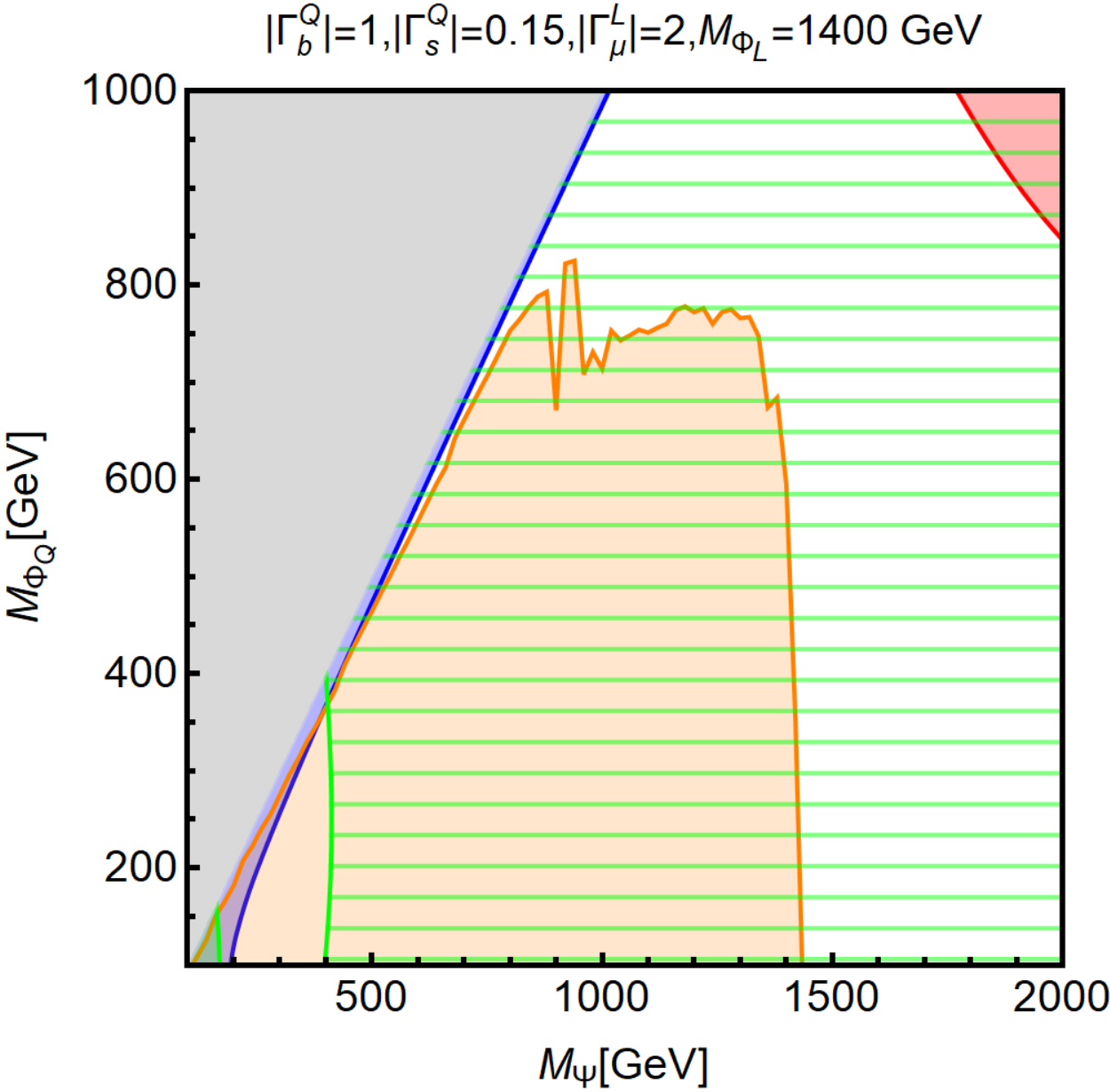}}
\subfloat{\includegraphics[width=0.33\linewidth]{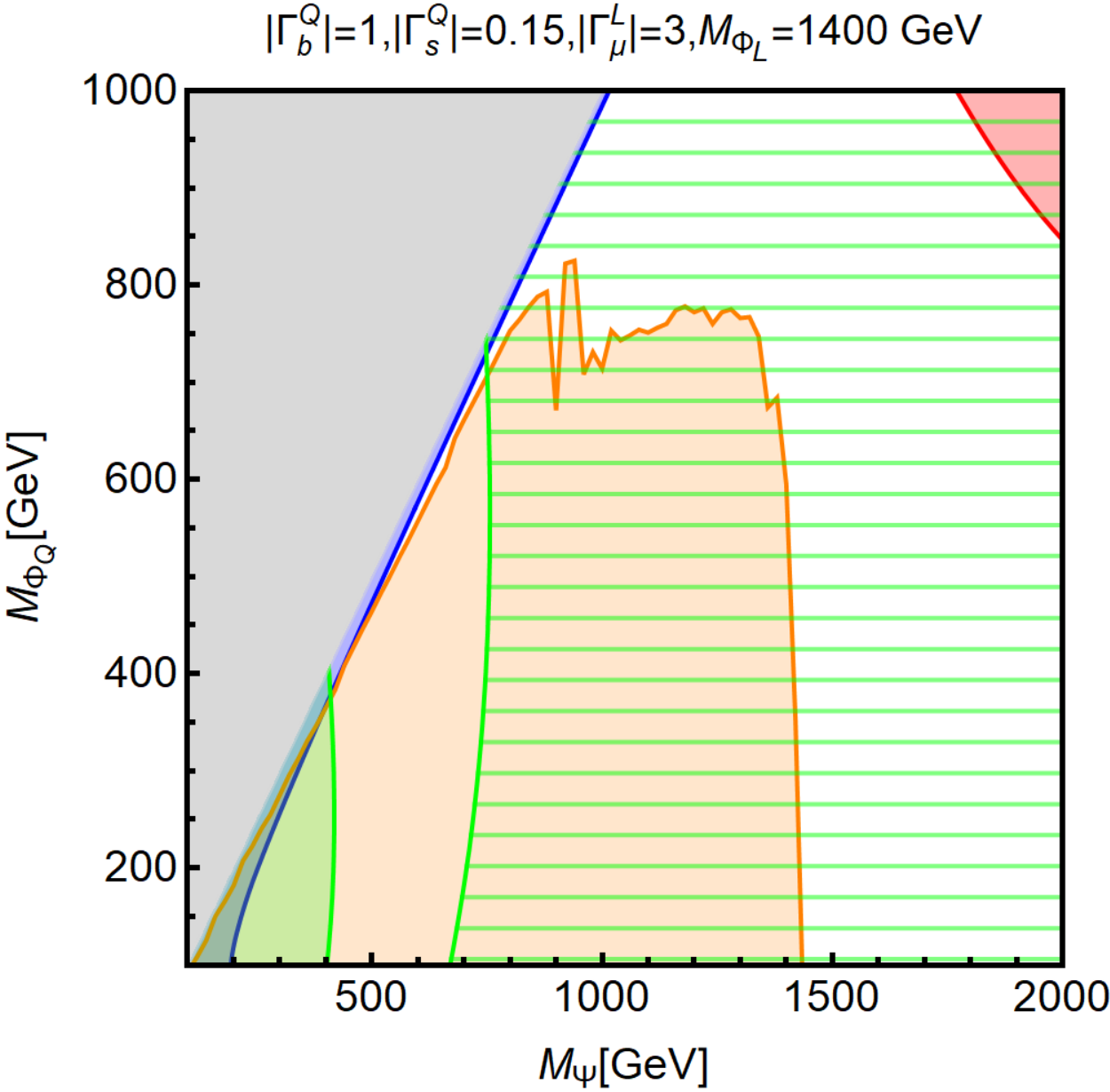}}\\
\subfloat{\includegraphics[width=0.33\linewidth]{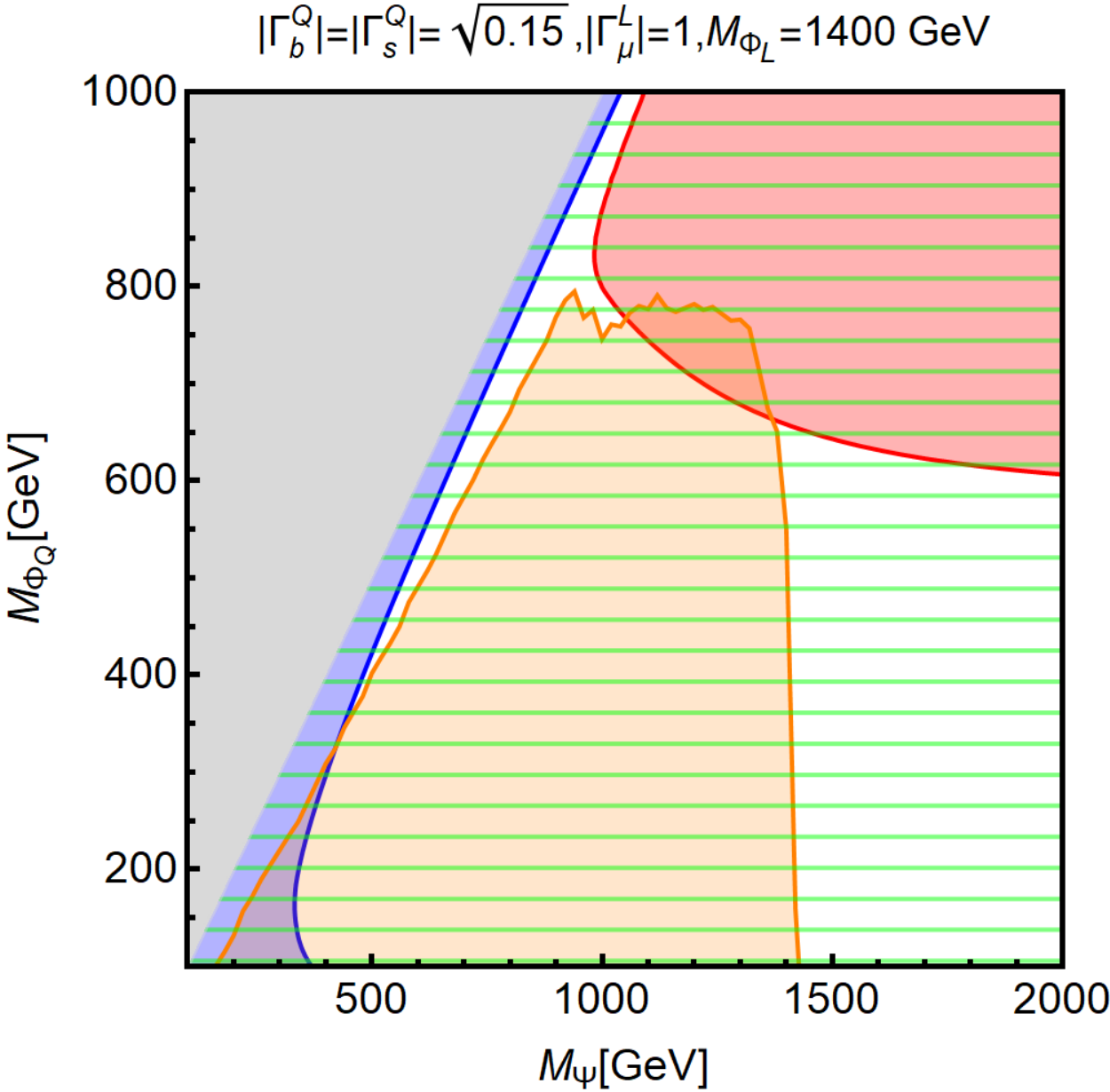}}
\subfloat{\includegraphics[width=0.33\linewidth]{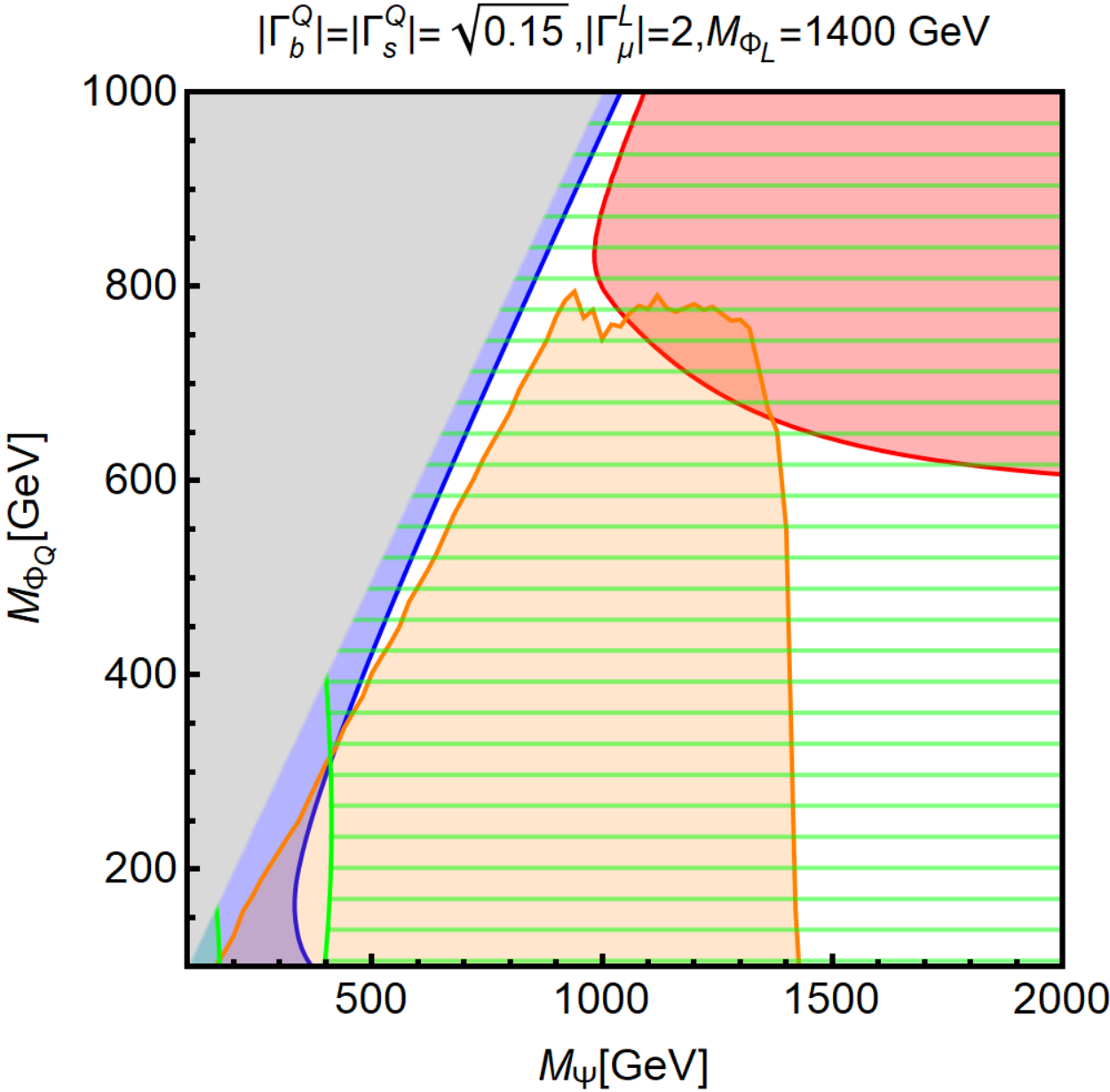}}
\subfloat{\includegraphics[width=0.33\linewidth]{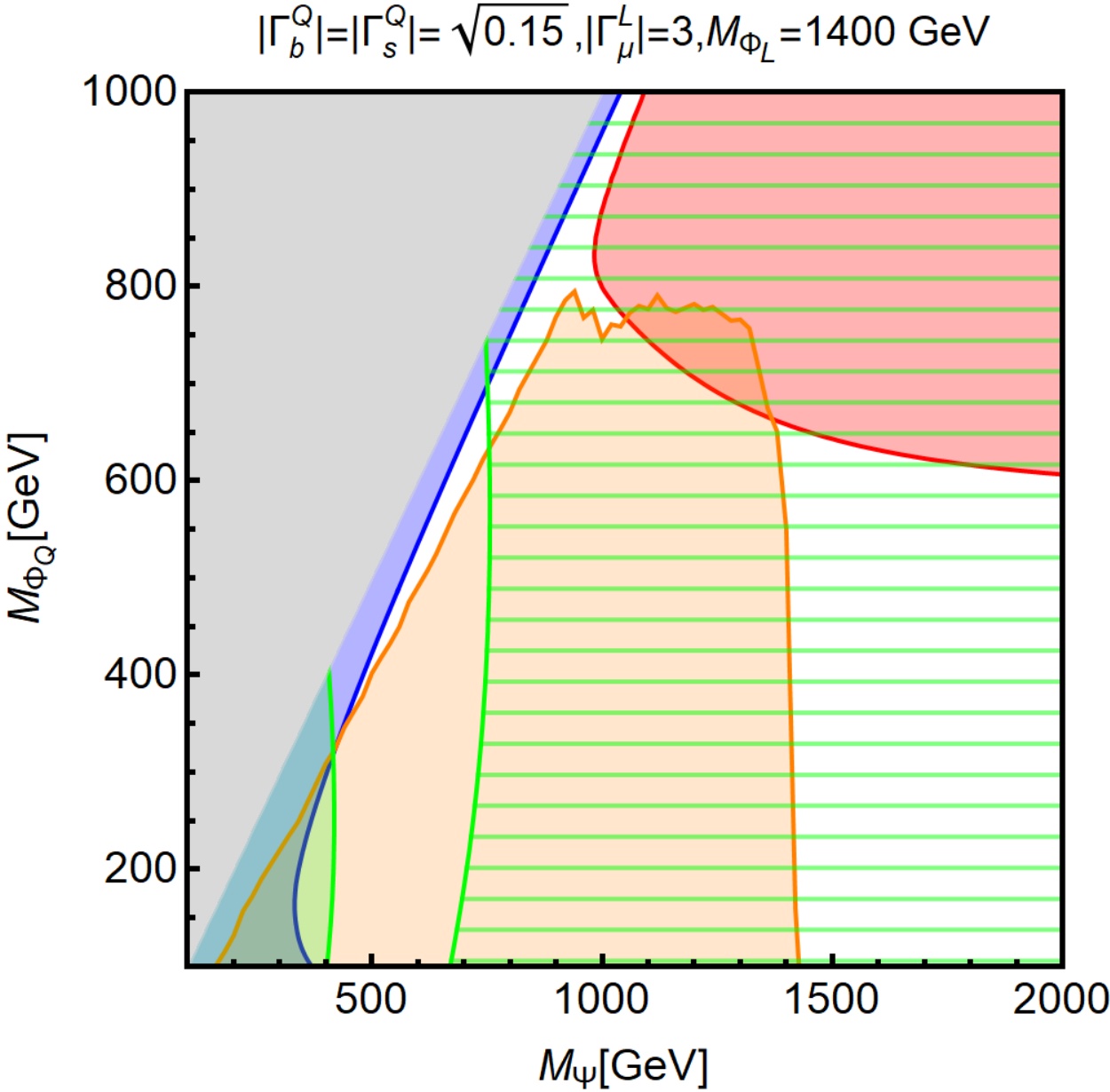}}
\caption{Summary of the constraints for the model $\mathcal{F}_\text{IB;\,-1/3}$ with $|\Gamma_b^Q|=1$, $|\Gamma_s^Q|=0.15$, and the three assignments $|\Gamma_\mu^L|=1,2,3$. The upper (lower) row correspond to configurations i) and ii) for the couplings of the NP fields. In all cases we have set $M_{\Phi_L}=1400\,\mbox{GeV}$.  The colour scheme is as defined in the caption of Figure~\ref{fig:F_IA_D}.}
\label{fig:F_IB_A_all}
\end{figure}
We now turn to consider an example with scalar DM.
In this case, DM is a (real) neutral state which is
part of the scalar $SU(2)$ doublet $\Phi_Q$, 
({\bf 1},\,{\bf 2},\,1/2).
The other NP fields $\Phi_L$ and $\Psi$ transform under the SM gauge group respectively as 
$({\bf\bar 3},\,{\bf 2},\,-1/6)$ and $({\bf 3},\,{\bf 1},\,-1/3)$. The Lagrangian resembles that of the previously-considered model:
\begin{eqnarray}
{\cal L}_{\rm int} =  {\Gamma^Q_i\bar Q_i}{P_R}{\Psi}{\Phi_Q} + \Gamma^L_i \bar L_i {P_R}{\Psi}{\Phi_L} +
{\rm{h}}{\rm{.c.}}
\label{eq:L_IIB_S}
\end{eqnarray}
 This is the simplest model, with our general classification, featuring the DM belonging to a $SU(2)$ doublet. As discussed in Section~\ref{sec:setup}, our minimality criteria only allow scalar DM to belong to an $SU(2)$ multiplet. 
 Notice also that, as usual, the DM field $\Phi_Q$ and the other NP scalar $\Phi_L$ could couple with the Higgs field and among each other through operators of the type $|\Phi_Q|^2 |\Phi_L|^2$, $|H|^2 |\Phi_Q|^2$ and $|H|^2 |\Phi_L|^2$. As already pointed out, we are assuming in this work that these quartic couplings are so small that have a negligible impact on phenomenology. The only exception is given by the coupling providing a mass splitting between real and imaginary parts of the neutral component of $\Phi_Q$, as discussed below.
 In fact, since the DM belongs to an $SU(2)$ doublet with non-zero hypercharge, it would feature as well tree level interactions with the $Z$-boson. In order to circumvent this possibility (which is already experimentally ruled out, see e.g.~Ref.~\cite{Arcadi:2017kky}), we assume that, similarly to the so-called inert doublet model~\cite{Barbieri:2006dq,LopezHonorez:2006gr,Honorez:2010re,LopezHonorez:2010tb}, the neutral component of $\Phi_Q$ can be separated into a CP-even state, which we assume here to be the DM candidate, and a CP-odd state with a sufficient mass splitting ($\gtrsim \mathcal{O}(100)$~keV) to avoid DD constraints. This can be achieved through a quartic operator involving the SM Higgs of the form 
$(|\Phi_Q^\dag H|^2 + \text{h.c.})$. Since a tiny coupling $\mathcal{O}(10^{-13})$ is enough
to induce an~$\mathcal{O}(100)$~keV mass splitting, we can safely assume that no other phenomenologically relevant effect follows from this (and other) Higgs-portal interactions.

The combined constraints on this model are shown, with the usual colour coding, in Figures~\ref{fig:F_IB_A_all} in the $(M_{\Psi},M_{\Phi_Q})$ plane (notice that what we label $M_{\Phi_Q}$ is just the DM mass). 
In contrast to the previous model, the strongest constraints come from flavour and LHC.\footnote{Here we have not considered bounds from disappearing tracks that would arguably have little sensitivity to the small production cross section and life-time of the charged states of the doublet.} 
First, notice that we set $M_{\Phi_L}=1400\,\mbox{GeV}$, in order to evade bounds on $\Phi_L$ production and cascade decay ($pp \to \Phi_L\Phi_L \to \mu^+\mu^- +\,\Phi\Phi \to   \mu^+\mu^- + qq^\prime  + \slashed{E}_T$)
from the LHC searches in Refs.~\cite{Sirunyan:2020tyy,ATLAS:2020dav}.
The dominant collider bound shown in the plots follows from the process $pp \to \Psi \Psi \to qq^\prime+  \Phi_Q \Phi_Q$, to which the CMS search of Ref.~\cite{Sirunyan:2019ctn} is sensitive.
On the other hand, the flavour anomalies are accounted for only for relatively light values of $M_\Psi$ which fall in the region excluded by the CMS search,\footnote{This bound extends to substantially larger masses than in the previous case $\mathcal{F}_\text{IA;\,0}$, as the produced particle here is a fermion, hence its production cross section is about one order of magnitude larger than that of a scalar of the same mass.} with the exception of narrow tuned strips on the $(M_{\Psi},M_{\Phi_Q})$ plane corresponding to a very compressed spectrum, where jets would be too soft for a substantial number of events to be selected by the experimental cuts. We remark again that, even if it is not capable of accounting for the $B$-anomalies compatibly with all experimental constraints, the model $\mathcal{F}_\text{IB;\,-1/3}$ is not strictly ruled out since LHC and DM constraints are evaded in large regions of the parameter space where the NP contributions to $B$-observables does not exceed, in a statistically relevant way, the SM ones. 
Moving to DM phenomenology, we can see from Figure~\ref{fig:F_IB_A_all} that direct detection constrains the model under consideration only poorly. This occurs because of the nature of DM: besides excluding tree-level interactions between DM pairs and the $Z$ boson, the small mass splitting we assumed also sets to zero the $\Phi_Q^\dagger i \overset{\leftrightarrow}{\partial_\mu} \Phi_Q$ operator in the DD effective Lagrangian in Eq.~\eqref{Scalar:leff}. Finally, concerning DM relic density, we note that, similarly to what occurs in the case of the inert doublet model~\cite{Barbieri:2006dq,LopezHonorez:2006gr,Honorez:2010re,LopezHonorez:2010tb}, very efficient annihilation processes into gauge bosons dominate DM production, as long as they are kinematically accessible.  Thus relatively large values of the DM mass ($\approx 500-600$~GeV) outside the region fitting the $B$-anomaly are required to have the correct relic density.


\subsection{\texorpdfstring{$\mathcal{S}_\text{IA}$}{SIA}, Complex-scalar singlet DM}
\begin{figure}[!t]
\centering
\subfloat{\includegraphics[width=0.33\linewidth]{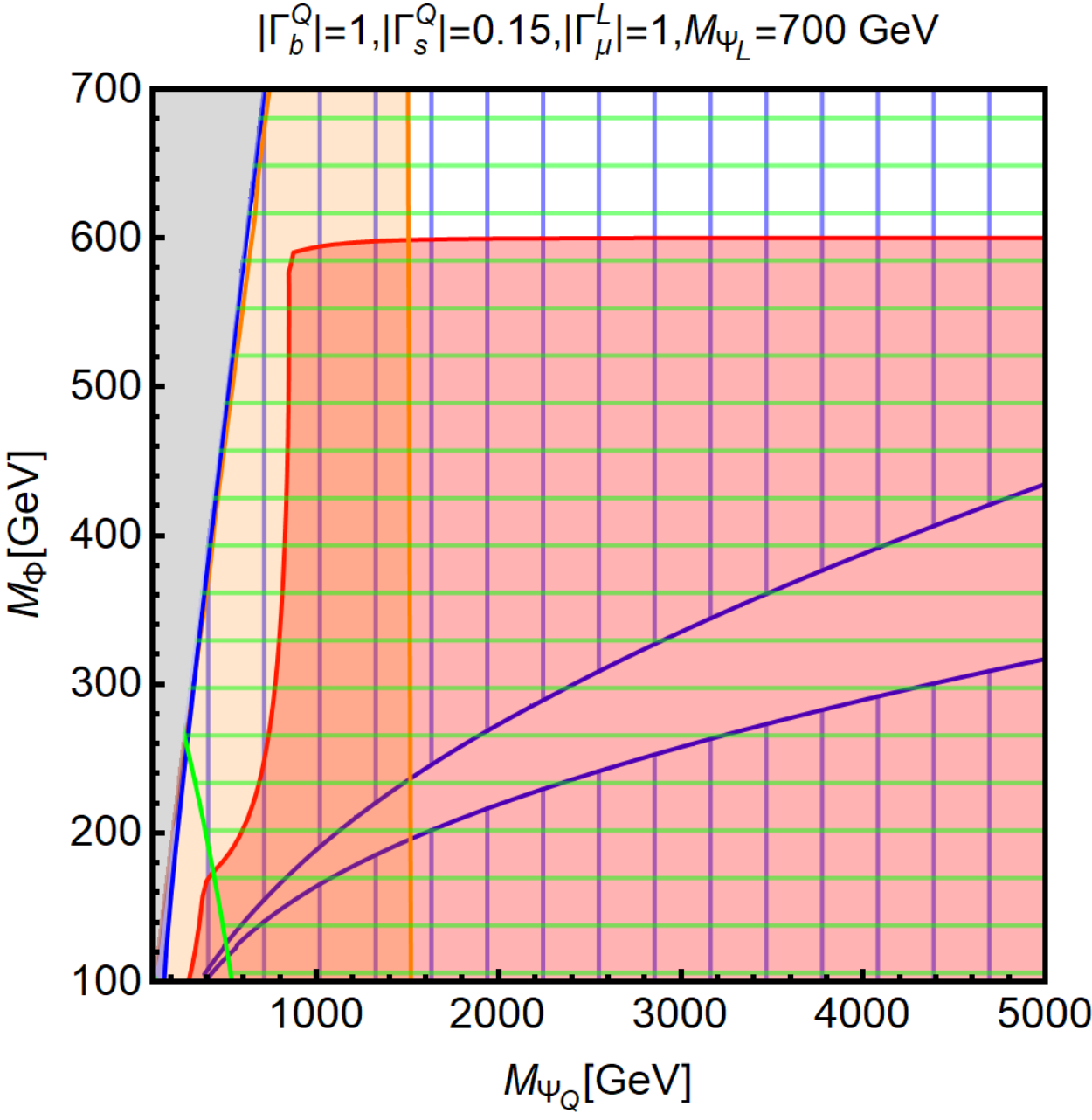}}
\subfloat{\includegraphics[width=0.33\linewidth]{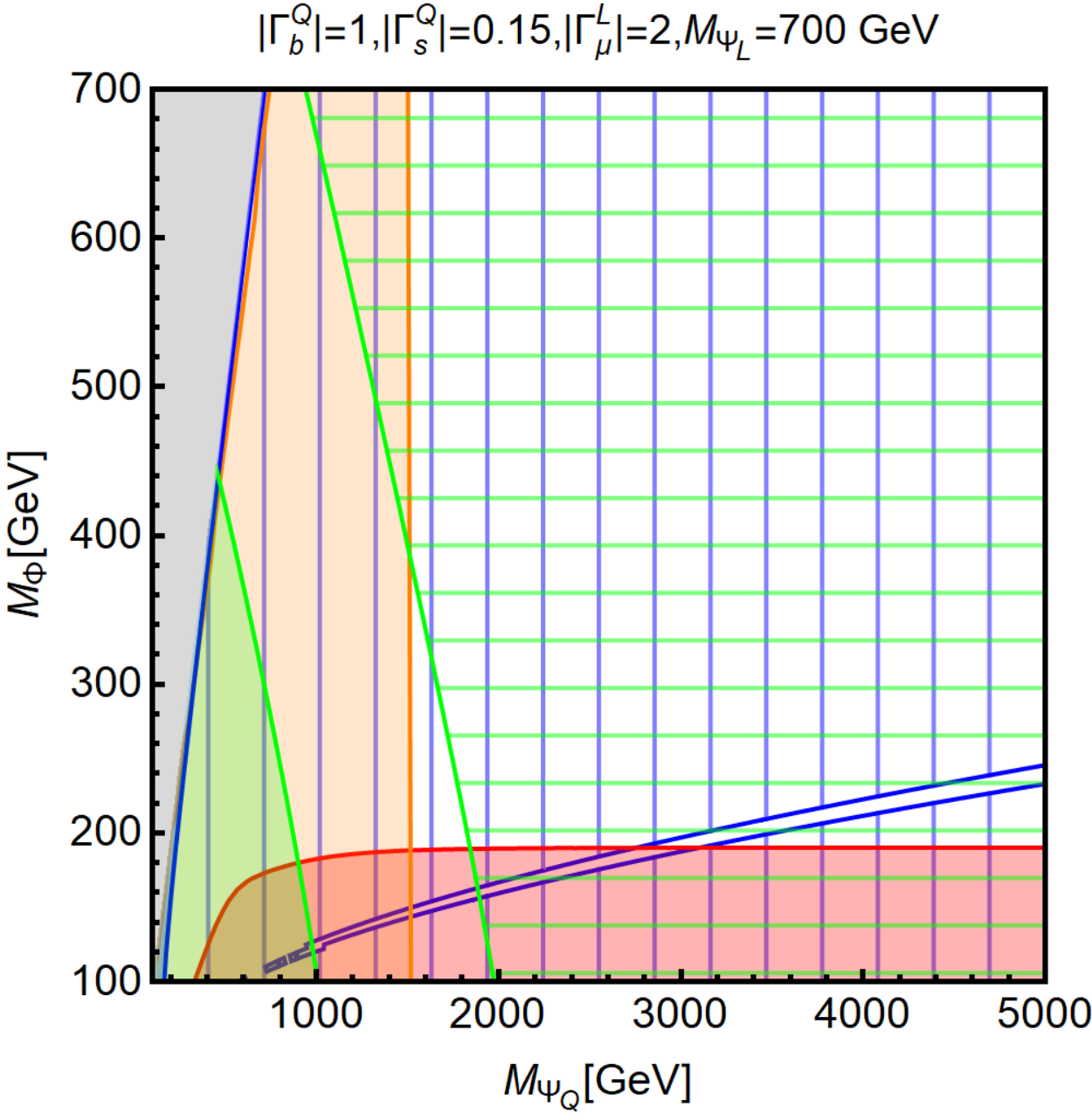}}
\subfloat{\includegraphics[width=0.33\linewidth]{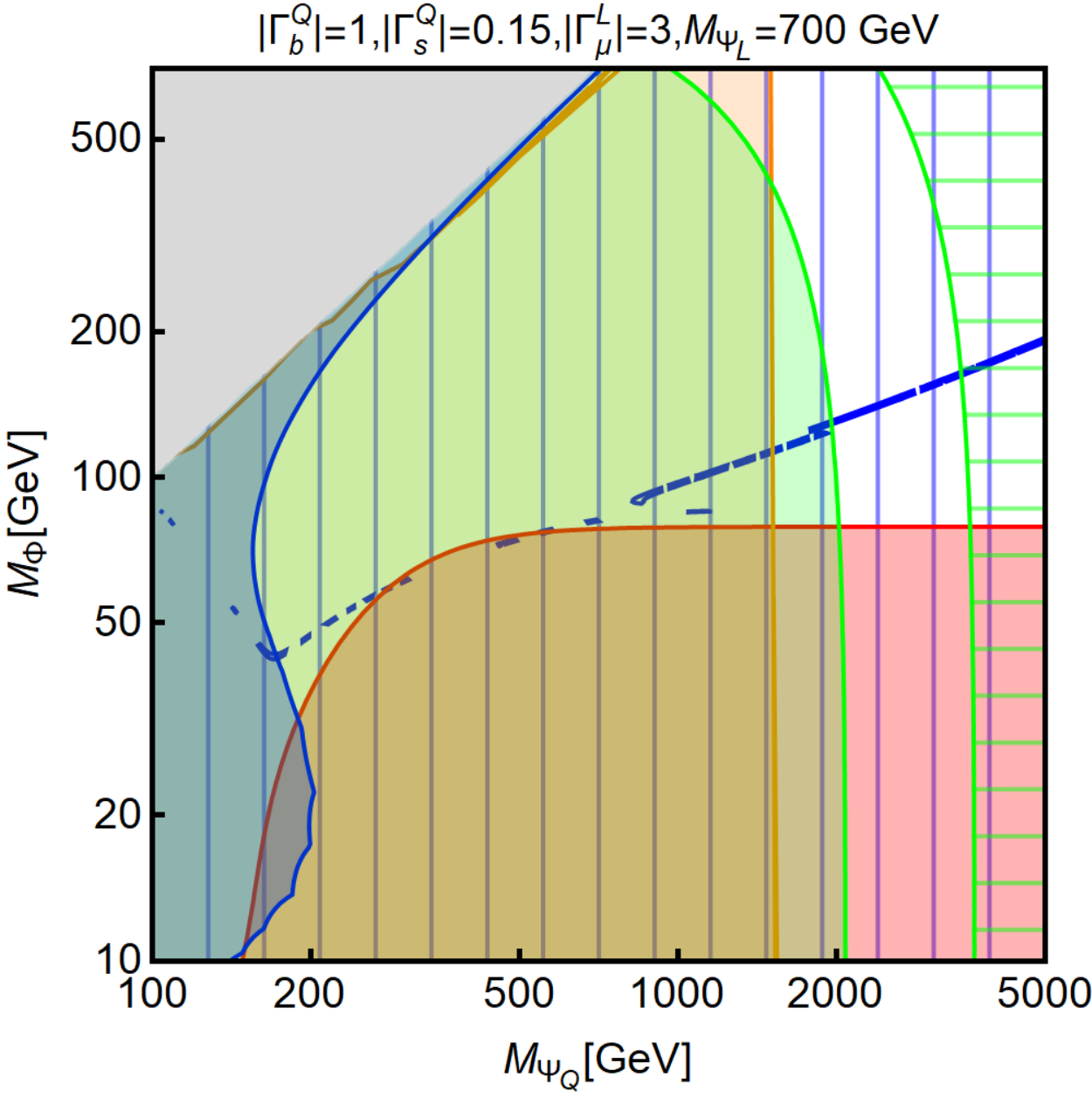}}\\
\subfloat{\includegraphics[width=0.33\linewidth]{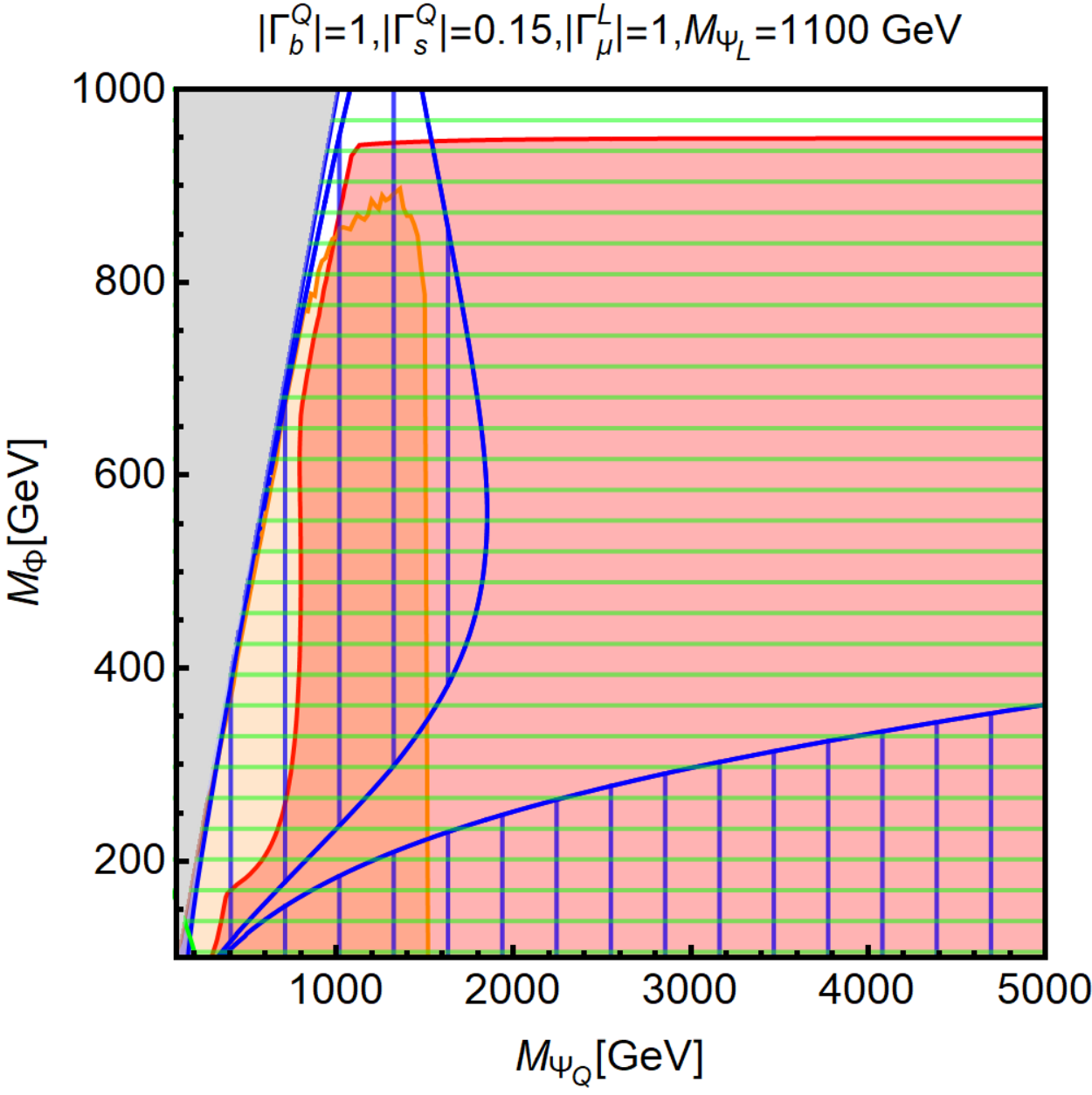}}
\subfloat{\includegraphics[width=0.33\linewidth]{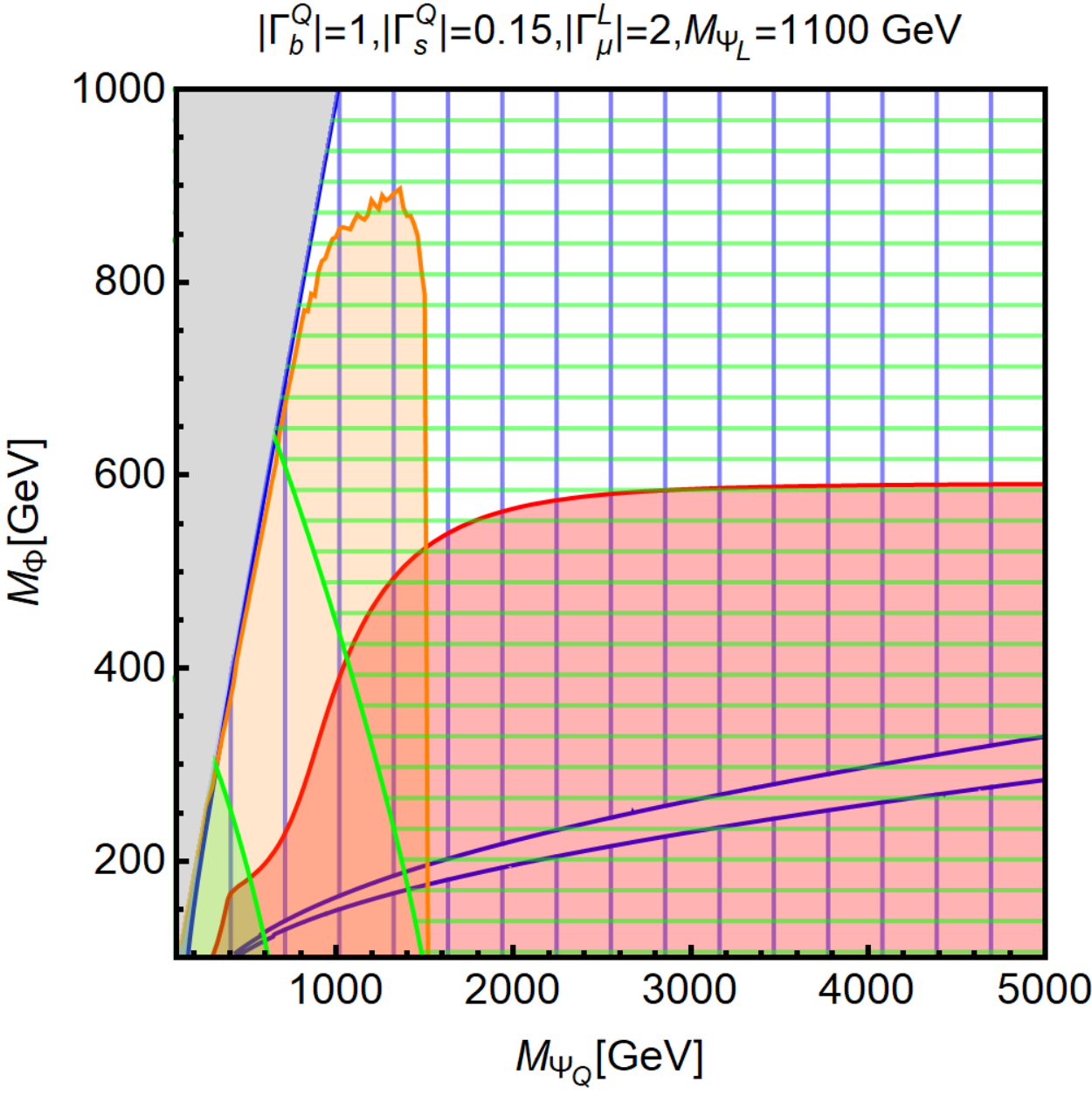}}
\subfloat{\includegraphics[width=0.33\linewidth]{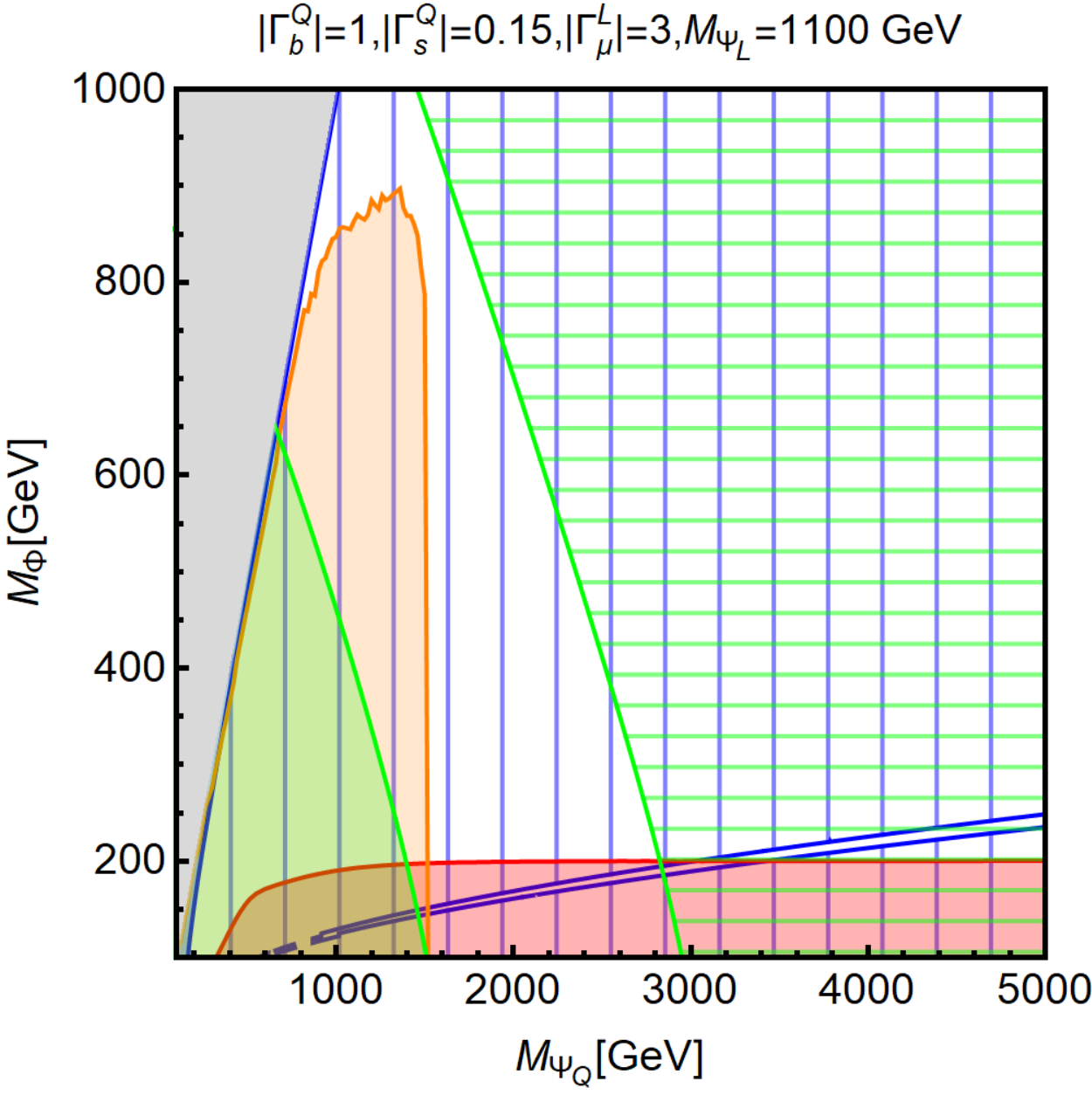}}
\caption{Summary of the constraints for the model $\mathcal{S}_\text{IA}$ with $|\Gamma_b^Q|=1$, $|\Gamma_s^Q|=0.15$, and the three assignments $|\Gamma_\mu^L|=1,2,3$. The upper (lower) row corresponds to $M_{\Phi_L}=700~(1100)\,\mbox{GeV}$. The colour scheme is as in Figure~\ref{fig:F_IA_D}. In particular, the regions strictly excluded by DD are filled in blue, while the areas denoted by a vertical blue hatching are excluded only if the two components of the DM singlet are degenerate, see the text for details.}
\label{fig:S_IA_A_all}
\end{figure}
This model is the counterpart of $\mathcal{F}_\text{IA;\,0}$ with scalar and fermion fields exchanging roles.
It is then of primary interest due to the high degree of correlation among our observables. 
While a similar study of this scenario has been presented already in Ref.~\cite{Kawamura:2017ecz}, our results are notably different, as discussed below.

The DM candidate belongs to the complex field $\Phi$, which is a complete singlet under the SM gauge group. 
Since $\Phi$ also plays the role of the flavour mediator in the diagram of Fig.~\ref{fig:boxes}, it is not possible in this case to assume that it is a real scalar field, otherwise an additional ``crossed'' box diagram would exactly cancel the effect on $b\to s \ell \ell$, as can be seen from~Eq.~\eqref{eq:C9_I,II}.
The quantum numbers of the other two fields, $\Psi_Q$ and $\Psi_L$, are respectively $({\bf 3},\,{\bf 2},\,1/6)$, $({\bf 1},\,{\bf 2},\,-1/2)$, and the Lagrangian of the model reads:
\begin{eqnarray}
{\cal L}_{\rm int} =  {\Gamma^Q_{i}\bar Q_{i}}{P_R}{\Psi_Q}{\Phi} + \Gamma_\mu^L \bar L_\mu {P_R}{\Psi_L}{\Phi} +
{\rm{h}}{\rm{.c.}}
\label{eq:L_IA_PHI}
\end{eqnarray}
Given our results for the $\mathcal{F}_\text{IA;\,0}$ model discussed in Section~\ref{sec:FIA_D}, 
since the $\Phi^\dagger i \overset{\leftrightarrow}{\partial_\mu} \Phi$ operator behaves, in the non-relativistic limit relevant for DM DD, in an analogous way as the Dirac DM operator $\bar{\Psi} \gamma^\mu \Psi$, we can expect this model to be completely ruled out by DM direct detection as well. Furthermore, as we mentioned above, we can not assume $\Phi$ to be a real scalar if we want the model to address the $B$-physics anomalies. 
To circumvent this problem, we can however assume that the two components of $\Phi$ (the real part and the imaginary part) have a small mass splitting $>\mathcal{O}(100)$~keV.\footnote{If DM is stabilised by a $\bf Z_2$ symmetry, this can be achieved for instance through an invariant term  in the scalar potential of the form $\mu^2(\Phi^2+\Phi^{*\,2})$~\cite{Huang:2020ris}.} In this ``non-degenerate'' case the DM candidate would be a real scalar, the lightest  of the two states.
While this would ensure the needed suppression for the scattering cross-section with nuclei, a small mass splitting would avoid the above-mentioned cancellation of $\delta C^{9,10}_\mu$ (the DM is the flavour mediator in this case) and maintain efficient DM annihilations. In fact, the annihilation cross-section into SM fermions would be $d$-wave suppressed for real DM, hence the correct relic density would follow from coannihilations of the two components of $\Phi$.

\begin{figure}[!t]
\centering
\subfloat{\includegraphics[width=0.33\linewidth]{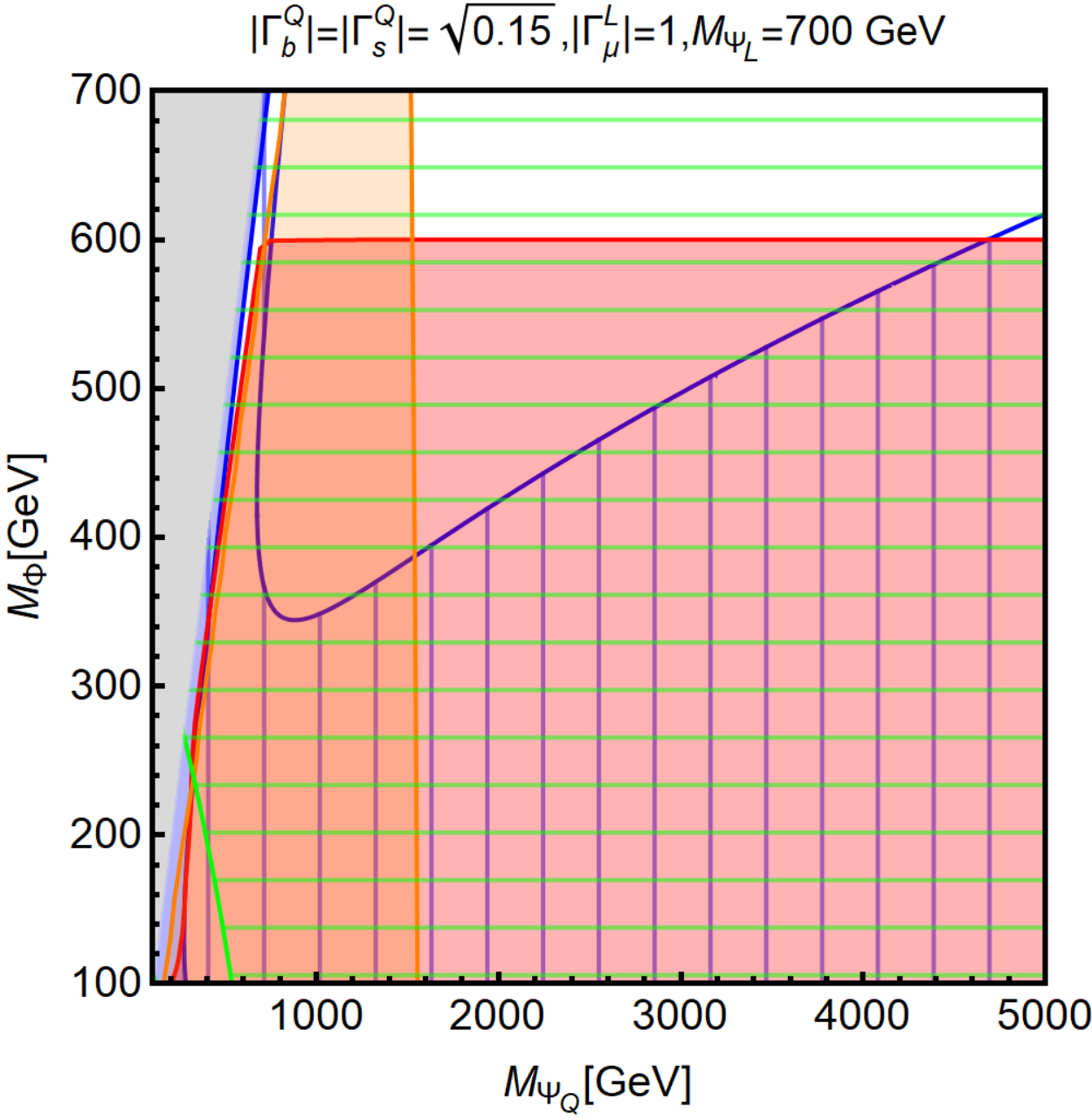}}
\subfloat{\includegraphics[width=0.33\linewidth]{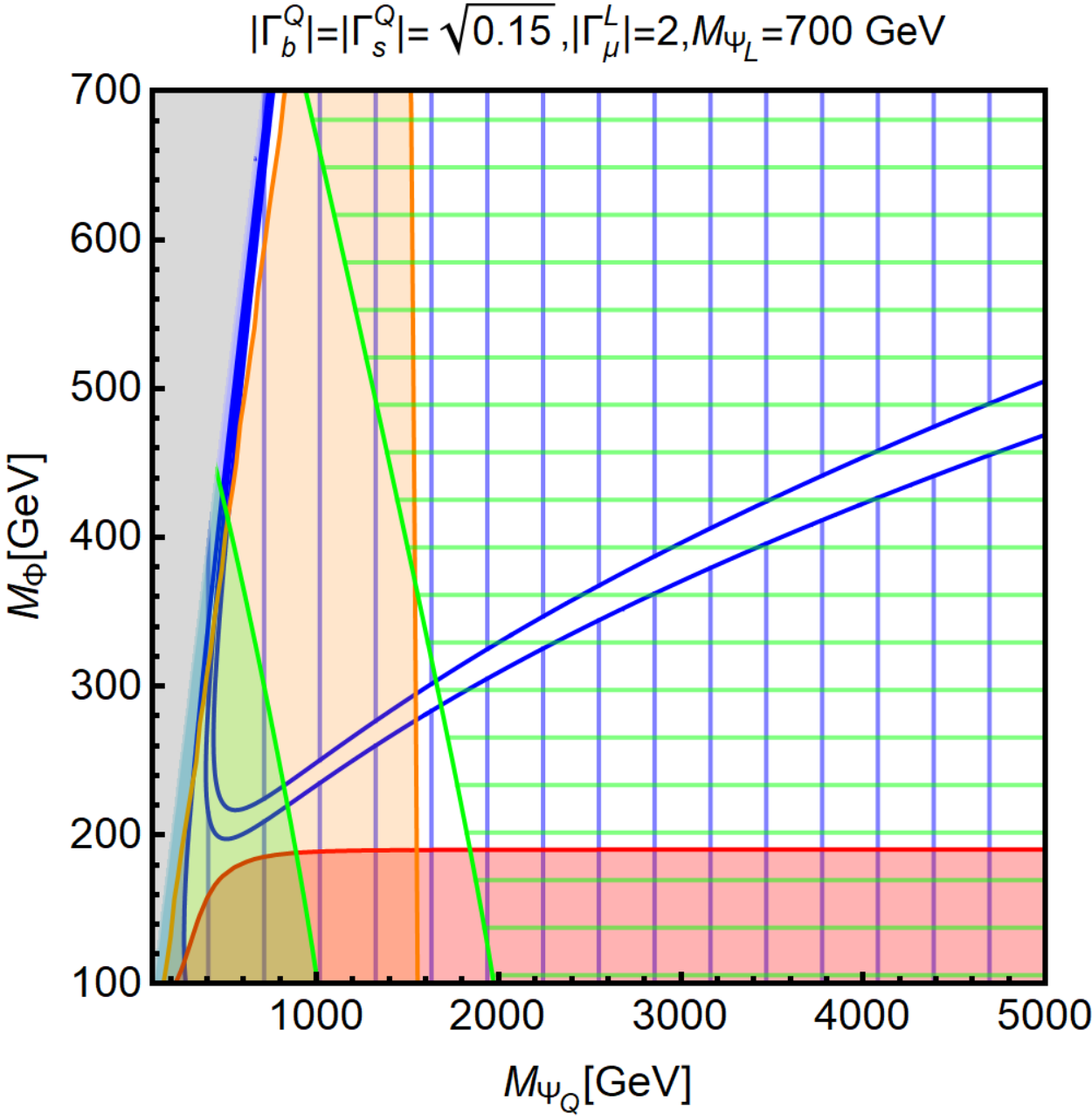}}
\subfloat{\includegraphics[width=0.33\linewidth]{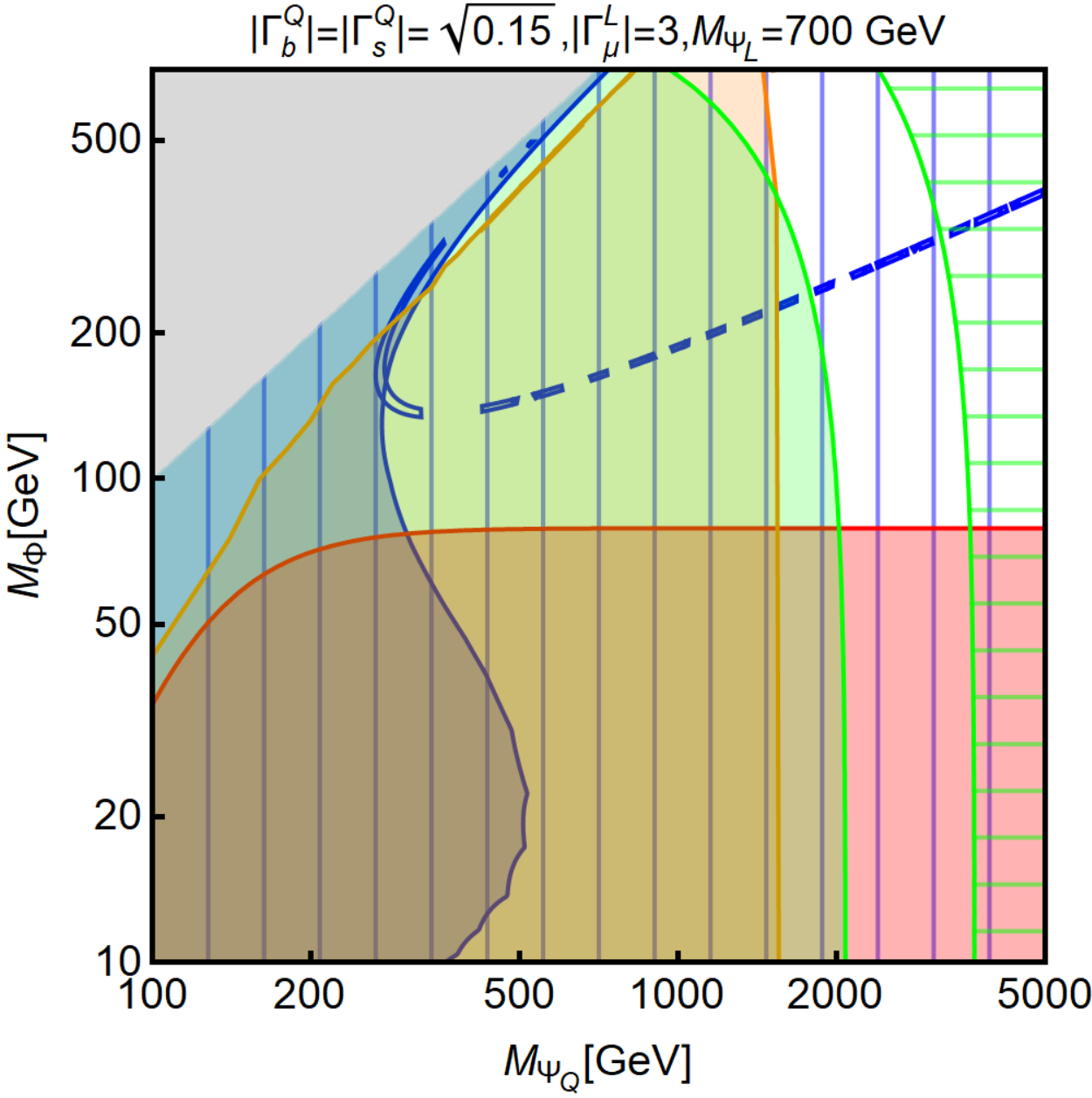}}\\
\subfloat{\includegraphics[width=0.33\linewidth]{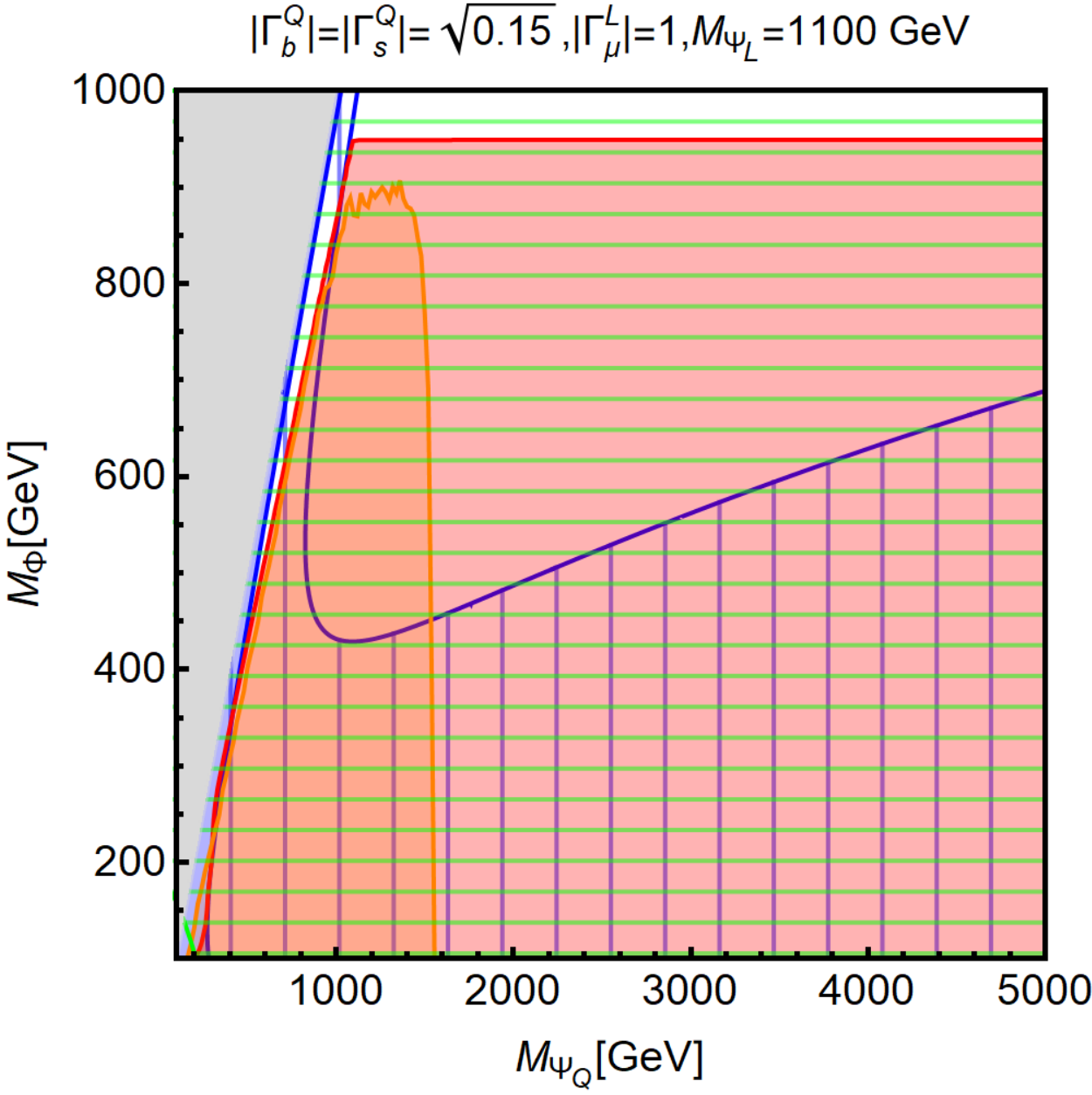}}
\subfloat{\includegraphics[width=0.33\linewidth]{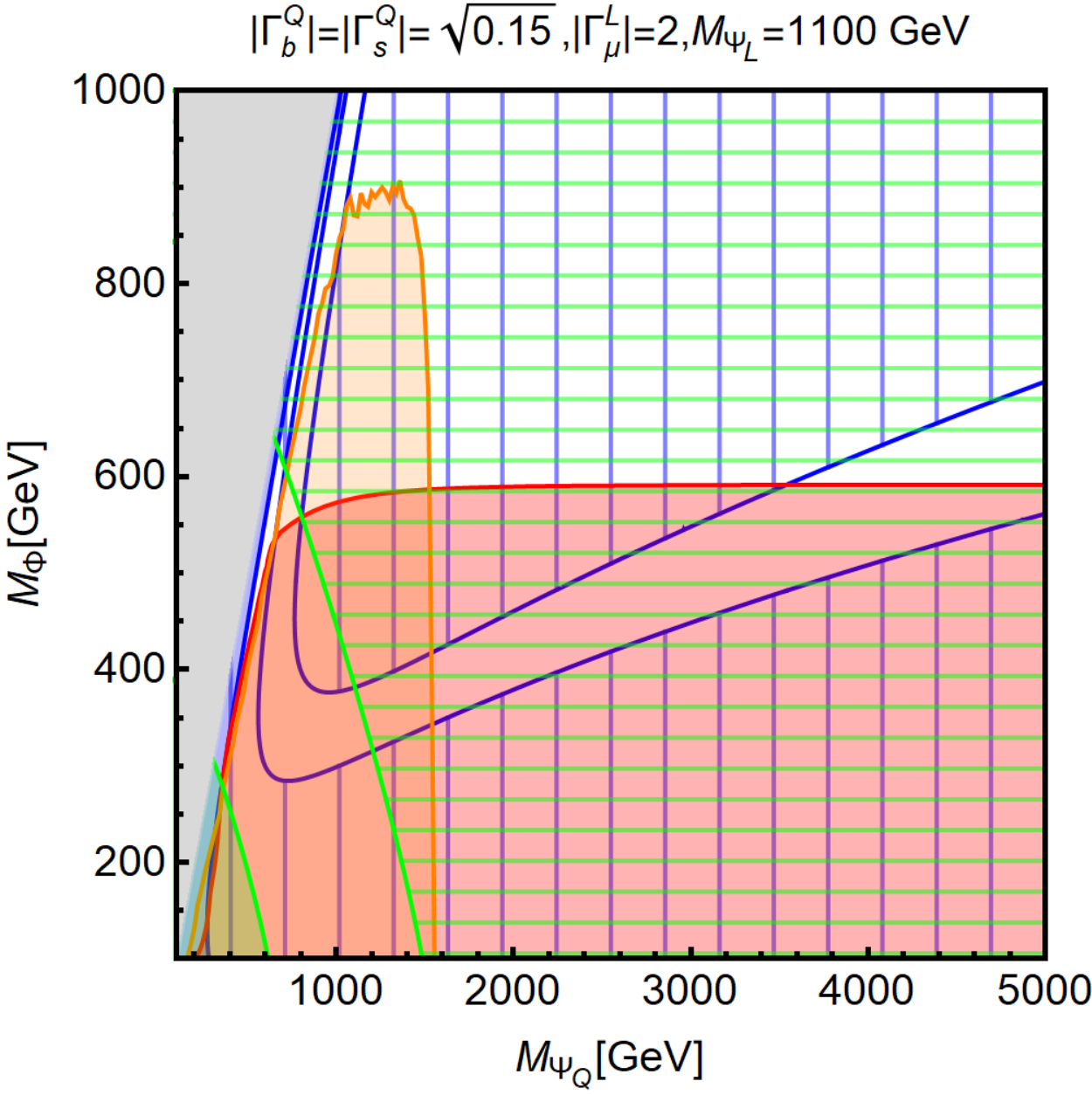}}
\subfloat{\includegraphics[width=0.33\linewidth]{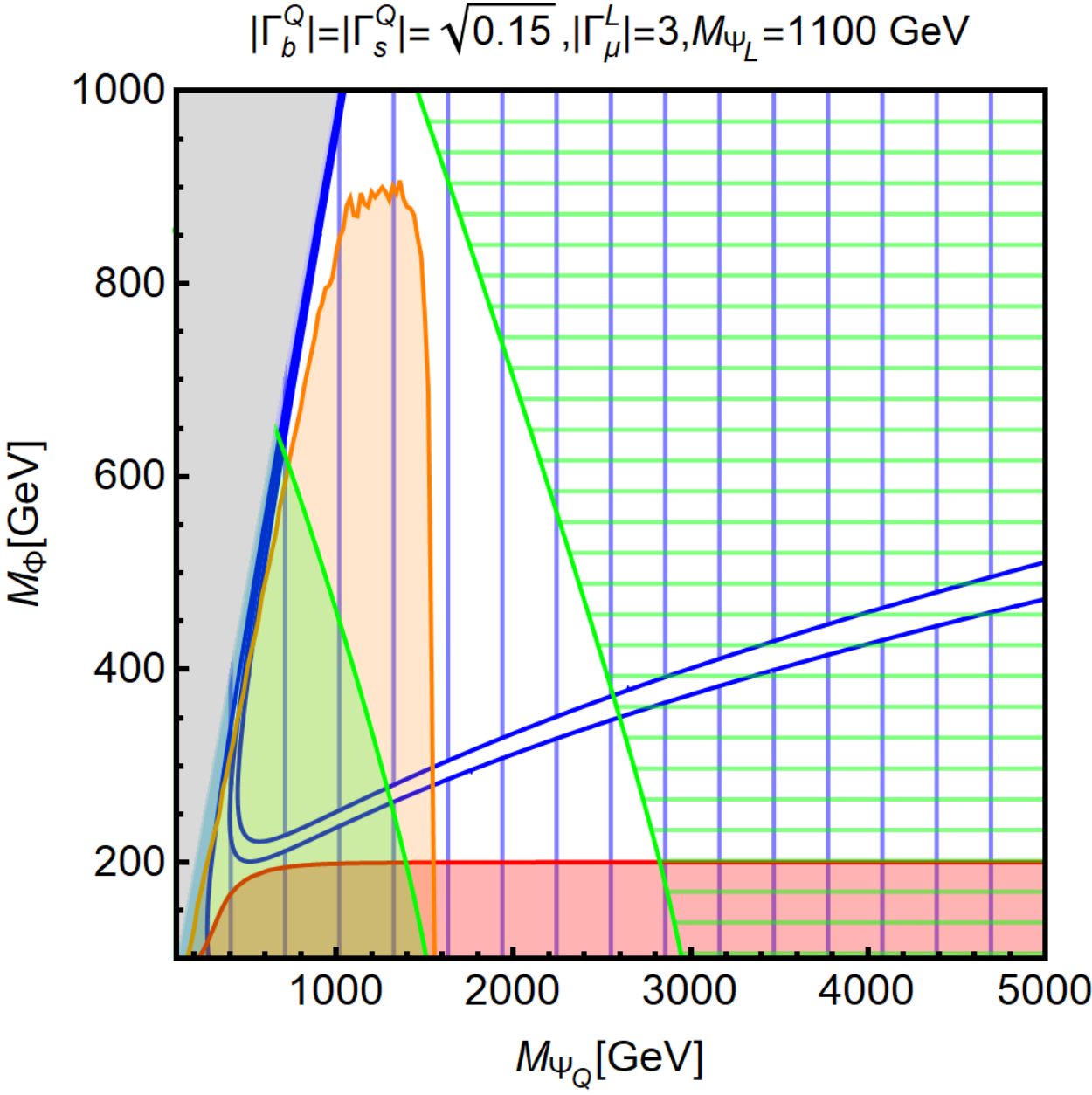}}
\caption{Summary of the constraints for the model $\mathcal{S}_\text{IA}$ with $|\Gamma_b^Q|=|\Gamma_s^Q|=\sqrt{0.15}$, and the three assignments $|\Gamma_\mu^L|=1,2,3$. The upper (lower) row corresponds to $M_{\Phi_L}=700~(1100)\,\mbox{GeV}$. The colour scheme is as in Figure~\ref{fig:F_IA_D}. In particular, the regions strictly excluded by DD are filled in blue, while the areas denoted by a vertical blue hatching are excluded only if the two components of the DM singlet are degenerate, see the text for details.}
\label{fig:S_IA_B_all}
\end{figure}

The scenario with two degenerate states has been studied in Ref.~\cite{Kawamura:2017ecz}.\footnote{A slightly more complicated variant of this model, featuring an additional scalar field, has been studied in Ref.~\cite{Grinstein:2018fgb}. Furthermore, adding to the model a scalar doublet mixing with the singlet through a coupling with the Higgs introduces interactions to right-handed fermions as well, allowing in particular a chirally-enhanced contribution to $(g-2)_\mu$ and thus a natural explanation of the observed anomaly~\cite{Calibbi:2019bay}.} 
Considering also the non-degenerate case will allow us to open a large viable region of the parameter space.
Besides this important point, Ref.~\cite{Kawamura:2017ecz} differs from our study also from the fact that LHC data available at the time lead to less stringent bounds for couplings as in scenario \emph{ii}) compared to the limits in scenario \emph{i}). This is however not the case anymore, since the two scenarios are now no longer significantly different once the recasting of the CMS search of Ref.~\cite{Sirunyan:2019ctn} is employed.

The combined constraints for this model are shown in Figures~\ref{fig:S_IA_A_all} and~\ref{fig:S_IA_B_all}. 
In these plots, the DD constraint for the degenerate case
are shown as regions denoted by a vertical blue hatching.
As we can see, very efficient interactions mediated by the photon and the $Z$-boson penguins make the model subject to very strong constraints from XENON1T, such that only narrow tuned regions are compatible with a good fit of the flavour anomalies.
On the other hand, assuming non-degenerate singlet states drastically reduces the exclusions from DD to the blue-filled areas, especially in the $M_{\Psi_Q} \simeq M_\Phi$ regime, where QCD operators for DD are relevant. In absence of strong bounds from direct detection we see that $\mathcal{S}_\text{IA}$ model can account for the $B$-anomalies in wide regions of the parameter space. In addition, the correct DM relic density can be achieved for $\Gamma_\mu^L \gtrsim 2$ and $M_{\Psi_L}=700\,\mbox{GeV}$ while for $M_{\Psi_L}=1100\,\mbox{GeV}$ the coupling needs to be larger, $\Gamma_\mu^L\approx 3$. Thermal DM and a good fit of the flavour anomalies are thus simultaneously possible with DM masses approximately in the 50-200~GeV range.

%
\subsection{\texorpdfstring{$\mathcal{S}_\text{IIB}$}{SIIB}, Dirac singlet DM}
We now discuss another model with fermion singlet DM.
The gauge quantum numbers of the NP fields are $({\bf 1},\,{\bf 1},\,0)$, $({\bf \bar3},\,{\bf 1},\,-2/3)$ and $({\bf 3},\,{\bf 2},\,1/6)$ for $\Psi_Q$, $\Psi_L$ and $\Phi$, respectively. The DM candidate is the Dirac field $\Psi_Q$, and the flavour mediator $\Phi$ is a complex scalar: 
\begin{eqnarray}
{\cal L}_{\rm int} =  {\Gamma^Q_i\bar Q_i}{P_R}{\Psi_Q}{\Phi} + \Gamma^L_i \bar L_i {P_R}{\Psi_L}{\Phi} +
{\rm{h}}{\rm{.c.}}\,.
\label{eq:L_IIB_PHI}
\end{eqnarray}
This model is characterised by the fact that the DM field only couples to quarks and not to muons. 
It then allows for an interesting comparison with the $\mathcal{F}_\text{IA;\,0}$ model. Similarly to the latter scenario we will consider, respectively here and in the next subsection, both cases of Dirac and Majorana DM.
\begin{figure}[!t]
\centering
\subfloat{\includegraphics[width=0.33\linewidth]{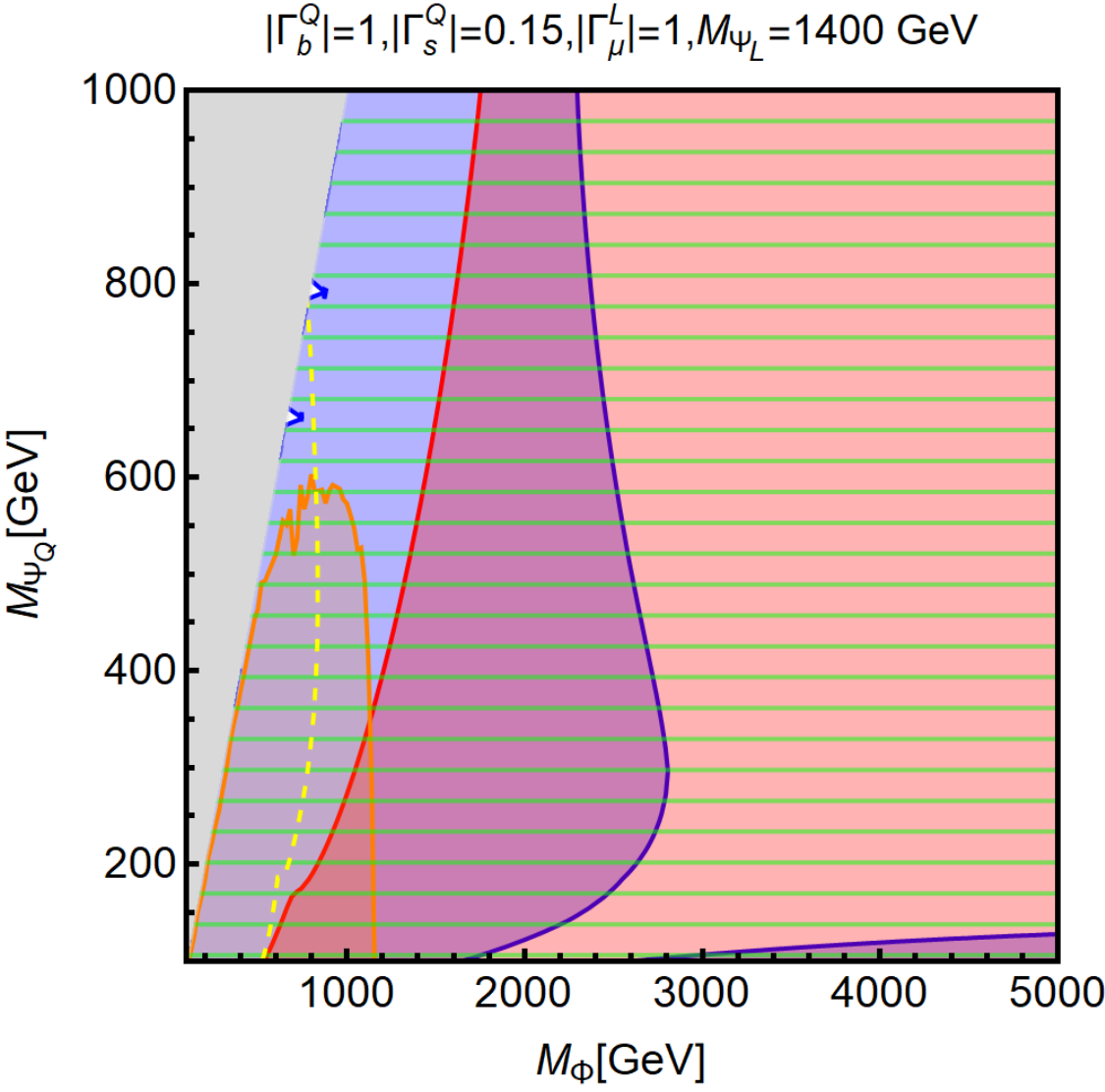}}
\subfloat{\includegraphics[width=0.33\linewidth]{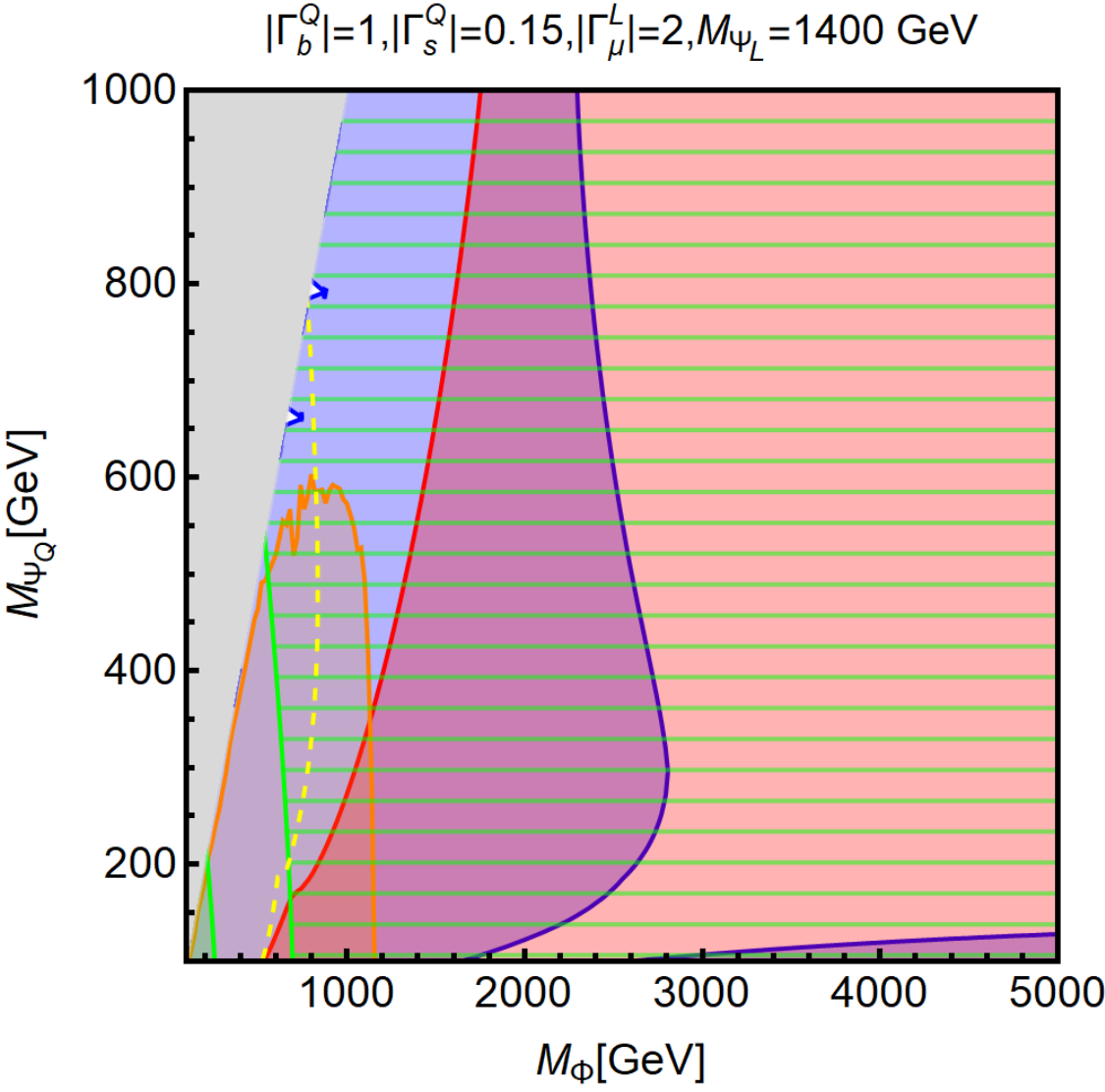}}
\subfloat{\includegraphics[width=0.33\linewidth]{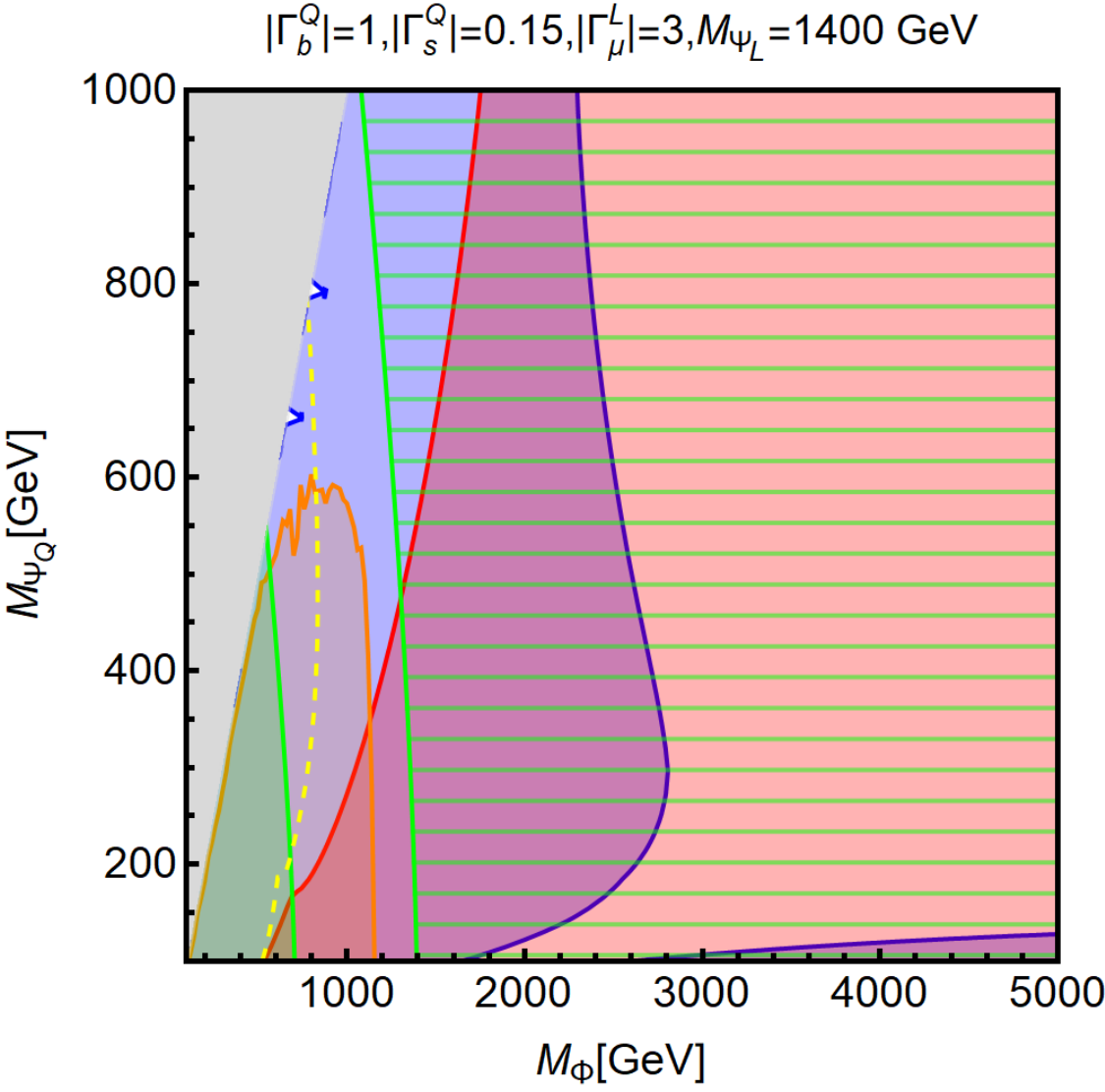}}\\
\subfloat{\includegraphics[width=0.33\linewidth]{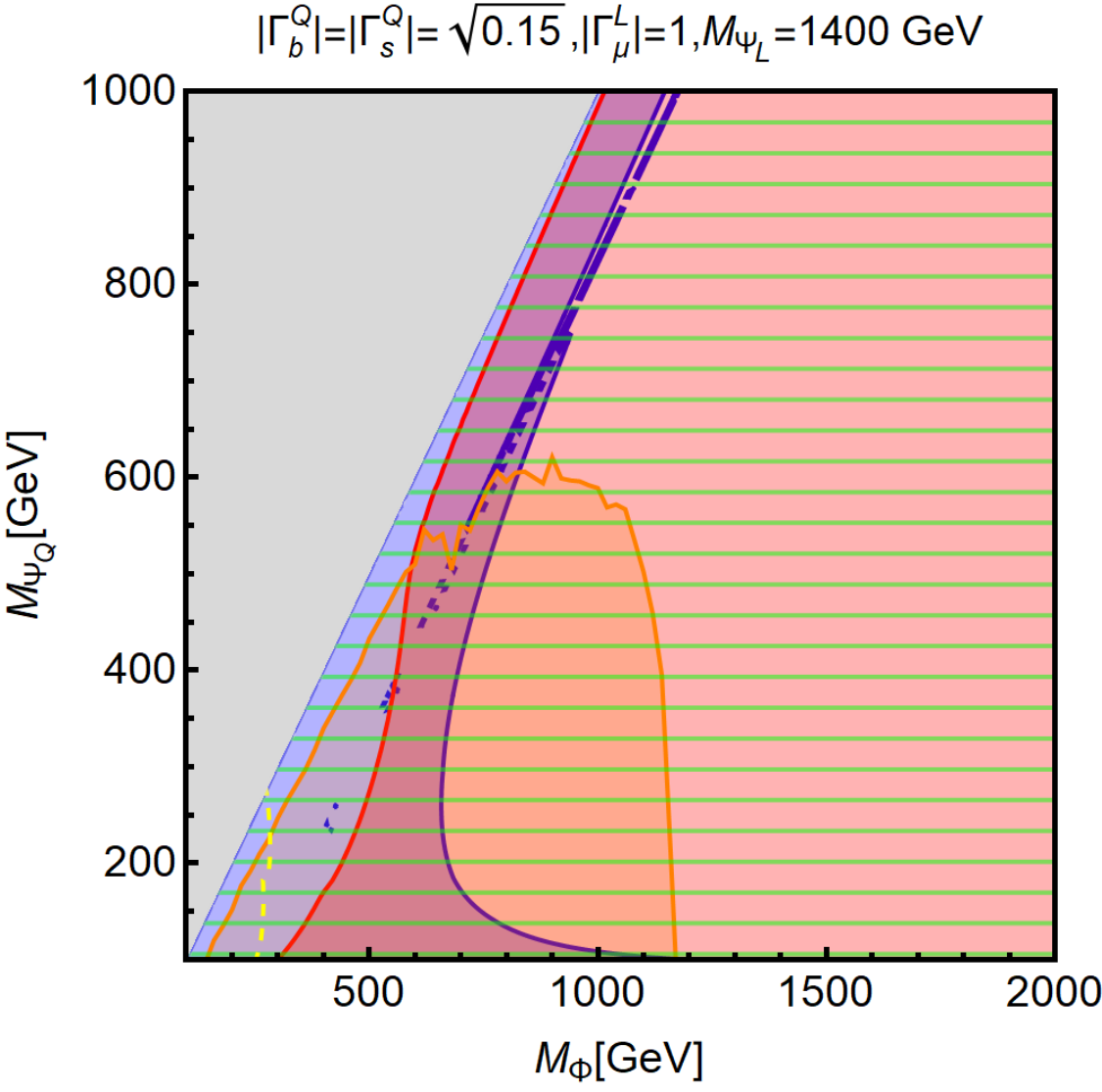}}
\subfloat{\includegraphics[width=0.33\linewidth]{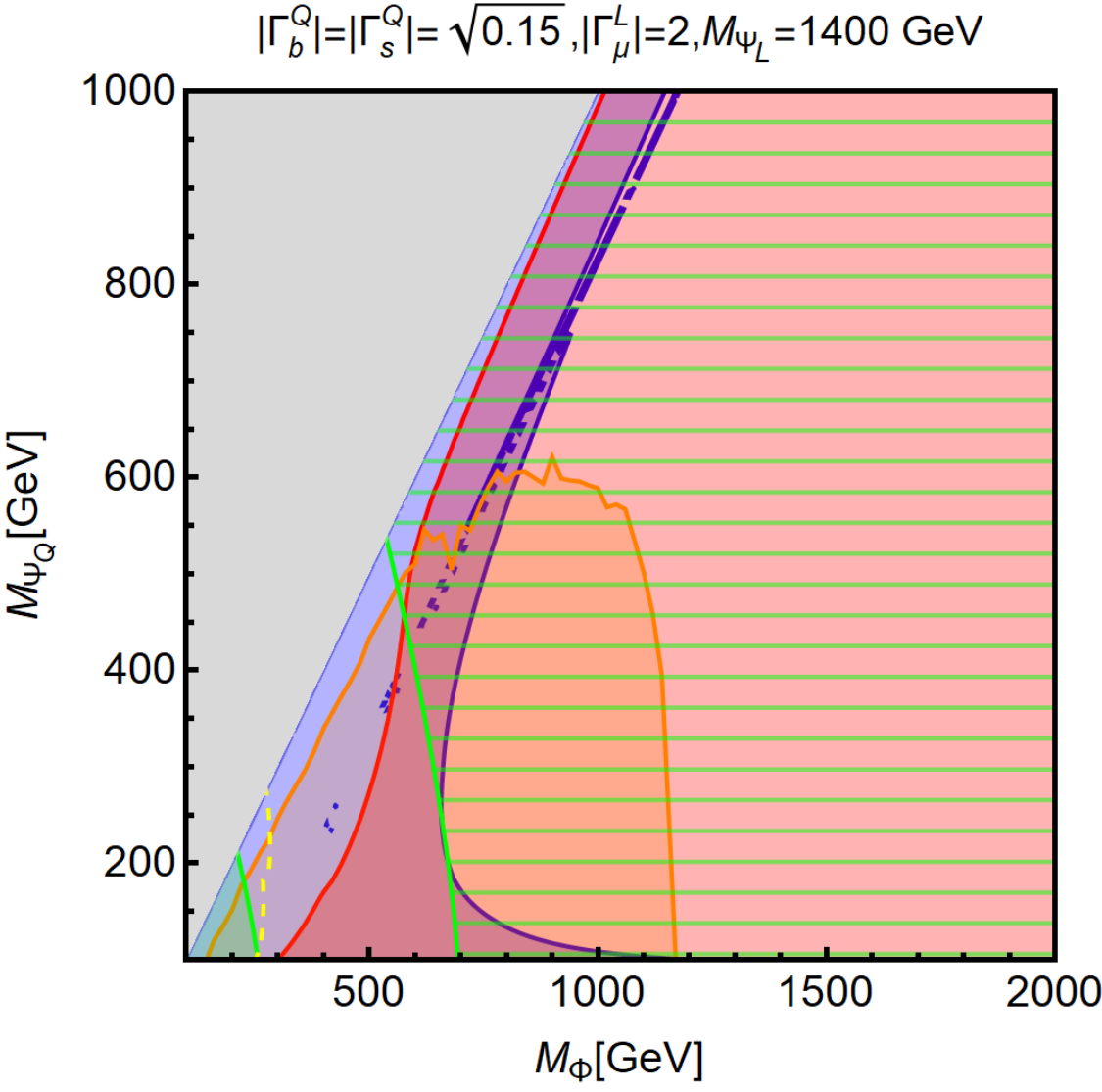}}
\subfloat{\includegraphics[width=0.33\linewidth]{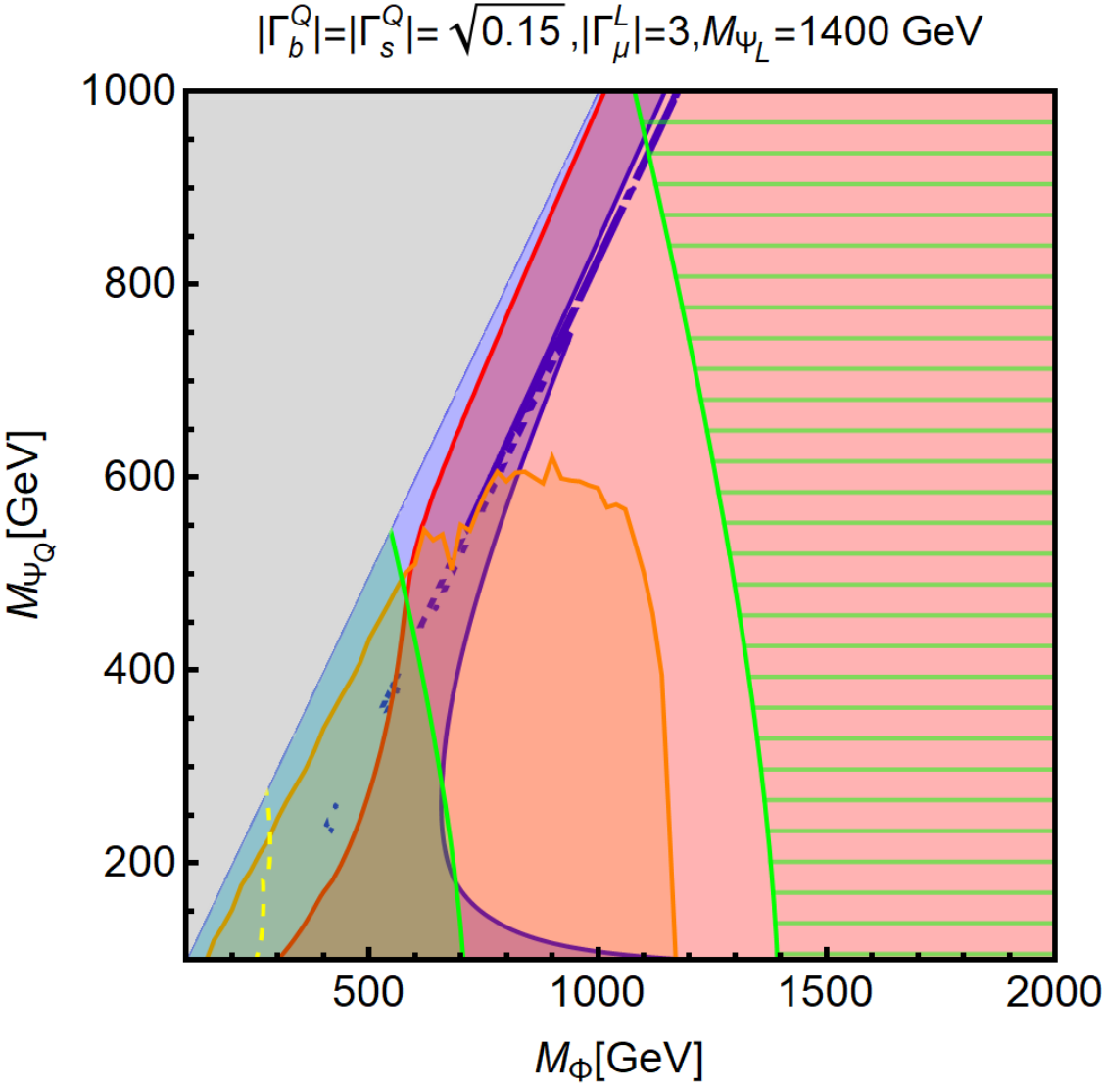}}
\caption{Summary of the constraints for the model $\mathcal{S}_\text{IIB}$ with Dirac DM. The upper row refers to the assignments $|\Gamma_{s}^Q|=1$, $|\Gamma_{s}^Q|=0.15$, while the lower row corresponds instead to $|\Gamma_{s}^Q|=|\Gamma_{b}^Q|=\sqrt{0.15}$. In all cases we have set $M_{\Psi_L}=1400\,\mbox{GeV}$ and  $|\Gamma_{\mu}^L|=1,2,3$. The colour scheme is as defined in the caption of Figure~\ref{fig:F_IA_D}.
}
\label{fig:S_IIB_D_all}
\end{figure}

The results of our analysis are shown in Figure~\ref{fig:S_IIB_D_all} and confirm the tendency, already observed for $\mathcal{F}_\text{IA;\,0}$, of models with Dirac DM of being already experimentally excluded. Compared to the $\mathcal{F}_\text{IA;\,0}$ scenario we notice some differences though. First of all the relic density constraint is much stronger and, moreover, is not sensitive to the value of the $\Gamma_\mu^L$ coupling. In fact, all DM phenomenology does not depend on this latter parameter (neither on $M_{\Psi_L}$), as it can be seen by comparing the three columns of plots of Figure~\ref{fig:S_IIB_D_all}.
This is due to the fact that the DM field is now only coupled with the field $\Phi$ and left-handed quarks. DM observables hence depend only on the $\Gamma_{s,b}^Q$ couplings which are fixed by the fit of flavour observables. Furthermore, being the coupling with muons absent, DM annihilation cross-section is more suppressed than in the $\mathcal{F}_\text{IA;\,0}$ model, especially for the configuration \emph{ii}) of the couplings. Looking instead at direct detection, while being still the most challenging constraint for the assignment \emph{i}) of the couplings, its impact is strongly reduced, relative to the requirement of viable relic density, for the assignment \emph{ii}). This is because the most relevant contribution to the SI cross-section here come from loop diagrams involving the $Z$ bosons, whose effect is enhanced by the mass of the top quark. Since the coupling of the DM field with the top is reduced, once moving from the assignment \emph{i}) to \emph{ii}), the DM scattering rate is consequently reduced. We have not observed such an outcome in the model $\mathcal{F}_\text{IA;\,0}$ since a comparable or even larger contribution to the DM scattering rate on nucleons was coming from interactions with the photon, controlled by the $\Gamma_{\mu}^L$ coupling.

\begin{figure}[!t]
\centering
\subfloat{\includegraphics[width=0.33\linewidth]{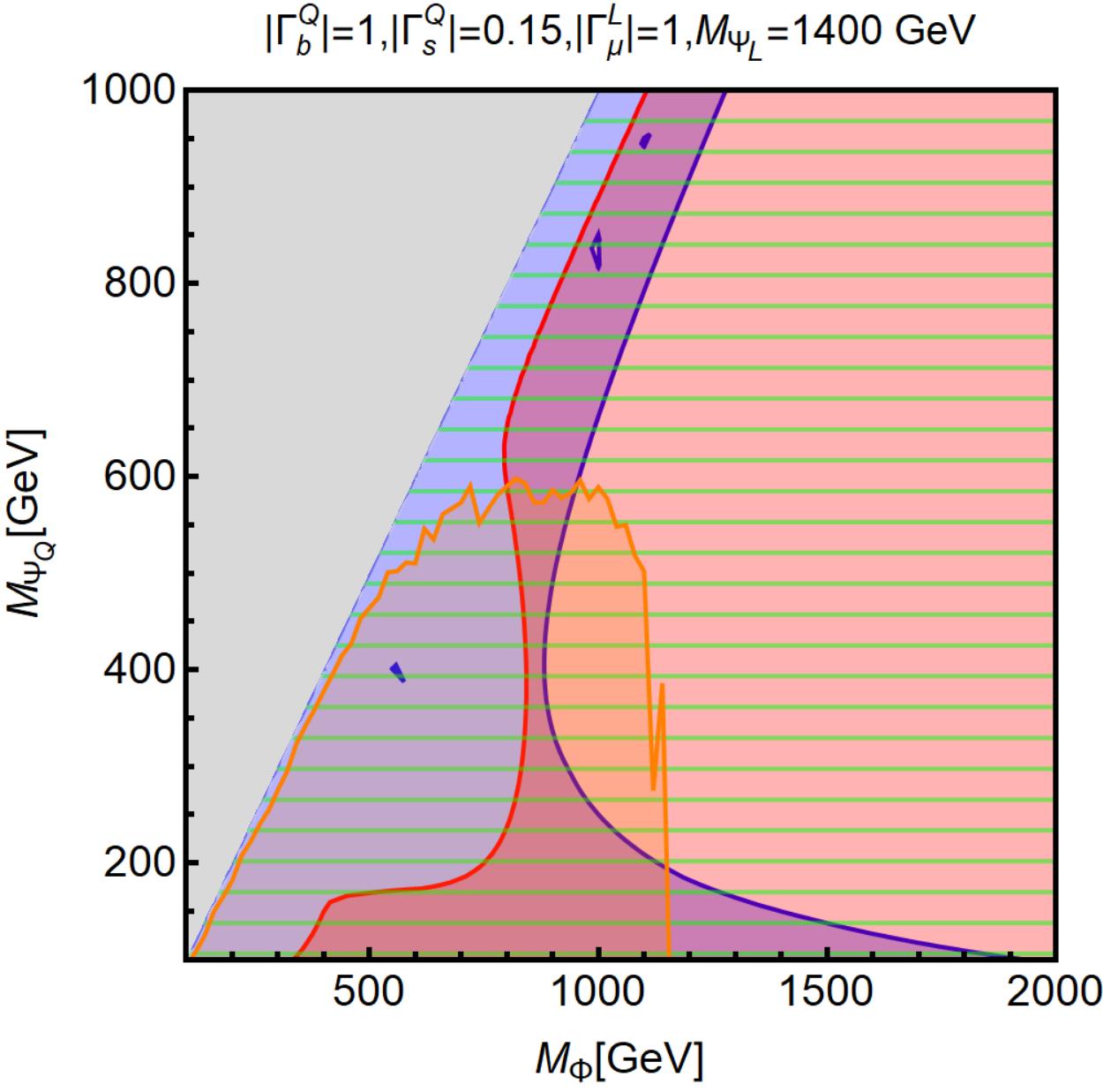}}
\subfloat{\includegraphics[width=0.33\linewidth]{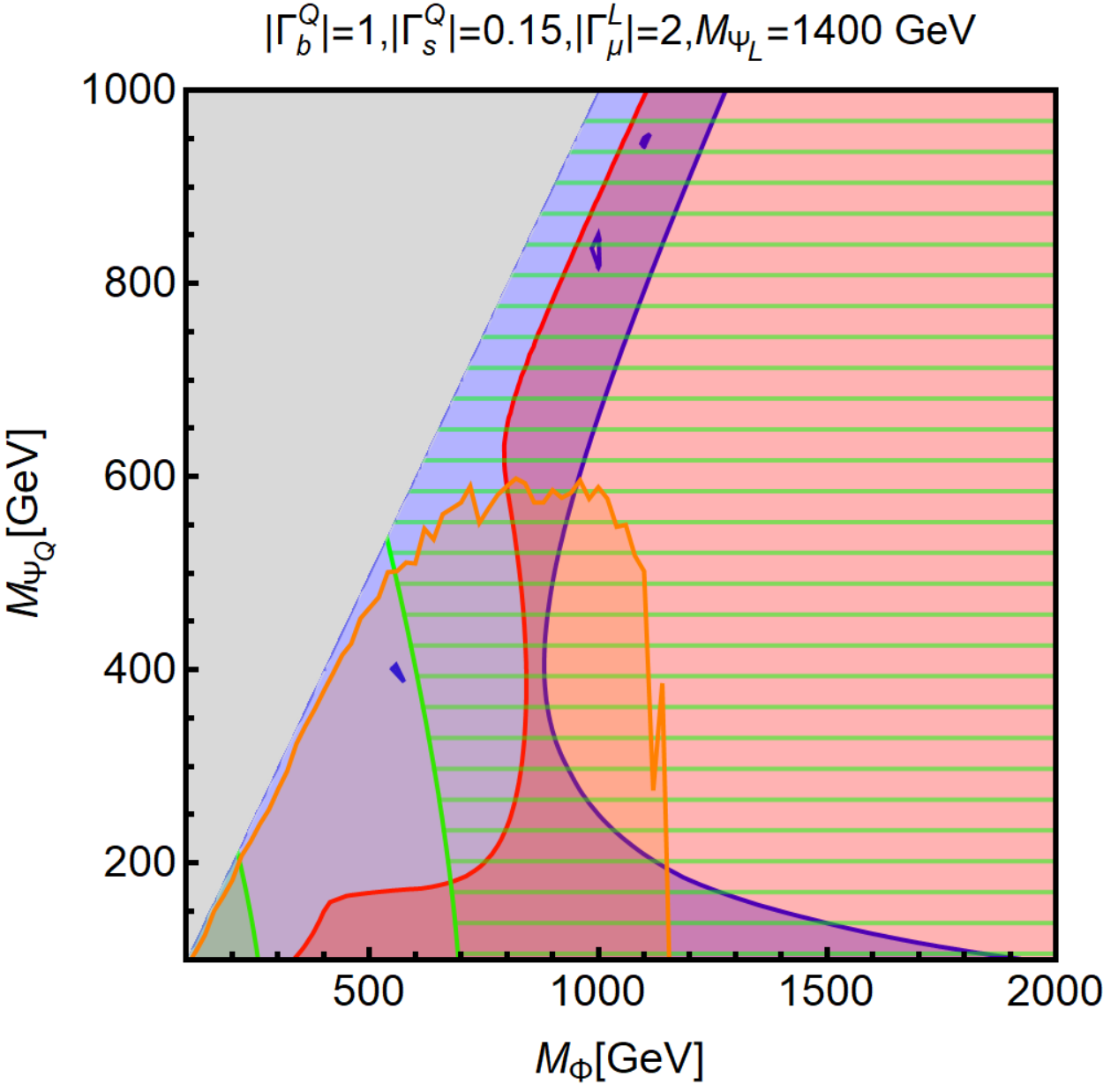}}
\subfloat{\includegraphics[width=0.33\linewidth]{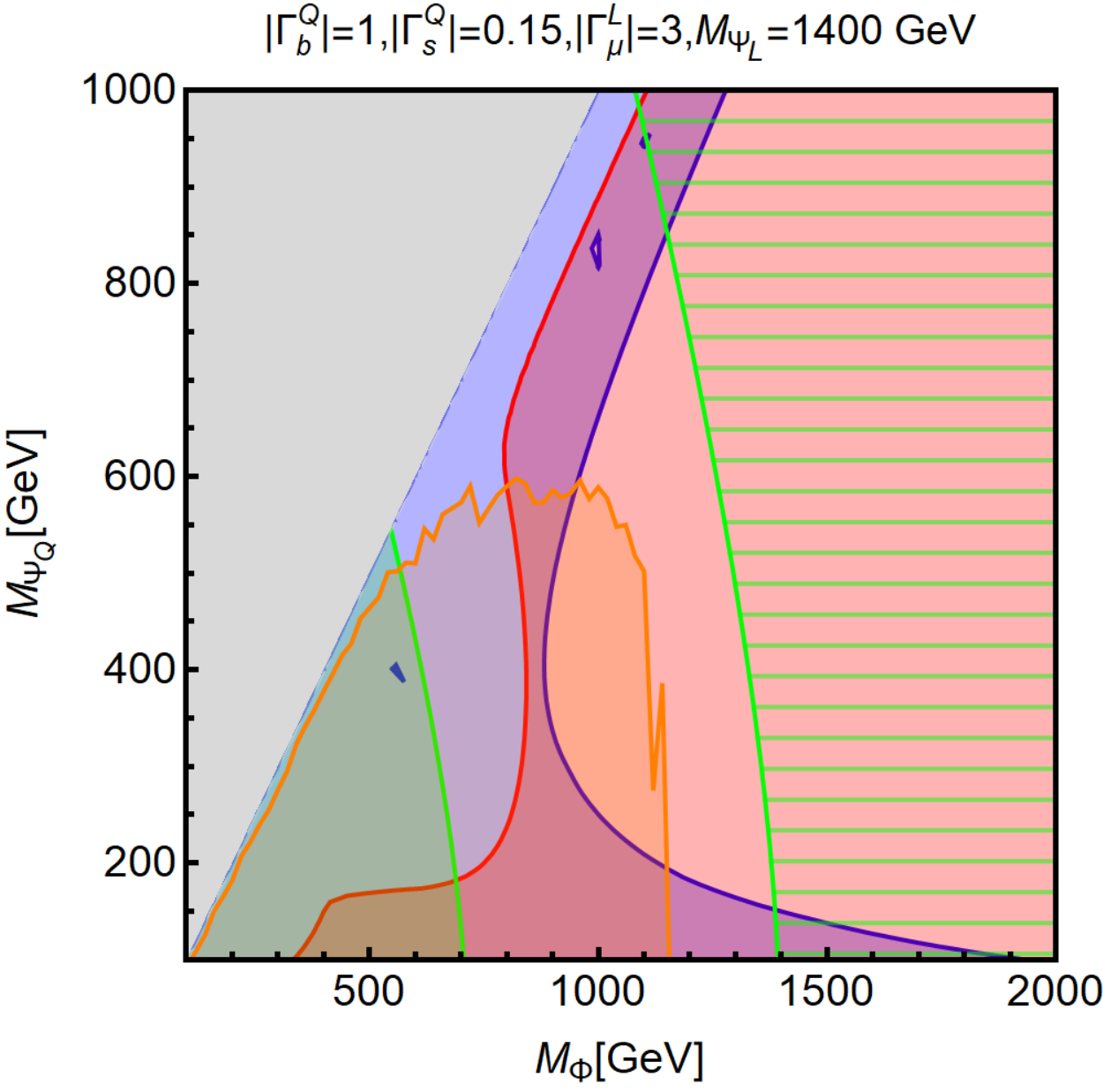}}\\
\subfloat{\includegraphics[width=0.33\linewidth]{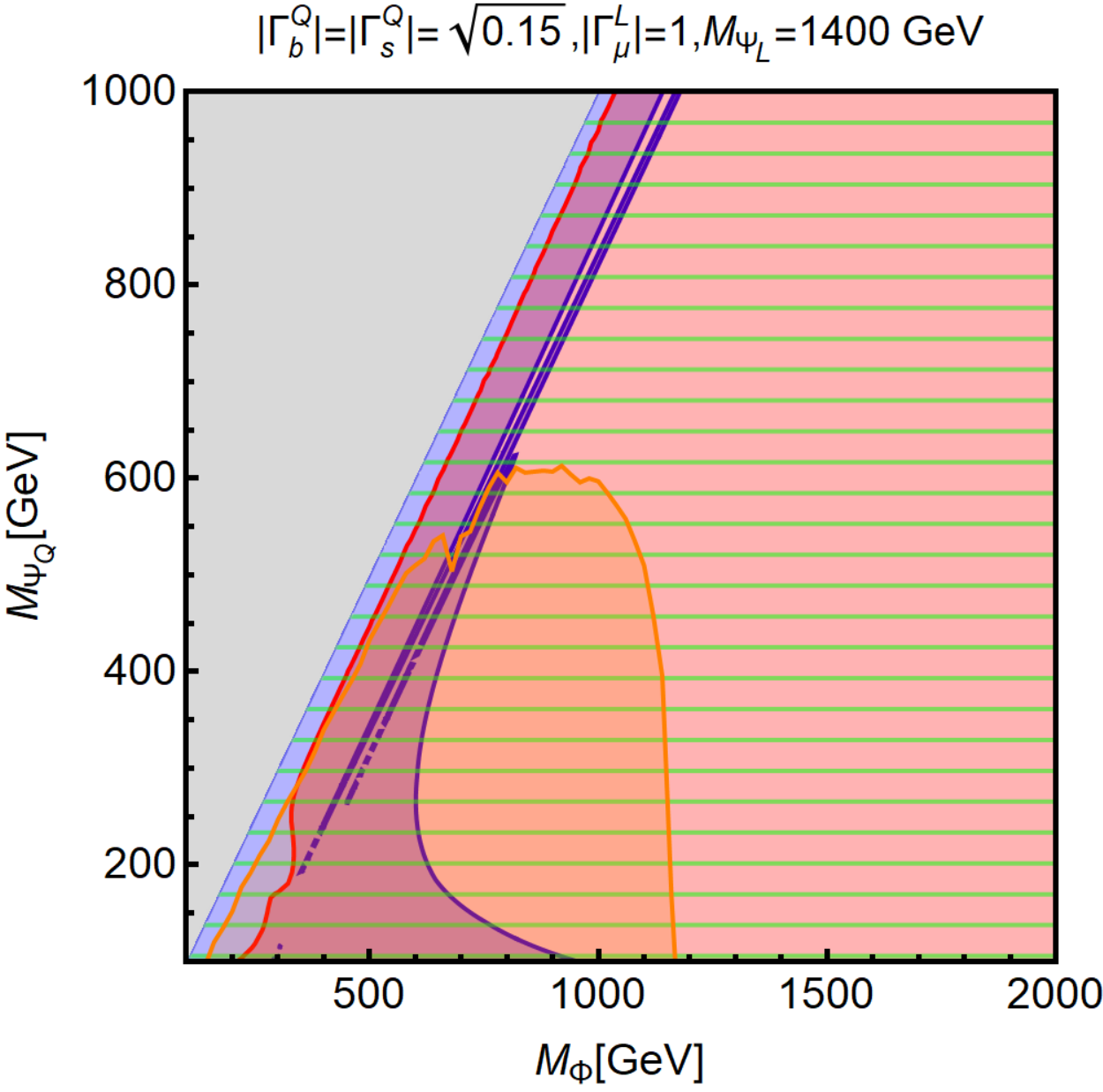}}
\subfloat{\includegraphics[width=0.33\linewidth]{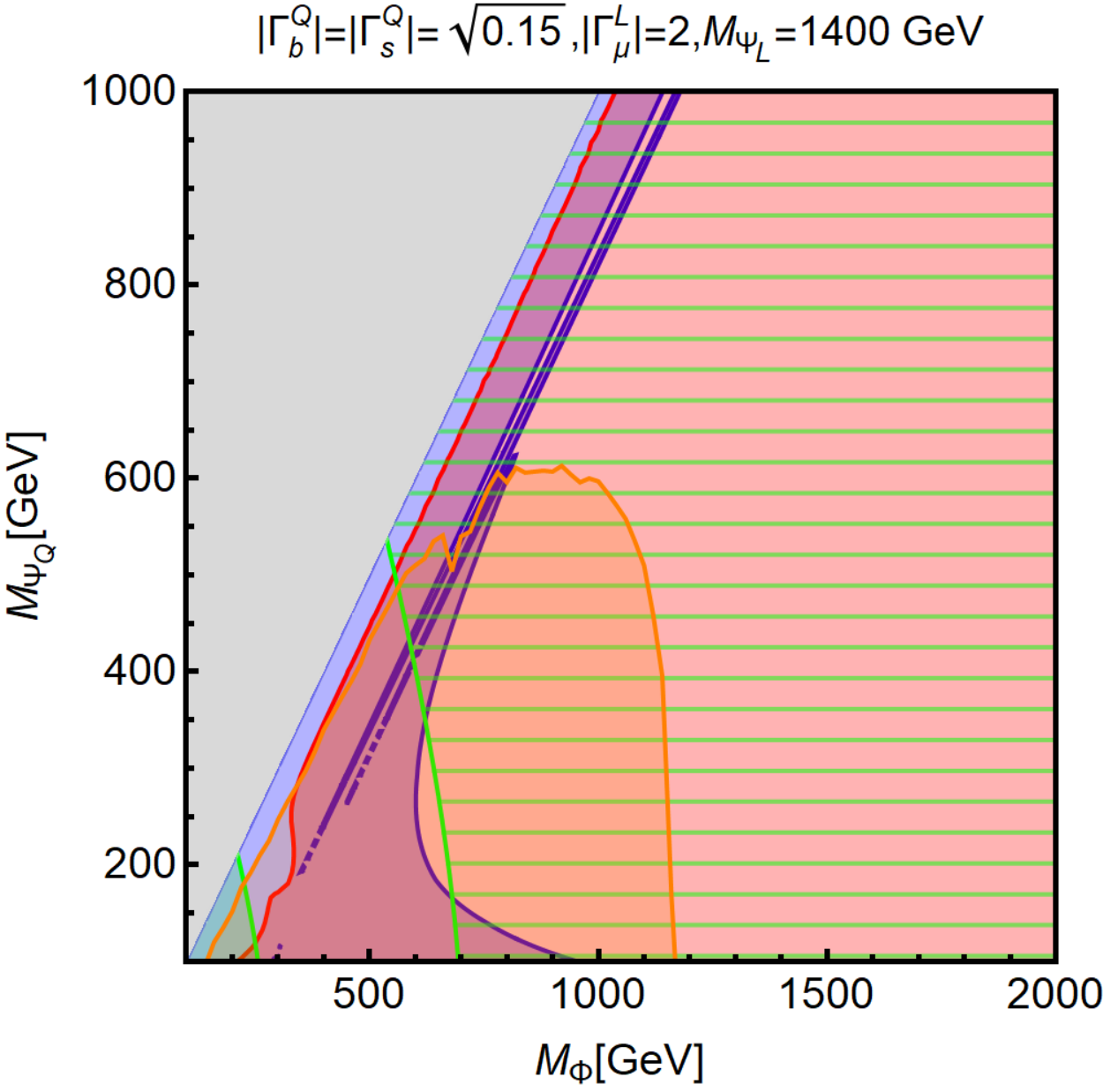}}
\subfloat{\includegraphics[width=0.33\linewidth]{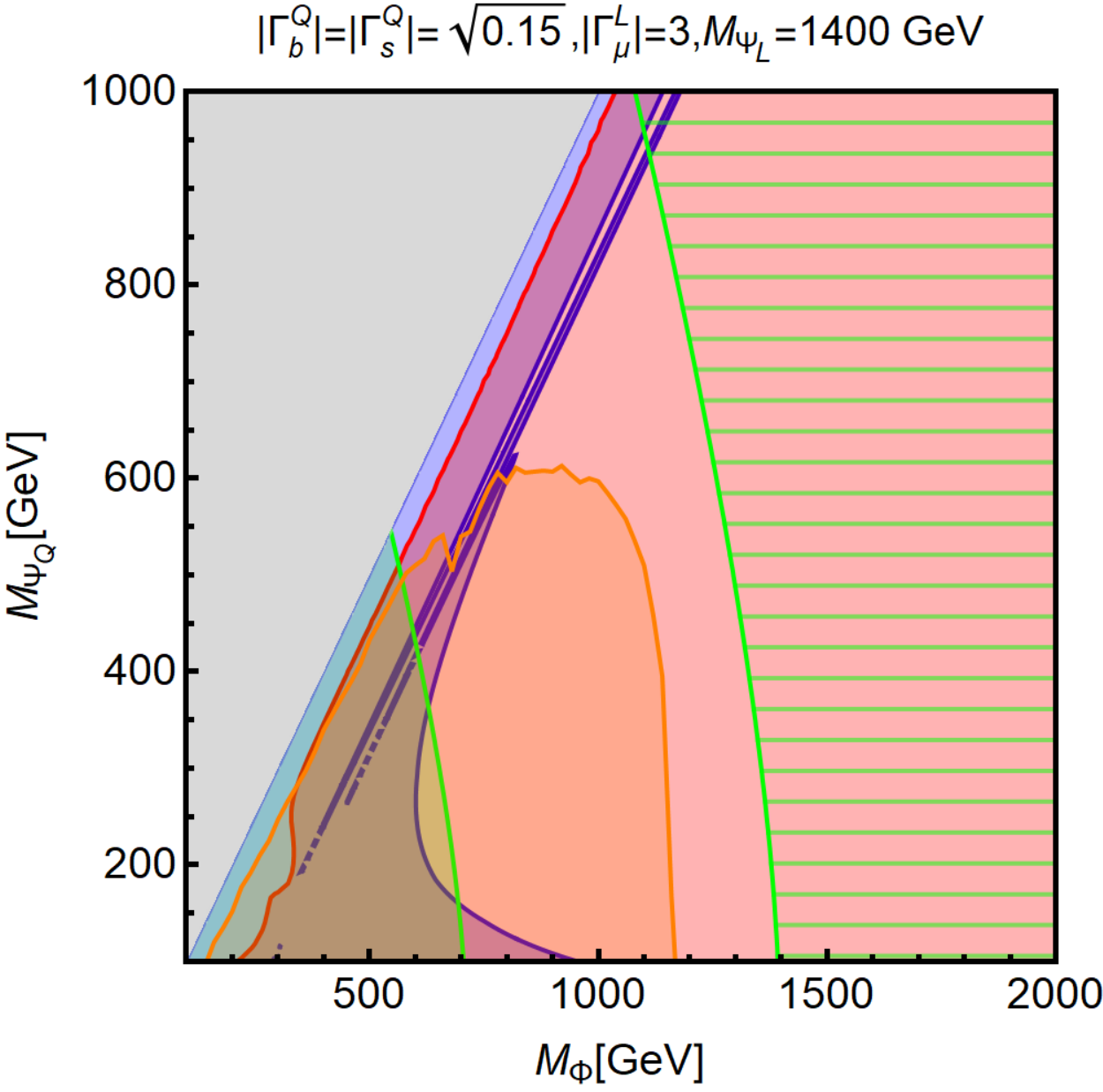}}
\caption{Same as Figure~\ref{fig:S_IIB_D_all} but for $\mathcal{S}_\text{IIB}$ with Majorana DM.}
\label{fig:S_IIB_M_all}
\end{figure}

As mentioned above, in the $\mathcal{S}_\text{IIB}$ models, all the phenomenological constraints are independent on the $M_{\Psi_L}$ and $\Gamma_\mu^L$ parameters (as long as $\Psi_L$ is heavy enough to evade searches for cascade decays as in Eq.~\eqref{eq:mumuqq_decaychannel}) with the exception of the region favoured by $B$-physics. 
Moreover we have found that the latter changes only marginally for different values of $M_{\Psi_L}$. 
Hence we have set for all plots $M_{\Psi_L}=1400\,\mbox{GeV}$, a value that allows to evade the constraint on $pp \to \Psi_L\Psi_L \to \mu^+\mu^-+\,\Phi\Phi \to  \mu^+\mu^-+ qq^\prime +  \slashed{E}_T$ of the searches  in Refs.~\cite{Sirunyan:2020tyy,ATLAS:2020dav}.
Figure~\ref{fig:S_IIB_D_all} shows that, mainly due to the combined effect of relic density and DD constraints, this model is ruled out way beyond the region favoured by the flavour anomalies.
One could possibly overcome this problem by assuming a non-standard cosmological history of the early universe, see e.g.~\cite{Gelmini:2006pq,Gelmini:2006pw,Arcadi:2011ev,Drees:2017iod,Arias:2019uol}, such that the DM abundance is diluted to the extent
that the region compatible with the flavour anomalies in the last plot of Figure~\ref{fig:S_IIB_D_all} becomes partly viable.


\subsection{\texorpdfstring{$\mathcal{S}_\text{IIB}$}{SIIB}, Majorana singlet DM} 
Here, we consider a variant of the previous model featuring Majorana rather than Dirac DM. As shown in Figure~\ref{fig:S_IIB_M_all}, this time moving from Dirac to Majorana DM does not open new viable regions of the parameter space.
The Majorana nature of the DM particle eliminates the $Z$-penguin contribution to the scattering with nuclei discussed in the previous subsection. This
noticeably relaxes the DD constraints for the case of large couplings to the top as in our scenario \emph{i}), shown in the first line of Figure~\ref{fig:S_IIB_M_all}. However, irreducible QCD contributions to DD still impact the parameter space even for Majorana DM.
Furthermore, the region compatible with the flavour constraints typically correspond to overabundance of DM. In the Majorana case, the DM annihilation cross-section is even more suppressed, because of velocity dependence, with respect to the case of Dirac DM. As a result, this model is still not viable unless a non-standard cosmology provides additional DM dilution. As already pointed out, this outcome follows from to the fact that the DM is coupled only with one NP state. This is the main difference with e.g.~model $\mathcal{F}_\text{IA;\,0}$ that, as we showed in Section~\ref{sec:FIA_M}, easily fulfills all constraints in the case of Majorana DM.


\subsection{\texorpdfstring{$\mathcal{S}_\text{IIIA;\,-1/2}$}{SIIIA}, Majorana triplet DM}

To conclude our overview of scenarios with distinct phenomenology, we illustrate two models with DM belonging to a $SU(2)_L$ triplet. In fact notice that,
for all models in Tables~\ref{tab:fmodels} and~\ref{tab:smodels} featuring a complete singlet, gauge invariance allows to substitute to the singlet an $SU(2)_L$ triplet with zero hypercharge. As we will see below, the consequent change in the phenomenology of our models is dramatic.

We start with the model $\mathcal{S}_\text{IIIA;\,-1/2}$,
whose DM candidate is part of the fermion triplet $\Psi_L$, $({\bf 1},\,{\bf 3},\,0)$.
The field $\Psi_Q$ also transforms as an $SU(2)_L$ triplet, as well as a colour triplet: $({\bf 3},\,{\bf 3},\,2/3)$. The mediator is a complex scalar doublet $({\bf 1},\,{\bf 2},\,-1/2)$. The Lagrangian reads:
\begin{eqnarray}
{\cal L}_{\rm int} =  {\Gamma^Q_{i}\bar Q_{i}}{P_R}(\tau^a{\Psi_Q^a}){\Phi} + \Gamma_\mu^L \bar L_\mu {P_R}(\tau^a{\Psi_L^a}){\Phi} +
{\rm{h}}{\rm{.c.}}\,,
\label{eq:L_IA_PHIb}
\end{eqnarray}
 As mentioned above, the DM candidate is the neutral component of the Majorana field $\Psi_L$.
\begin{figure}[!t]
\centering
\subfloat{\includegraphics[width=0.33\linewidth]{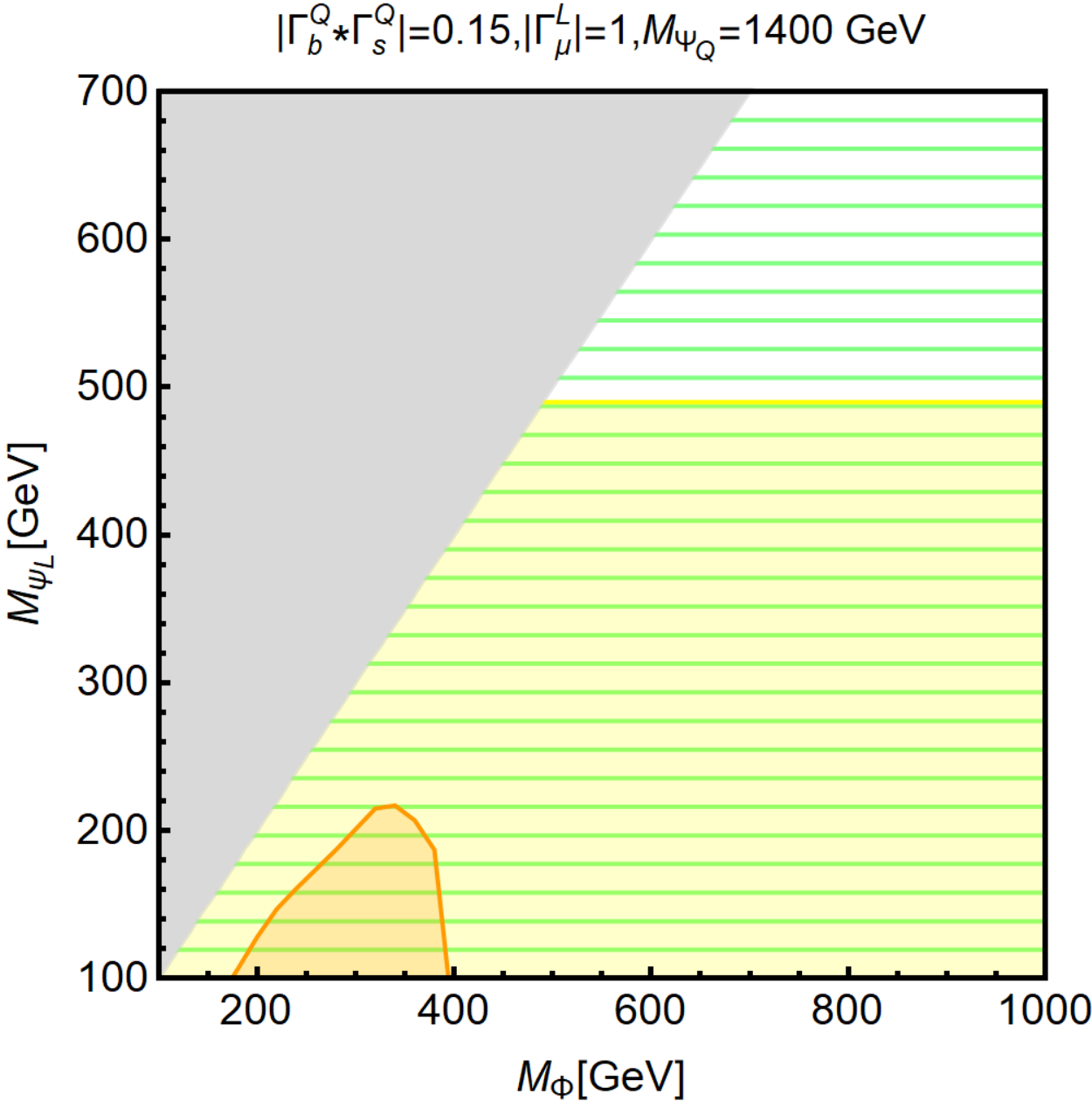}}
\subfloat{\includegraphics[width=0.33\linewidth]{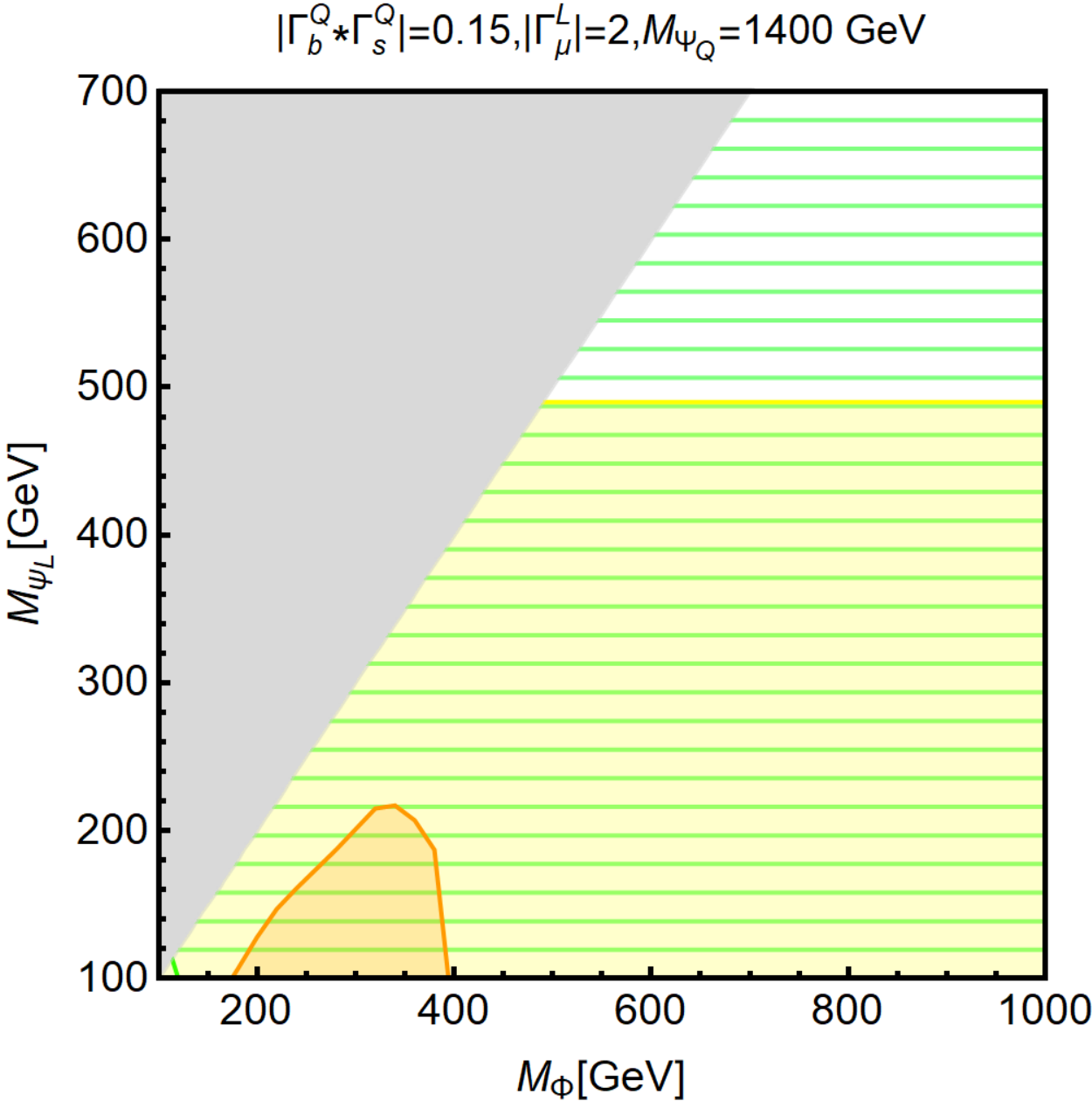}}
\subfloat{\includegraphics[width=0.33\linewidth]{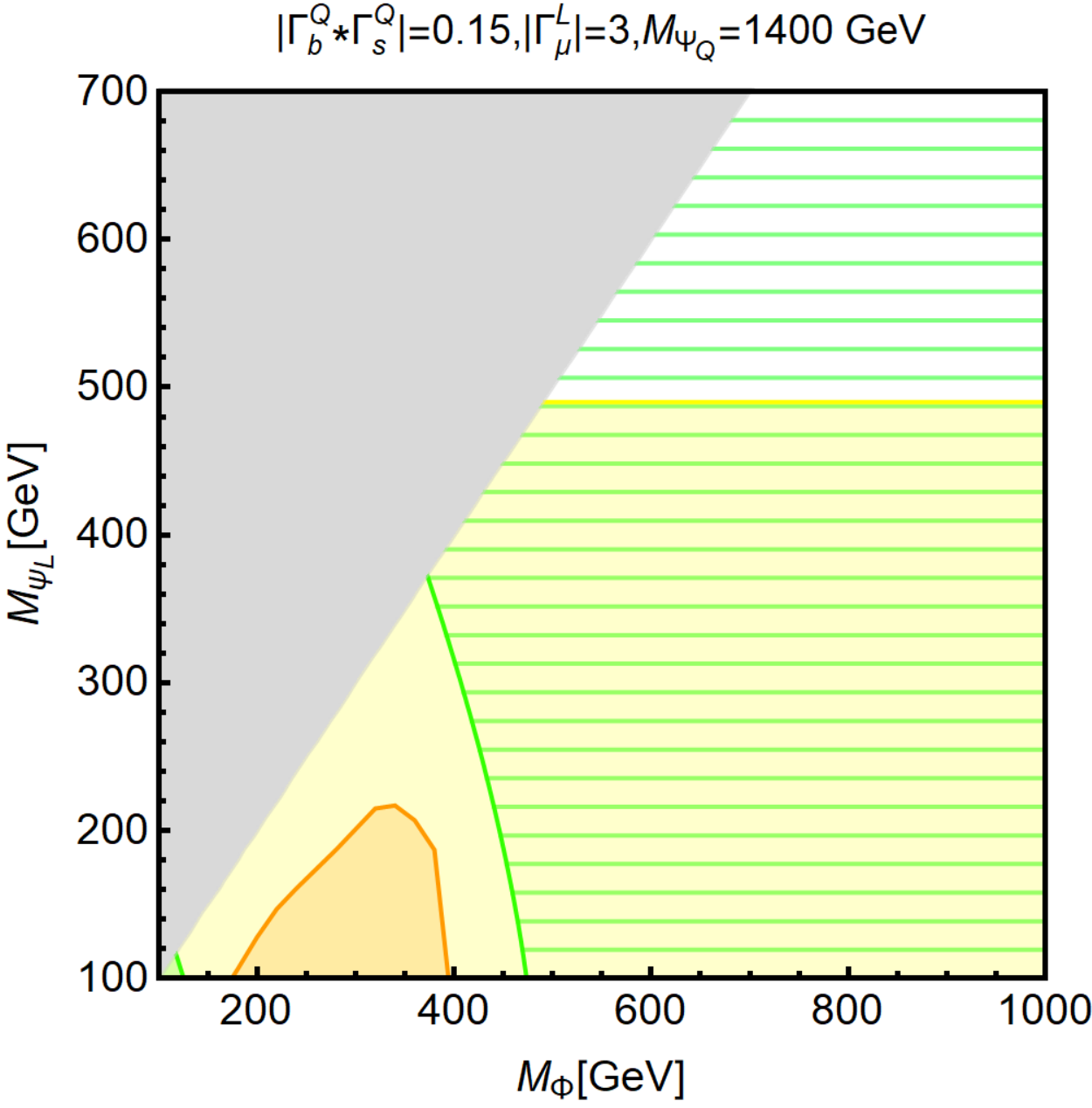}}
\caption{Summary of the constraints for the model $\mathcal{S}_\text{IIIA;\,-1/2}$ with triplet Majorana DM. We have considered the values $1,2,3$ for the $|\Gamma_\mu^Q|$ coupling. Contrary to the other models, DM constraints do not depend on the individual values of the $\Gamma_s^Q$ and $\Gamma_b^Q$ couplings, hence have not made the usual distinction between the configurations \emph{i}) and \emph{ii}). The colour scheme is as defined in Figure~\ref{fig:F_IA_D}. In addition, the yellow region is excluded by LHC searches of disappearing tracks, see the text for details.}
\label{fig:S_IIIA_M_all}
\end{figure}

The combined constraints on this model are shown in Figure~\ref{fig:S_IIIA_M_all}. We consider again the single assignment $M_{\Psi_Q}=1400\,\mbox{GeV}$, as our usual benchmark values $700,\,1100\,\mbox{GeV}$ are mostly excluded by recasting the LHC searches in Refs.~\cite{Sirunyan:2020tyy,ATLAS:2020dav} in terms of the process $pp \to \Psi_Q\Psi_Q \to qq^\prime +\,\Phi\Phi \to  qq^\prime +  \mu^+\mu^- + \slashed{E}_T$. 

The $\mathcal{S}_\text{IIIA;\,-1/2}$ model features the weakest correlation among flavour/LHC and DM observables. As we can see from the Lagrangian, the DM is coupled only with the colour singlet $\Phi$. Being in addition a Majorana fermion, the only contribution to SI interactions comes from loop diagrams involving the charged components of the DM multiplet $\Psi_L$ as well as the $W,Z$ bosons~\cite{Hisano:2010fy,Hisano:2011cs}. This kind of interactions lead to cross-sections still below current experimental sensitivity~\cite{Cirelli:2014dsa,Hisano:2015rsa}. For this reason no DD exclusion region appears in Figure~\ref{fig:S_IIIA_M_all}. For what concerns the relic density, belonging the DM candidate to a triplet, it features a very efficient and possibly Sommerfeld-enhanced annihilation cross-section into gauge boson pairs (cf.~e.g.~the third diagram in Figure~\ref{fig:DMann}) such that the CMB bound $\Omega_\text{DM}h^2\simeq 0.12$ is saturated only for DM masses of the order of 3~TeV, far from the region compatible with the fit of flavour observables. In the regime shown in the plots, the DM is always underabundant (unless some non-thermal production mechanism is assumed), irrespective of the values of the masses and couplings of the NP fields and, hence, no relic density exclusion appears.  
From Figure~\ref{fig:S_IIIA_M_all}, we can also see that the region excluded by LHC searches for events featuring missing energy (in this case $pp \to \Phi\Phi \to \mu^+\mu^- +  \slashed{E}_T$~\cite{Aad:2019vnb}) is not very pronounced. 
However, compared to the previous models, these plots feature a new type of excluded region, filled in yellow. This bound corresponds to the negative results from LHC searches~\cite{Aaboud:2017mpt,Sirunyan:2020pjd} for the disappearing charged tracks that, in the model under consideration, would be associated to the pair production of the electrically charged components of the electroweak multiplet the DM belongs to, see the related discussion in Section~\ref{sec:lhc}. In the minimal setup considered in this work, the different states composing the DM multiplet have loop-suppressed $\mathcal{O}(100)$~MeV mass splitting determined by electroweak gauge interactions. As a consequence, the charged DM partner is long-lived and decays into final state particles which are too soft to be detected or would, hence leading to disappearing track events rather than prompt jets or leptons and missing energy events. For the $\mathcal{S}_\text{IIIA;\,-1/2}$ model we can directly apply the disappearing-track bound obtained for the case of a supersymmetric Wino~\cite{Sirunyan:2020pjd} which translates into a limit of the triplet mass $M_{\Psi_L}\ge490$~GeV. As we can see, this latter bound completely covers the region fitting the flavour anomalies. 


\subsection{\texorpdfstring{$\mathcal{F}_\text{IIIA;\,-1/2}$}{FIIIA}, Real-scalar triplet DM}

\begin{figure}[!t]
\centering
\subfloat{\includegraphics[width=0.33\linewidth]{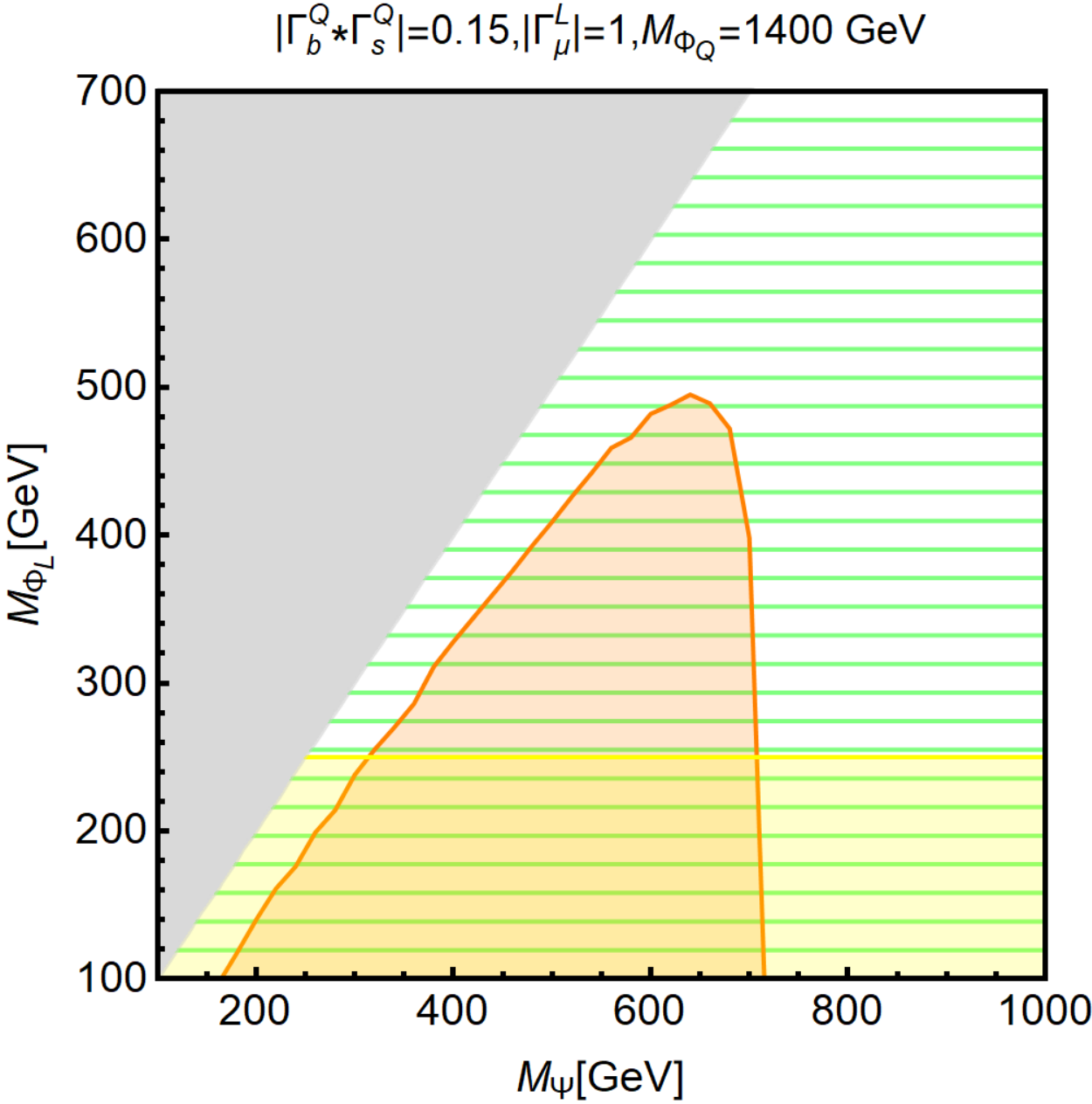}}
\subfloat{\includegraphics[width=0.33\linewidth]{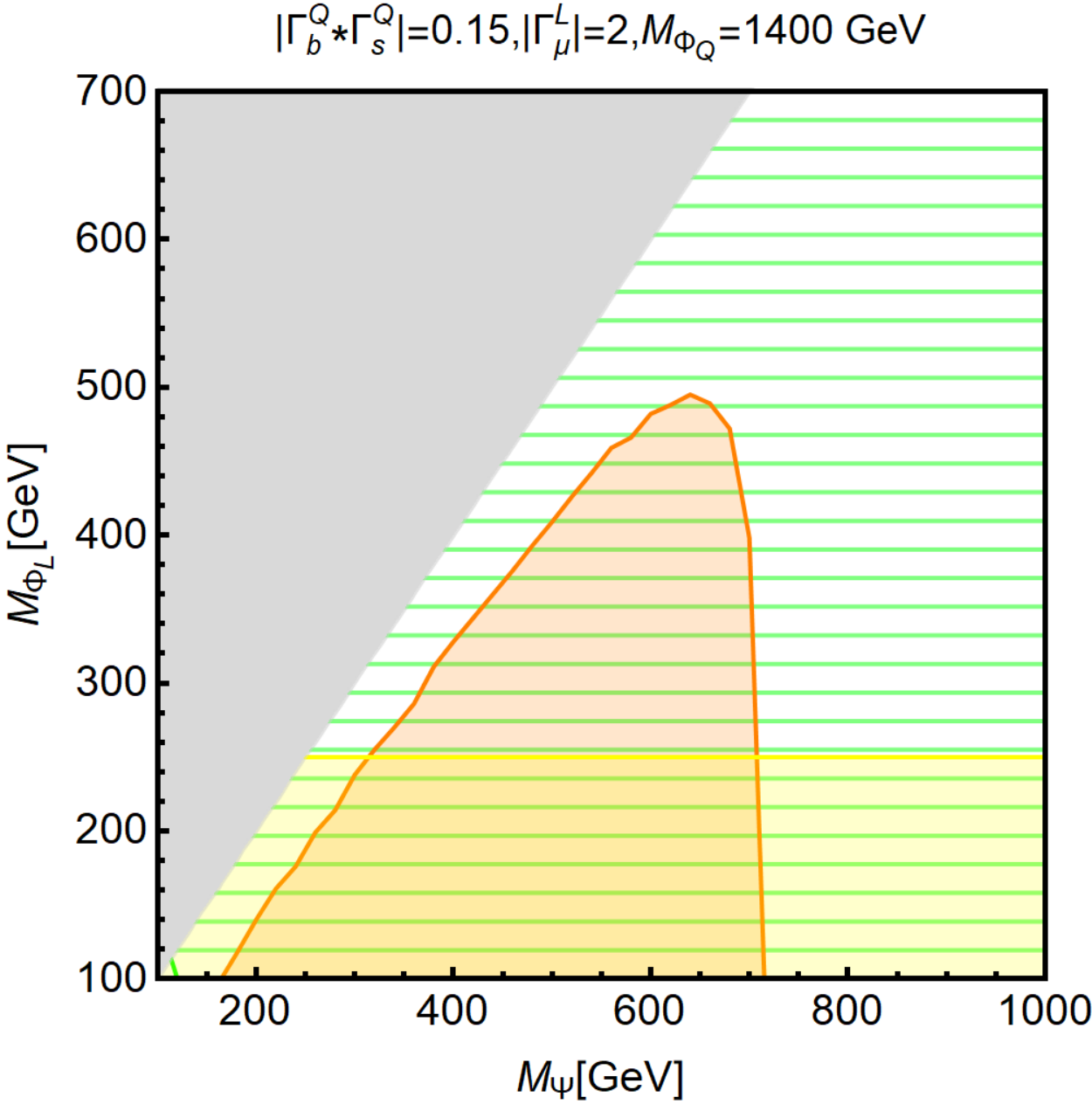}}
\subfloat{\includegraphics[width=0.33\linewidth]{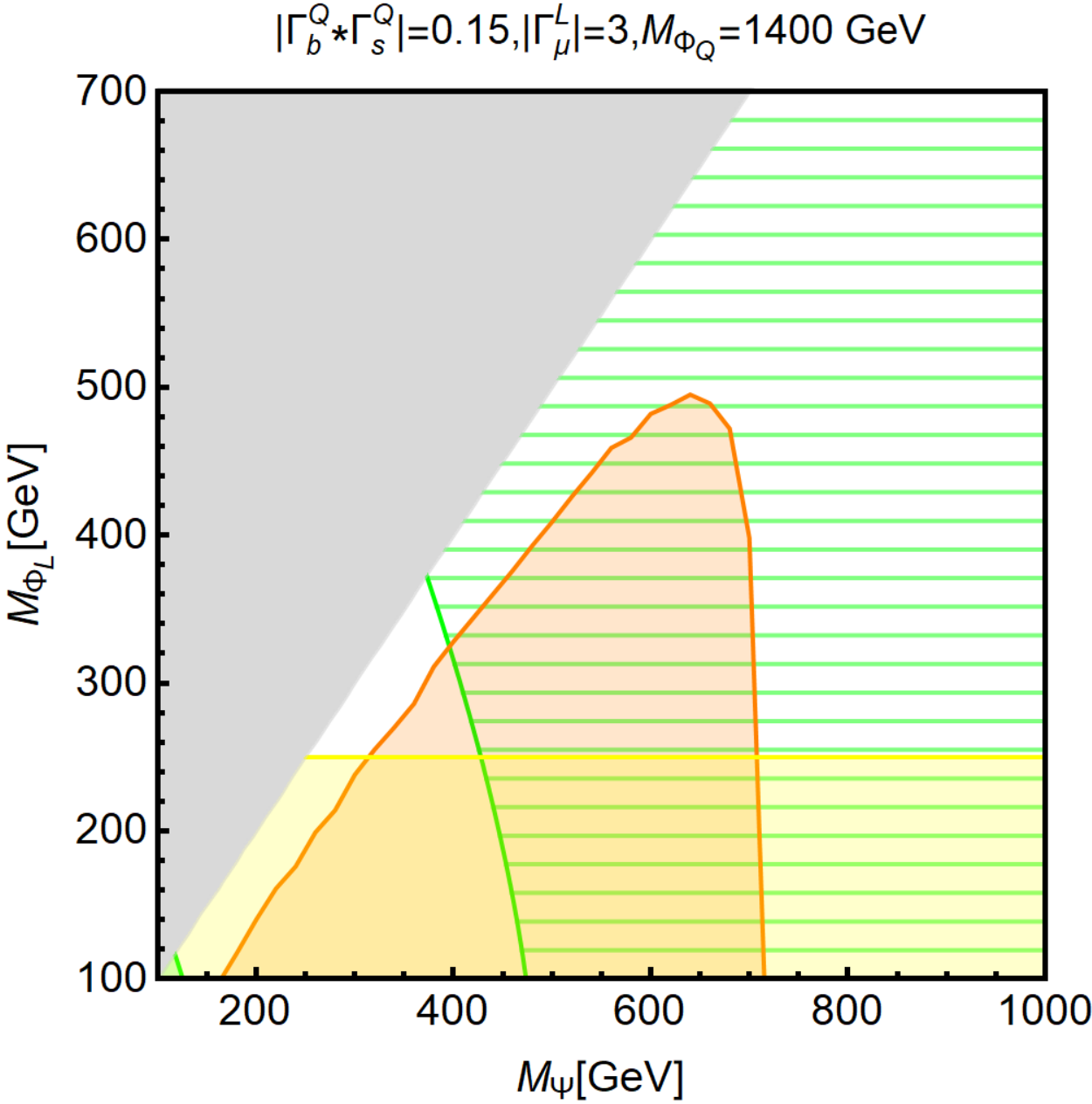}}
\caption{Summary of the constraints for the $\mathcal{F}_\text{IIIA;\,-1/2}$ model with scalar triplet DM. Similarly to $\mathcal{S}_\text{IIIA;\,-1/2}$, the model is constrained only by flavour and LHC and the constraints do not depend on the individual values of $\Gamma_s^Q$, $\Gamma_b^Q$. The colour scheme is as defined in Figure~\ref{fig:F_IA_D}. The yellow region is excluded by LHC searches of   disappearing tracks.}
\label{fig:F_IIIA_all}
\end{figure}

The last model we consider is analogous to the previous one with the role of fermion and scalar fields reversed. It is described by the following Lagrangian:
\begin{eqnarray}
{\cal L}_{\rm int} =  {\Gamma^Q_{i}\bar Q_{i}}{P_R}(\tau^a{\Phi_Q^a}){\Psi} + \Gamma_\mu \bar L_\mu {P_R}(\tau^a{\Phi_L^a}){\Psi} +
{\rm{h}}{\rm{.c.}}\,,
\label{eq:L_FIIIA_PHI}
\end{eqnarray}
and the  $SU\!\left(3 \right)$, $SU\!\left( 2 \right)_L$ and $U{{\left( 1 \right)}_Y}$ quantum numbers of the NP fields are $({\bf 3},\,{\bf 3},\,2/3)$, $({\bf 1},\,{\bf 3},\,0)$ and $({\bf 1},\,{\bf 2},\,-1/2)$ for $\Phi_Q$, $\Phi_L$ and $\Psi$, respectively. The DM candidate is the neutral component of the scalar triplet $\Phi_L$, and the mediator is the fermion doublet $\Psi$.

The effect of the combined constraints on the $\mathcal{F}_\text{IIIA;\,-1/2}$ model is shown in Figure~\ref{fig:F_IIIA_all}.
For analogous reasons to those illustrated in the previous subsection, also in the case of scalar triplet DM, we notice the absence of bounds coming from DM direct detection and relic density. Detailed studies of the DM phenomenology of real scalar triplets have been conducted e.g.~in Refs.~\cite{Chiang:2020rcv,Arakawa:2021vih}. A notable difference with respect to the previous model $\mathcal{S}_\text{IIIA;\,-1/2}$ emerges, on the contrary, for what concerns LHC bounds. Indeed, the bounds from missing energy events (specifically on $pp \to \Psi\Psi \to \mu^+\mu^- + \slashed{E}_T$~\cite{Aad:2019vnb}) impact a larger (orange) region of the parameter space, compared to the analogous model with fermionic DM. This is again due to the fact that the Drell-Yan production cross-section of the fermion pair $\Psi$ is substantially larger than the one of a scalar pair with the same mass and the same quantum numbers under the SM gauge group. On the contrary the bound from disappearing tracks 
(here we show the limit as recasted for the case of a scalar triplet in Ref.~\cite{Chiang:2020rcv})
is weaker for scalar DM, again due to the different production cross section. As a consequence, we notice the presence of (narrow) regions of the parameter space compatible with the flavour anomalies, provided that $|\Gamma_\mu^L| \gtrsim 3$.
We expect that these unconstrained regions can be tested employing future LHC data by a combination of searches for disappearing tracks and searches for events with soft leptons and missing energy like those in Refs~\cite{Aad:2019qnd,Sirunyan:2018iwl}.


\section{Summary and conclusions}
\label{sec:conclusions}
In this work, we have presented a systematic study of minimal scenarios providing a viable fit of the observed anomalies in semileptonic $B$-meson decays and simultaneously solving the DM puzzle thanks to a particle candidate that can achieve, through the thermal freeze-out mechanism, a relic density compatible with the experimental determination from CMB anisotropies. In this minimal setup, the SM spectrum is extended by three new states, either two scalars and a fermion, or two fermions and a scalar, coupled, according to gauge invariance, with left-handed muons and quarks of the second and third generation. All these new fields, including the DM candidate, are present in the loop diagrams associated to the NP contributions to the rates of $B$-meson decays, as shown in Figure~\ref{fig:boxes}. This kind of models hence features an interesting connection between flavour and DM physics. To our knowledge, the present work shows for the first time a complete classification of the possible models of this kind which can be elaborated, depending on the quantum numbers of the new fields. The
details of the considered setup are given in Section~\ref{sec:setup} and the outcome of such a classification is summarised in Tables~\ref{tab:fmodels}
and~\ref{tab:smodels}.
In Section~\ref{sec:results}, we have studied in detail a selection of these models encompassing a large variety of scenarios. Among the models we chose, four possible natures for the DM candidate (namely real scalar DM, complex scalar DM, Dirac and Majorana fermionic DM) are represented. Furthermore, our selection includes examples with the DM field being a singlet of $SU(2)_L$, as well as cases of DM belonging to an $SU(2)_L$ doublet or triplet. 
Following the strategy described in Section~\ref{sec:strategy}, for each model we have performed a fit to the $B$-physics anomalies and used the results to define benchmark assignments for the couplings of the new particles with quarks as well as the mass of one of the (non-DM) NP fields. We have then studied, in terms of the remaining parameters, a broad range of constraints: bounds from searches for the new states at the LHC, DM relic density, DM direct detection and, when appropriate, DM indirect detection. 

The results of this analysis have been presented in detail in Section~\ref{sec:results}, and the general lessons that we can extract from it can be summarised as follows.
\begin{itemize}
    \item A good fit to the flavour anomalies is possible if the product of the couplings of the NP fields to bottom and strange quarks is moderate $\Gamma^Q \sim 0.15$ (larger values would be in conflict with constraints from $B_s$ mixing) and consequently the coupling to muons must be rather large $\Gamma^L_\mu \gtrsim 2$ (to fit the anomalies at the 2$\sigma$ level, cf.~Section~\ref{sec:fit}). This has important consequences for DM phenomenology: if DM belongs to one of the two fields coupled to muons, annihilations into muons are very efficient in depleting the DM abundance to (or below) the observed value. Moreover,  
    electroweak penguin diagrams like the one depicted in Figure~\ref{diag} can give a large contribution to the DM-nucleon scattering cross section relevant for DM direct detection. 
    \item As a consequence, in the cases with DM coupling to muons, especially if it belongs to the field ($\Phi$ or $\Psi$) that acts as ``flavour messengers'' in Figure~\ref{fig:boxes}, we observe a high degree of correlation among our observables, namely the couplings of the NP fields to SM fermions simultaneously control DM, flavour, and collider observables (this is the case for instance of models $\mathcal{F}_\text{IA;\,0}$ and $\mathcal{S}_\text{IA}$ featuring singlet DM). Furthermore, the relic density constraints can be easily satisfied in the region of the parameter space that fits the flavour anomalies.
    \item However, strong constraints from DM direct detection would substantially rule out these scenarios, unless DM is a Majorana fermion or a real scalar (cases for which the most relevant DM-nucleon operator vanishes) or it is a complex scalar with a mass splitting $>\mathcal{O}(100)$~keV between its two components (making the scattering inelastic). In fact, model $\mathcal{F}_\text{IA;\,0}$ with Dirac DM is completely excluded (cf.~Figure~\ref{fig:F_IA_D}) while it is among the most favourable scenarios from an experimental perspective if DM is Majorana, see~Figures~\ref{fig:F_IA_M_A_all},~\ref{fig:F_IA_M_B_all}. Similarly, $\mathcal{S}_\text{IA}$ is a viable option and fits well both DM and the flavour anomalies only if the above-mentioned mass splitting is assumed, cf.~Figures~\ref{fig:S_IA_A_all},~\ref{fig:S_IA_B_all}.
    Interestingly, the viable regions of the parameter space of these models are already partially constrained by LHC searches for jets/muons and missing energy and by direct detection, hence they have good prospects of being tested by next generation detectors like XENONnT~\cite{Aprile:2020vtw}
    and future runs of the LHC.
    \item
    For models where DM is still a singlet but couples only to quarks (such as in the example $\mathcal{S}_\text{IIB}$), DM annihilation is typically not efficient enough and the fit of the flavour anomalies points toward regions of the parameter space where DM is overproduced. These cases are then typically excluded by the relic density constraint independent of whether DM is a Dirac or a Majorana field,~cf.~Figure~\ref{fig:S_IIB_D_all} and~\ref{fig:S_IIB_M_all}, but they 
    could be possibly viable within modified cosmological histories of the early universe providing additional DM dilution.
    \item Our analysis also shows that a combined fit of DM and flavour anomalies favours scenarios where DM is a singlet of the SM gauge group. If DM is instead part of an $SU(2)_L$ multiplet the correlation between relic density and flavour observables is lost, as DM annihilation mainly proceeds through gauge interactions, thus independently on the couplings with quarks and leptons.
    Furthermore, one should rely on a non-thermal DM production mechanism since, in the regions of parameter space where a viable fit of the $B$-anomalies is achieved, the DM is always underabundant in light of its very efficient annihilations into gauge bosons, see models $\mathcal{F}_\text{IB;\,-1/3}$, $\mathcal{S}_\text{IIIA}$ and $\mathcal{F}_\text{IIIA}$.  The main challenge to this kind of models comes from LHC searches. In particular, in case of DM belonging to an $SU(2)_L$ triplet the interesting signature of disappearing charged tracks
    excludes or drastically restricts the regions of the parameter space compatible with the flavour physics anomalies, see~Figures~\ref{fig:S_IIIA_M_all} and~\ref{fig:F_IIIA_all}.
\end{itemize}

As mentioned in the introduction, the present exercise did not aim at proposing ``realistic'' BSM  scenarios, rather at highlighting the minimal ingredients that a more fundamental theory may need to include if the new physics (possibly) behind the $B$-physics anomalies is indeed related to the DM sector. The above analysis
studied the role and the phenomenological consequences of such minimal building blocks. These scenarios could be easily extended to include more particles and interactions. In particular, additional vectorlike fermions or scalars, mixing through a SM Higgs vev with the fields considered here, would also induce operators involving right-currents that may provide an even better fit to the $b\to s\ell \ell$ data. Similarly, this would introduce couplings to right-handed muons that can realise chirally-enhanced contributions to the muon $g-2$ and thus a natural fit of the observed anomaly, see e.g.~\cite{Calibbi:2018rzv}.
Moreover, a more realistic flavour structure of the couplings (rather than our ad hoc assignment) could be considered, possibly following from some flavour symmetry or other models explaining the observed hierarchies of fermion masses and mixing. Within frameworks of such kind, one could find correlations between our observables and flavour processes in other sectors (e.g.~$s-d$ transitions), and thus additional constraints and handles to test the scenarios we considered.

\paragraph {Acknowledgments.}
The authors wish to thank Mauro Valli and Javier Virto for useful discussions in the early stages of the project.
LC~thanks the University of Barcelona\,---\,where this project was initiated\,---\,for hospitality and financial support.
LC~is partially supported by the National Natural Science Foundation of China under the grant No.~12035008. FM acknowledges financial support from the State Agency for Research of the Spanish Ministry of Science and Innovation through the ``Unit of Excellence Mar\'ia de Maeztu 2020-2023'' award to the Institute of Cosmos Sciences (CEX2019-000918-M) and  from PID2019-105614GB-C21 and  2017-SGR-929 grants. The work of MF is supported by the Deutsche Forschungsgemeinschaft (DFG, German Research Foundation) under grant  396021762 - TRR 257, ``Particle Physics Phenomenology after the Higgs Discovery''.

\FloatBarrier

\bibliography{bibliography}
\bibliographystyle{utphys}

\end{document}